\documentclass[manuscript]{aastex63}

\usepackage{amsmath}
\usepackage{breakcites}
\usepackage{rotating}
\usepackage{algorithm}
\usepackage{algorithmic}
\usepackage{tikz}
\usetikzlibrary{fit,positioning}

\shorttitle{Bayes-based orbital elements estimation in triple hierarchical stellar systems}
\shortauthors{Villegas et al.}

\graphicspath{{./}{figures/}}

\begin{document}

\title{Bayes-based orbital elements estimation in triple hierarchical stellar systems\footnote{Released on June, 10th, 2019}\footnote{Based partially on observations obtained at the Southern
    Astrophysical Research (SOAR) telescope, which is a joint project
    of the Minist\'{e}rio da Ci\^{e}ncia, Tecnologia, e
    Inova\c{c}\~{a}o (MCTI) da Rep\'{u}blica Federativa do Brasil, the
    U.S. National Optical Astronomy Observatory (NOAO), the University
    of North Carolina at Chapel Hill (UNC), and Michigan State
    University (MSU).}}

\correspondingauthor{Constanza Villegas}
\email{constanza.villegas@ug.uchile.cl}

\author{Constanza Villegas}
\affiliation{Department of Electrical Engineering, Facultad de Ciencias F\'isicas y Matem\'aticas, Universidad de Chile \\
Tupper 2007, Santiago, Santiago, Chile}

\author{Rene A. Mendez}
\affiliation{Department of Astronomy, Facultad de Ciencias F\'isicas y Matem\'aticas, Universidad de Chile \\
Camino El Observatorio 1515, Las Condes, Santiago, Chile}

\author{Jorge F. Silva}
\affiliation{Department of Electrical Engineering, Facultad de Ciencias F\'isicas y Matem\'aticas, Universidad de Chile \\
Tupper 2007, Santiago, Santiago, Chile}

\author{Marcos E. Orchard}
\affiliation{Department of Electrical Engineering, Facultad de Ciencias F\'isicas y Matem\'aticas, Universidad de Chile \\
Tupper 2007, Santiago, Santiago, Chile}

\newcommand{\bigCI}{\mathrel{\text{\scalebox{1.07}{$\perp\mkern-10mu\perp$}}}}


\begin{abstract}
Under certain rather prevalent conditions (driven by dynamical orbital evolution), a hierarchical triple stellar system can be well approximated, from the standpoint of orbital parameter estimation, as two binary star systems combined. Even under this simplifying approximation, the inference of orbital elements is a challenging technical problem because of the high dimensionality of the parameter space, and the complex relationships between those parameters and the observations (astrometry and radial velocity). In this work we propose a new methodology for the study of triple hierarchical systems using a Bayesian Markov-Chain Monte Carlo-based framework. In particular, graphical models are introduced to describe the probabilistic relationship between parameters and observations in a dynamically self-consistent way. As information sources we consider the cases of isolated astrometry, isolated radial velocity, as well as the joint case with both types of measurements. Graphical models provide a novel way of performing a factorization of the joint distribution (of parameter and observations) in terms of conditional independent components (factors), so that the estimation can be performed in a two-stage process that combines different observations sequentially. Our framework is tested against three well-studied benchmark cases of triple systems, where we determine the inner and outer orbital elements, coupled with the mutual inclination of the orbits, and the individual stellar masses, along with posterior probability (density) distributions for all these parameters. Our results are found to be consistent with previous studies. We also provide a mathematical formalism to reduce the dimensionality in the parameter space for triple hierarchical stellar systems in general.
\end{abstract}




\section{Introduction} \label{sec:intro}

Quantitative models of stellar formation, structure, and evolution require precise estimates of the physical properties of stars. This holds also for the empirical validation of fundamental predictions from stellar astrophysics, such as the mass-luminosity relationship \citep{muterspaugh2010phases, kohler2012orbits}. One of the key ingredients in this regard are the stellar masses, which have been traditionally determined by studying the motion of stars that are bound by their mutual gravitational attraction, i.e., binary stars \citep{pourbaix1994trial, czekala2017architecture}. More recently, microlensing events \citep{WyrMan2020}, and circumstelar disks around young stars \citep{Pegueset2021} have become viable and promising methods for mass determination as well. In the case of microlensing events, the mass of the lens can be determined only in limited cases, because it requires a knowledge of  both the source and lens distances, as well as their relative proper motions. The second method relies on the existence of a purely Keplerian disk\footnote{Usually found only on young stars.} (i.e., in a steady-state configuration, and not subject to magneto-hydrodynamical effects), which enables a purely dynamical mass determination. In the case of binary stars, the subject of this paper, a mass determination requires a determination of the so-called orbital elements that completely define the projected orbit in terms of the true intrinsic orbital parameters.

The problem of estimating orbital elements in binaries has been widely studied in the literature, in particular - more recently - with algorithms that involve a Bayesian methodology \citep[e.g.:][]{ford2005quantifying, sahlmann2013astrometric, lucy2014mass, blunt2017orbits, mendez2017orbits, lucy2018binary}. Bayesian approaches have a probabilistic nature, their final objective being a precise approximation of the expected distributions of orbital elements conditioned to the observations. It is important to have an indicator that characterizes the confidence regarding the estimated observational parameters after analyzing the star motion data: the aforementioned distributions capture the uncertainties and all the information that can be inferred from the data. Bayesian orbit fitting has also been found useful to determine the optimal placement of future observations, and thus reduce the uncertainty in the computed distributions \citep{blunt2017orbits}.

Multiple stellar systems (triple, quadruple systems), viewed as the end-point evolution of small, dissolving, star clusters ($N_{\mbox{members}} < 10$), are also important astrophysical laboratories to test stellar formation and evolution at small scales, including the formation, evolution, and survival of exoplanets. Advances in stellar interferometry are now allowing us to measure the astrometric wobble of stars induced by low-mass companions. This information, combined with available high-precision radial velocities (RVs hereafter), permit us to infer important dynamical parameters. For example, there is great interest in determining the relative orbit orientation, because it provides information about the formation and evolution of the stars and planets involved in the system \citep{muterspaugh2010phases, tokovinin2017}. In the case of triple systems in particular, to derive stellar masses, luminosities, and radii, along with determining the system's coplanarity, both visual and RV data are required \citep{muterspaugh2010phases, tokovinin2017}.  Nevertheless, visual-only orbits coupled with parallax measurements, can be used to measure the total mass of the system. 

The so-called hierarchical approximation\footnote{For a precise definition of a hierarchical stellar system see Appendix~\ref{app:triple}.} is in many cases useful when applied to the study of triple or quadruple systems, because it describes the whole system as two binary systems (each with their own time-independent orbital elements) interacting between themselves. In particular, this configuration allows to obtain fully analytic expressions, since they follow Keplerian motion. There are plenty of research works that exploit this approximation and handle the estimation by disconnecting the inner and the outer orbits. They treat each system as a binary case, where they perform the estimation of orbital elements either through the optimization of a suitable chosen merit function, or through a geometric procedure \citep[e.g.,][]{docobo2008methodology, kohler2012orbits,  tokovinin2018dancing}. In general, the long-term stability of the system, or the adequacy of the assumptions embedded in the hierarchical approximation, has to be tested using more detailed dynamical models (for a recent example of this, see \citet{Docoetal2021}),

Various approaches have been proposed, in the context of the hierarchical approximation, to combine astrometry with spectroscopic measurements. For example, \citet{muterspaugh2010phases} combine both data sets and minimize the $\chi^2$ statistic. \citet{czekala2017architecture} determine the parameters using cross-correlation peaks, while \citet{torres2002spectroscopic} use Markov-Chain Monte Carlo (MCMC hereafter), combining RV data with archival astrometry, and assessing convergence using the Gelman-Rubin statistic \citep{gelman1992inference}. Finally, \citet{tokovinin2017}  propose to perform the estimation in consecutive sequential stages, alternating between visual and spectroscopic data from the inner and outer systems, while minimizing the $\chi^2$ statistic. 

To the best of our knowledge, a Bayesian approach has not been explored for considering the combination of RV data and astrometry on hierarchical stellar systems. While Bayesian methods have been widely adopted in exoplanets research \citep[e.g.,:][]{gregory2005bayesian, ford2006bayesian, gregory2009detecting, gregory2010bayesian, retired2010stars, gregory2011bayesian}, they only consider RV measurements. However, relevant information pertaining to several orbital elements is present in both types of data sets \citep{lucy2018binary}, which are thus complementary. In this \textit{combined} scenario the problem of estimating the orbital elements becomes a very challenging technical problem, because of the highly non-linear and intricate inter-relationship between the different orbital \& dynamical parameters and the data, the large dimensionality of the parameter space to properly describe the orbits ($N_{\mbox{params}} > 10$), the increased number of observational sources\footnote{In principle we could have astrometry for the inner and outer systems as well as RV measurements for each of the bodies involved.}, and the presence of observational uncertainty and partially missing data. For these reasons, the physical analysis of these systems require advanced statistical estimation techniques using as much prior information and data as possible: a fertile ground for Bayesian-driven estimation methodologies.

To cope with the challenges mentioned in the previous paragraph, we address the task of estimating the orbital elements in triple hierarchical stellar systems by obtaining the conditional distribution over the full parameter space, where visual and spectroscopic information is taken into account. We propose (generative) models that capture the distribution (probabilistic relationship) between parameters and observations. To accomplish this, we resort to graphical model tools to portrait the probabilistic relationships considering the underlying dynamical model that characterizes a triple hierarchical stellar system. These graphical models provide a novel way to perform the factorization of the joint distribution (of parameter and observations) in terms of conditional independent components (factors). Taking into account these factorizations of the joint distribution, certain probabilistic relationships between parameters and observations get disconnected, and, as a result, the estimation of the posterior distribution can be performed in a multiple-stage process \citep{jordan1998learning}. This process combines different sources of observations sequentially to update the posterior distribution of parameters, given the observations. 

To compute the distributions mentioned above, we adopt the well-known MCMC simulation-based scheme \citep{robert2004monte}. MCMC has been widely exploited to provide an empirical but precise approximation of the posterior distribution when the exact expression is intractable \citep[e.g.,][]{gamerman2006markov, liu2008monte}. In the present work, we develop a new MCMC-based code to compute the orbital configuration of a triple hierarchical stellar system, partially constructed upon the binary case described by \citet{mendez2017orbits} and \citet{claveria2019visual}. As a result of this computation, we can obtain the most likely solution of orbital elements (in this particular work we adopt the so-called Maximum a Posteriori solution, MAP afterwards\footnote{MAP is defined as the value that maximize the joint posterior distribution, for more details see \citet{Gelmanet2013}.}), as well as confidence intervals measured in terms of the variance of the posterior distributions.

The structure of our paper is as follows: in Section~\ref{sec:graphmodels} we introduce the concept of graphical models; Section~\ref{sec:orbitscalc} describes the observational model for triple hierarchical systems, along with our probabilistic modeling and our orbit calculation. Then, Section~\ref{sec:benchmarks} compares our model to three well-studied systems selected from \citet{tokovinin2017} and \citet{tokovinin2018dancing}, in different scenarios that serve as benchmarks. Finally, in Section~\ref{sec:conclusions} we present our summary, conclusions, and outlook. The Appendices give full details about some relevant aspects of our methodology, which are described only succinctly in the main body of the paper. 


\section{Graphical models}
\label{sec:graphmodels}

Before we analyze the interactions between the observations and the parameters, let us introduce some basic concepts from the theory of graphical models. They are a powerful tool that comes from statistics, graph theory, and computer science, which seek to represent statistical relationships and dependencies between variables through a graphical representation. Every node in the graph is a (scalar) random variable and the arcs (connecting lines) capture concrete independence properties of the joint distribution of the problem. Therefore a graph is induced by a joint distribution and its structure encodes key conditional dependencies within the variables. Furthermore, graphs can be directed or undirected depending on the system being graphically represented, and the direction of the arrows represent influence. 

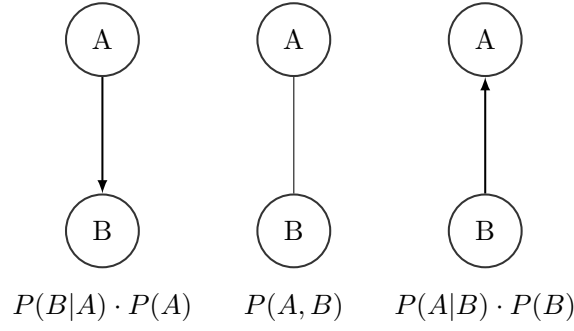
\begin{figure}[htp]
  \centering
        \resizebox{0.45\textwidth}{!}{
        \begin{tikzpicture}
        \tikzstyle{main}=[circle, minimum size = 10mm, thick, draw =black!80, node distance = 16mm]
        \tikzstyle{connect}=[-latex, thick]
        \tikzstyle{box}=[rectangle, draw=black!100]
          \node[main] (A1) [] {A};
          \node[main] (B1) [below=of A1] {B};
          \node[main] (A2) [right=of A1] {A};
          \node[main] (B2) [below=of A2] {B};
          \node[main] (A3) [right=of A2] {A};
          \node[main] (B3) [below=of A3] {B};
          \path 
                (A1) edge [connect] (B1)
                (A2) edge[-] (B2)
        		(B3) edge [connect] (A3)
        		;
        	\node[rectangle, inner sep=2mm, fit= (A1) (B1),label=below : $P(B|A) \cdot P(A)$ ] {};
        	\node[rectangle, inner sep=2mm, fit= (A2) (B2),label=below : {$P(A,B)$} ] {};
        	\node[rectangle, inner sep=2mm, fit= (A3) (B3),label=below : $P(A|B) \cdot P(B)$ ] {};
        \end{tikzpicture}
        }
        \caption{Graphical model representation of the Bayes Theorem. Left and right graphs are directed graphs, while the center graph is an undirected one.}
        \label{fig:graphModelBayes}
\end{figure}

To illustrate, in the context of the Bayes Theorem $\left( P(A | B) = \frac{P(B|A)\cdot P(A)}{P(B)} \right)$, we can encode the prior-to-posterior inference process as the diagram depicted in Figure~\ref{fig:graphModelBayes}, where a joint distribution (center) can be factorized in conditional components (left and right). 

Bayesian networks are a specific configuration of directed and acyclic graphical models, which portray the joint distribution of a set of variables in terms of conditional and prior probabilities. They allow us to simplify the whole distribution in terms of factorization, due to the following basic principles \citep{jordan1998learning, bishop2006pattern}:
\begin{enumerate}
    \item The graph $Z \leftarrow X \rightarrow Y$ means that $Z \bigCI Y | X $, i.e. Z is conditionally independent from Y given X.
    \item Given any node $X_i$, let us denote by  pa$(X_i)$ its parents variables from the directed graph. Then its basic conditional probability (or predictive model) is: $P(X_i|\text{pa}(X_i))$.
    If a particular node $X_i$ has no parents (pa$(X_i)~=~\{\phi\}$), the marginal probability is computed as: $P(X_i|\text{pa}(X_i)) = P(X_i)$.
    \item Given a graph, the joint density over the set of variables $U=\left\{X_i, i=1,..,L \right\}$ follows a recursive factorization:
    \begin{equation}
    P(U) = \prod_{i=1}^L P(X_i|\text{pa}(X_i))
    \end{equation}
    Assuming that a variable X is independent of its non-descendants given its parents : $X \bigCI \text{nd}(X) | \text{pa}(X)$, where $\text{nd}(X)$ is a short-hand for all the variables not contained in $\text{pa}(X)$ and not-including $X$. This is called the directed Markov property. 
\end{enumerate}

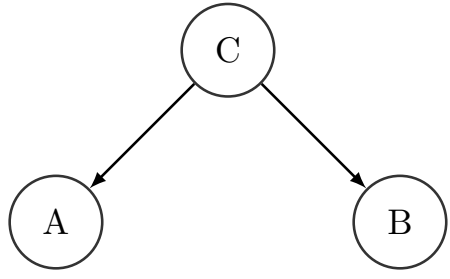
\begin{figure}[htp]
  \centering
        \resizebox{0.35\textwidth}{!}{
        \begin{tikzpicture}
        \tikzstyle{main}=[circle, minimum size = 10mm, thick, draw =black!80, node distance = 16mm]
        \tikzstyle{connect}=[-latex, thick]
        \tikzstyle{box}=[rectangle, draw=black!100]
          \node[main] (C) [] {C};
          \node[main] (A) [below left=of C] {A};
          \node[main] (B) [below right=of C] {B};
          \path 
                (C) edge [connect] (B)
        		(C) edge [connect] (A)
        		;
        \end{tikzpicture}
        }
        \caption{Three-variable Bayesian network.}
        \label{fig:exampleGraphModel}
\end{figure}

Another example can be seen in Figure~\ref{fig:exampleGraphModel}, where a Bayesian network with three variables is shown. In this case, the joint probability is given by $P(A,B,C)~=~P(A|C)\cdot P(B|C) \cdot P(C)$.

Therefore, graphical models (and Bayesian Networks in particular) can answer \textit{conditional probability queries} $P(\Theta~|~Z~=~z)$, where $Z$ is the evidence and $\Theta$ corresponds to some random variables in the network. As $P(\Theta~|~Z=~z)~=~\frac{P(\Theta, z)}{P(z)}$, $P(\Theta, z)$ can be obtained through the factorization in conditional independent components \citep{koller2007graphical}. Then, in this context, this tool provides us a way to conduct the estimation in several processes that combine different observations in a sequential fashion. 


\section{Orbital elements estimation}
\label{sec:orbitscalc}

The final objective in Bayesian inference is the computation of the predictive model of the parameters in the form of probability density functions (PDFs hereafter), $P_{\Theta|Z}(\cdot|z)$. This posterior distribution captures all the information inferred from the data $z$, allowing us to derive estimators of the parameters $\hat{\theta}$ given the evidence. Besides, it takes into account the uncertainty in the estimate, instead of basing the prediction just on the most likely value. The purpose of this section is to present some analysis of the inference problem to simplify the computation of the predictive model $P_{\Theta|Z}(\cdot|z)$. 

\begin{figure}[htp]
\centering
\includegraphics[scale=0.7]{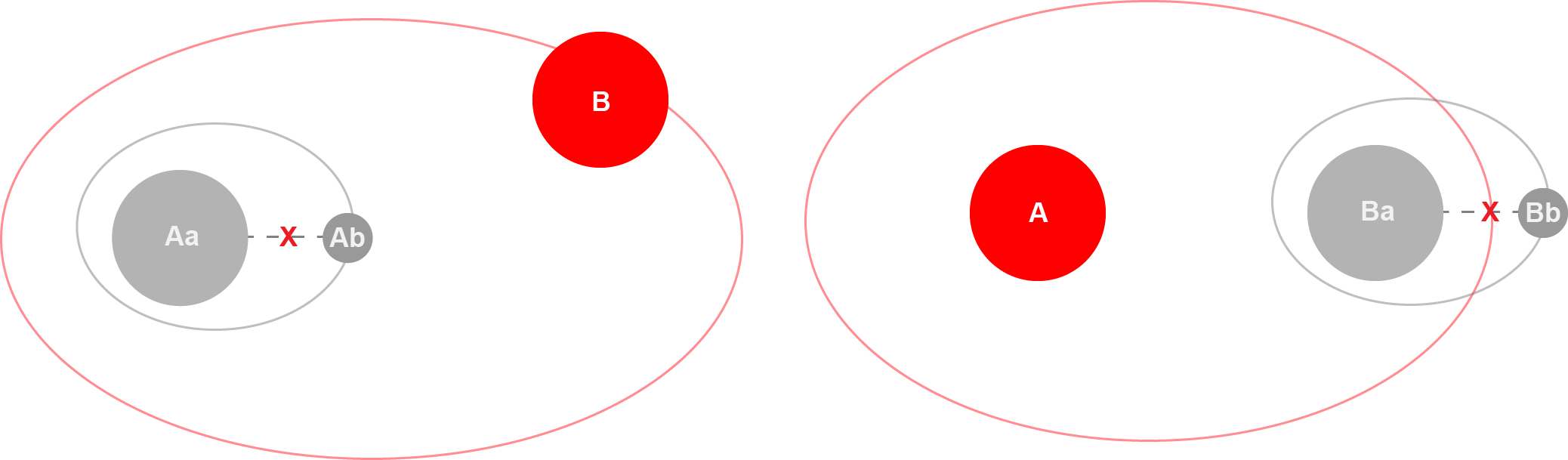}
\caption{Possible configuration of triple hierarchical stellar systems (not to scale). They consist of a tighter binary $A$ (the inner pair $A_{a}$ and $A_{b}$) orbited by a single secondary $B$, as in the left panel (following the classical convention on visual binary research, the A component is the brighter (more massive) star, while B is the the fainter (less massive) component). Of course, it could also be that the inner (secondary) binary ($B_{a}$ and $B_{b}$) orbits a primary ($A$) (right panel).}
\label{fig:tripleSystemVectors}
\end{figure}

Triple hierarchical stellar systems consist of two binary systems bounded together, as the ones illustrated in Figure~\ref{fig:tripleSystemVectors}. For each subsystem, we have access to certain sets of measurements. In all cases, we will assume the observation model from Equation~(\ref{eq:obsmodel}), which maps the $n_{\theta}$-dimensional parameter vector $\theta$ into the $n_{z}$-dimensional measurement vector $z$. Besides, $f(\theta, \tau)$ corresponds to a $n_{z}$-dimensional function, and $\epsilon(\tau)$ is assumed to be an additive white Gaussian noise. 
\begin{equation}
    z(\tau) =   f(\theta, \tau) + \epsilon(\tau)  
\label{eq:obsmodel}    
\end{equation}

In this context, the main practical goal consists of obtaining the relative orbit orientation along with the individual stellar masses, so we need the predictive model of the orbital elements ${a, P, q, i, \Omega}$ from the inner \textit{and} the outer subsystems (see Appendix \ref{sec:relevant_quant} for more details). For both of them, we have access to astrometry and RV measurements; however, they are not always available for both orbits simultaneously. Therefore, based on the type of data that are typically available, the following scenarios have been considered: \textit{astrometric observations alone}, \textit{radial velocity observations alone} and \textit{both sources combined}. Now, we will proceed to discuss each of these cases in detail.

\subsection{Astrometry alone}

In this scenario, we will consider the following notation:
\begin{eqnarray} \label{eq:notatal}
\Theta_1 & = & \{T_{A_{a}A_{b}}, P_{A_{a}A_{b}}, e_{A_{a}A_{b}}, a_{A_{a}A_{b}}, \omega_{A_{a}A_{b}}, \Omega_{A_{a}A_{b}}, i_{A_{a}A_{b}}\} \\
\Theta_2 & = & \{T_{AB}, P_{AB}, e_{AB}, a_{AB}, \omega_{AB}, \Omega_{AB}, i_{AB}\}, \nonumber
\end{eqnarray}
where $A_{a}A_{b}$ and $AB$ denote the classical seven inner and outer orbital elements, respectively, where A stands for the center of mass ($CoM$ hereafter) of the inner pair $A_{a}A_{b}$.

An astrometric observation of the inner system corresponds to the relative position of the secondary $A_b$ with respect to the primary $A_a$, described by a Keplerian orbit, in Cartesian coordinates. On the other hand, it follows the general observation model from Equation~(\ref{eq:obsmodel}), with a highly non-linear function $f_1$ (presented in Appendix \ref{app:as_eq}):
\begin{equation}
\Vec{z}_1(\tau) = f_1(\Theta_1, \tau) + \epsilon_1(\tau).
\label{eq:z_inner}
\end{equation}

On the other hand, an astrometric observation of the outer system corresponds to the relative position of $B$ with respect to the inner primary $A_a$. However, as (the centre of mass of) $A$ and $B$ behave in a Keplerian way, and the primary $A_a$ is moving along with $A_b$, this introduces a \textit{wobble}, leading to the fact that these measurements also depend on the inner orbital elements. The position of $B$ with respect to $A_a$ can be described in Cartesian coordinates as well, and the observations also follow the general observation model from Equation~(\ref{eq:obsmodel}):
\begin{equation}
	\Vec{z}_2(\tau) =f_2(\Theta_1, \Theta_2, q_{A_{a}A_{b}}, \tau) + \epsilon_2(\tau). \label{eq:z_outer1}
\end{equation}

If we define the virtual observations $\vec{y_1}(\tau)$ and $ \vec{y_2}(\tau)$ as:
\begin{eqnarray}
    \vec{y_1}(\tau) &=& \frac{q_{A_{a}A_{b}}}{1 + q_{A_{a}A_{b}}} \cdot f_1(\Theta_1, \tau), \nonumber \\
    \vec{y_2}(\tau) &=& f_1(\Theta_2, \tau). \label{eq:virtualobs}
\end{eqnarray}
Then, given the form of $f_2$ (presented in Appendix \ref{app:as_eq}), the observation equation from Equation~(\ref{eq:z_outer1}) can be written as:
\begin{eqnarray}
	\Vec{z}_2(\tau) &=& \vec{y_1}(\tau) + \vec{y_2}(\tau) + \epsilon_2(\tau).  \label{eq:z_outer2}
\end{eqnarray}

Given the parameters involved in the observation model from Equation~(\ref{eq:z_outer1}), this scenario allows us to compute the mutual inclination and the sum of the stellar masses\footnote{The parallax of the system is required.}, but not the individual stellar masses \citep{lane2014orbits}.

We must note that the expressions derived above are valid only if all of the three stars are resolved, or the Ab component has a negligible flux compared to Aa. In unresolved observations of the AaAb system, we will instead be measuring the photocenter, which is not necessarily coincident with Aa. This requires an adaptation of the method presented here, to include the relative position of the center of light with respect to the center of mass, following, e.g., Equation~1 in Section~3 of \citet{Toko2013}. This kind of scenario is likely to be important in the coming years with all of the unresolved photocenter orbits that Gaia is going to measure, specifically those in hierarchical systems with a resolved distant companion. In a forthcoming paper we will extend our methodology to this case, and provide a few examples on actual systems where we have precisely this situation.

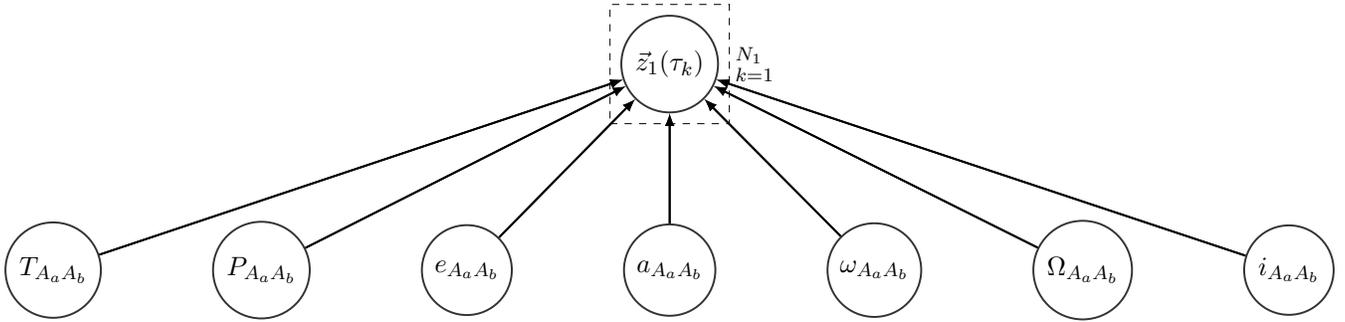
\begin{figure}[htp]
        \resizebox{\textwidth}{!}{
                \begin{tikzpicture}
        \tikzstyle{main}=[circle, minimum size = 10mm, thick, draw =black!80, node distance = 16mm]
        \tikzstyle{connect}=[-latex, thick]
        \tikzstyle{box}=[rectangle, draw=black!100, minimum size = 10mm,]      
          \node[main] (T12) [] {{$T_{A_{a}A_{b}}$}};
          \node[main] (P12) [right=of T12] {{$P_{A_{a}A_{b}}$}};
          \node[main] (e12) [right=of P12] {{$e_{A_{a}A_{b}}$}};
          \node[main] (a12) [right=of e12] {{$a_{A_{a}A_{b}}$}};
          \node[main] (omega12) [right=of a12] {{$\omega_{A_{a}A_{b}}$}};
          \node[main] (Omega12) [right=of omega12] {{$\Omega_{A_{a}A_{b}}$}};
          \node[main] (i12) [right=of Omega12] {{$i_{A_{a}A_{b}}$}}; 
          \node[main] (zi) [above=of a12] { $\Vec{z}_{1}(\tau_k)$};      
          
          \path (T12) edge [connect] (zi);
          \path (P12) edge [connect] (zi);
          \path (e12) edge [connect] (zi);
          \path (a12) edge [connect] (zi);
          \path (omega12) edge [connect] (zi);
          \path (Omega12) edge [connect] (zi);  
          \path (i12) edge [connect] (zi);
          
          \path (T12) edge [connect] (zi);
          \path (P12) edge [connect] (zi);
          \path (e12) edge [connect] (zi);
          \path (a12) edge [connect] (zi);
          \path (omega12) edge [connect] (zi);
          \path (Omega12) edge [connect] (zi);  
          \path (i12) edge [connect] (zi);  
          
		\node[rectangle, inner sep=1mm, fit= (zi), label=right :\textbf{$_{k=1}^{N_1}$}] {};	
          \node[rectangle, inner sep=1.5mm, draw=black!100, dashed, fit = (zi) ] {};                  
        
        \end{tikzpicture}
        }
    \caption{Graphical model representation of the observation model from inner astrometric measurements.}
    \label{fig:graphModelAstrometry:inner}
\end{figure}

\begin{figure}[htp]
        \resizebox{\textwidth}{!}{
                \begin{tikzpicture}
        \tikzstyle{main}=[circle, minimum size = 10mm, thick, draw =black!80, node distance = 16mm]
        \tikzstyle{connect}=[-latex, thick]
        \tikzstyle{box}=[rectangle, draw=black!100]

          \node[main] (T) [] {{$T_{AB}$}};   
          \node[main] (P) [right=of T] {{$P_{AB}$}};
          \node[main] (e) [right=of P] {{$e_{AB}$}};
          \node[main] (a) [right=of e] {{$a_{AB}$}};
          \node[main] (omega) [right=of a] {{$\omega_{AB}$}};
          \node[main] (Omega) [right=of omega] {{$\omega_{AB}$}};
          \node[main] (i) [right=of Omega] {{$i_{AB}$}};    
                
          \node[main] (zouter) [below =of e] {$\vec{y_2}(\tau_k)$};
          \node[main] (zo) [left =of zouter] { $\vec{z_2}(\tau_k)$};  
          \node[main] (zinner) [left =of zo] { $\vec{y_1}(\tau_k)$};  
          
          \node[main] (T12) [below = of zinner] {{$T_{A_{a}A_{b}}$}};
          \node[main] (q12) [left=of T12] {{$q_{A_{a}A_{b}}$}};                 
          \node[main] (P12) [right=of T12] {{$P_{A_{a}A_{b}}$}};
          \node[main] (e12) [right=of P12] {{$e_{A_{a}A_{b}}$}};
          \node[main] (a12) [right=of e12] {{$a_{A_{a}A_{b}}$}};
          \node[main] (omega12) [right=of a12] {{$\omega_{A_{a}A_{b}}$}};
          \node[main] (Omega12) [right=of omega12] {{$\Omega_{A_{a}A_{b}}$}};
          \node[main] (i12) [right=of Omega12] {{$i_{A_{a}A_{b}}$}}; 
          
          \path (T12) edge [connect, draw=red!100, dashed] (zinner);
          \path (P12) edge [connect, draw=red!100, dashed] (zinner);
          \path (e12) edge [connect, draw=red!100, dashed] (zinner);
          \path (a12) edge [connect, draw=red!100, dashed] (zinner);
          \path (omega12) edge [connect, draw=red!100, dashed] (zinner);
          \path (Omega12) edge [connect, draw=red!100, dashed] (zinner);  
          \path (i12) edge [connect, draw=red!100, dashed] (zinner);
          \path (q12) edge [connect, draw=red!100, dashed] (zinner);   
               
          \path (T) edge [connect, draw=red!100, dashed] (zouter);
          \path (P) edge [connect, draw=red!100, dashed] (zouter);
          \path (e) edge [connect, draw=red!100, dashed] (zouter);
          \path (a) edge [connect, draw=red!100, dashed] (zouter);
          \path (omega) edge [connect, draw=red!100, dashed] (zouter);  
          \path (Omega) edge [connect, draw=red!100, dashed] (zouter);    
          \path (i) edge [connect, draw=red!100, dashed] (zouter);     
          \path (omega) edge [connect, draw=red!100, dashed] (zouter);    
          
          \path (zouter) edge [connect] (zo);
          \path (zinner) edge [connect] (zo);          
          
		\node[rectangle, inner sep=1mm, fit= (zo), label=right :\textbf{$_{k=1}^{N_2}$}] {};	
        \node[rectangle, inner sep=1.5mm, draw=black!100, dashed, fit = (zo) ] {};  
        
		\node[rectangle, inner sep=1mm, fit= (zouter), label=right :\textbf{$_{k=1}^{N_2}$}] {};	
        \node[rectangle, inner sep=1.5mm, draw=black!100, dashed, fit = (zouter) ] {};    
        
		\node[rectangle, inner sep=1mm, fit= (zinner), label=left :\textbf{$_{k=1}^{N_2}$}] {};	
        \node[rectangle, inner sep=1.5mm, draw=black!100, dashed, fit = (zinner) ] {};            
                    
        \end{tikzpicture}
        }
    \caption{Graphical model representation of the observation model from outer astrometric measurements, shown in Equation~(\ref{eq:z_outer2}). Dashed red arrows represent deterministic (not probabilistic) relationships. Note that, in the lower row, in addition to the classical orbital elements, we now have the mass ratio of the inner system $q_{A_{a}A_{b}}$.}
    \label{fig:graphModelAstrometry:outer}
\end{figure}
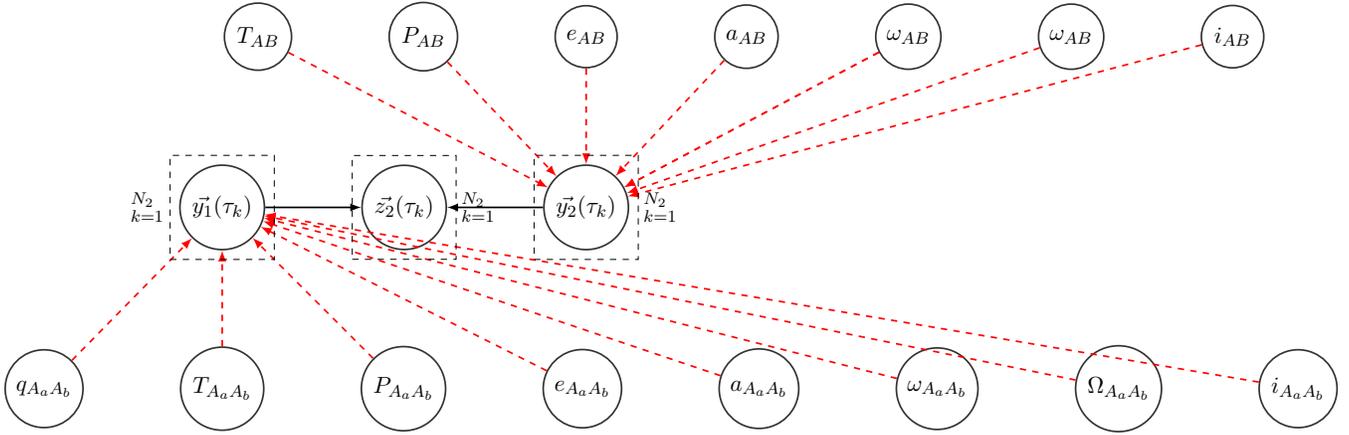

Based on the dependencies observed in Eqs.~(\ref{eq:z_inner}), (\ref{eq:z_outer1}), and (\ref{eq:z_outer2}), graphical models were designed expressing the relationships between parameters and the observations $\vec{z_1}$ (inner orbit) and $\vec{z_2}$ (outer orbit). They are shown in Figures \ref{fig:graphModelAstrometry:inner} and \ref{fig:graphModelAstrometry:outer}, where $N_1$ and $N_2$ are the number of observations of $\vec{z_1}$ and $\vec{z_2}$, respectively. Those networks allows us to factorize the joint distribution in conditional components and, therefore, arrange the inference process for this scenario. 

The graphical model of the inner system is simple, and can be expressed as the factorization in Equation~(\ref{eq:gm_asinner}).
\begin{equation}
    p(\vec{z}_1, \Theta_1) =  p(\vec{z}_1 | \Theta_1) \cdot \prod_{\theta \in \Theta_1} p(\theta). \label{eq:gm_asinner}
\end{equation}

By contrast, the graphical model of the outer system is more complex, because it involves more parameters and includes some virtual observations in it ($\vec{y_1}(\tau)$ and $ \vec{y_2}(\tau)$ from Equation~(\ref{eq:virtualobs})). In this case, the factorization can be written as:
\begin{equation}
    p(\vec{z}_2, \vec{y}_1, \vec{y}_2, \Theta_1, q_{A_{a}A_{b}}, \Theta_2) =  p(\vec{z}_1 | \vec{y}_1, \vec{y}_2) \cdot   p(\vec{y}_1 | \Theta_1, q_{A_{a}A_{b}}) \nonumber \cdot p(\vec{y}_2 | \Theta_2) \cdot p( q_{A_{a}A_{b}}) \cdot \prod_{\theta \in \Theta_1 \cup \Theta_2} p(\theta). \label{eq:gm_asouter1}
\end{equation}

The red arrows in Figure~\ref{fig:graphModelAstrometry:outer} indicate that the conditional distribution is a Dirac function, representing a deterministic relationship, i.e.: 
\begin{equation}
    p(\vec{z}_2, \vec{y}_1, \vec{y}_2, \Theta_1, q_{A_{a}A_{b}}, \Theta_2) =  p(\vec{z}_1 | \vec{y}_1, \vec{y}_2) \cdot   \delta(\vec{y}_1 - \frac{q_{A_{a}A_{b}}}{1 + q_{A_{a}A_{b}}} \cdot f_1(\Theta_1)) \cdot \delta(\vec{y}_2 - f_1(\Theta_2)) \cdot p( q_{A_{a}A_{b}}) \cdot \prod_{\theta \in \Theta_1 \cup \Theta_2} p(\theta). \label{eq:gm_asouter2}
\end{equation}

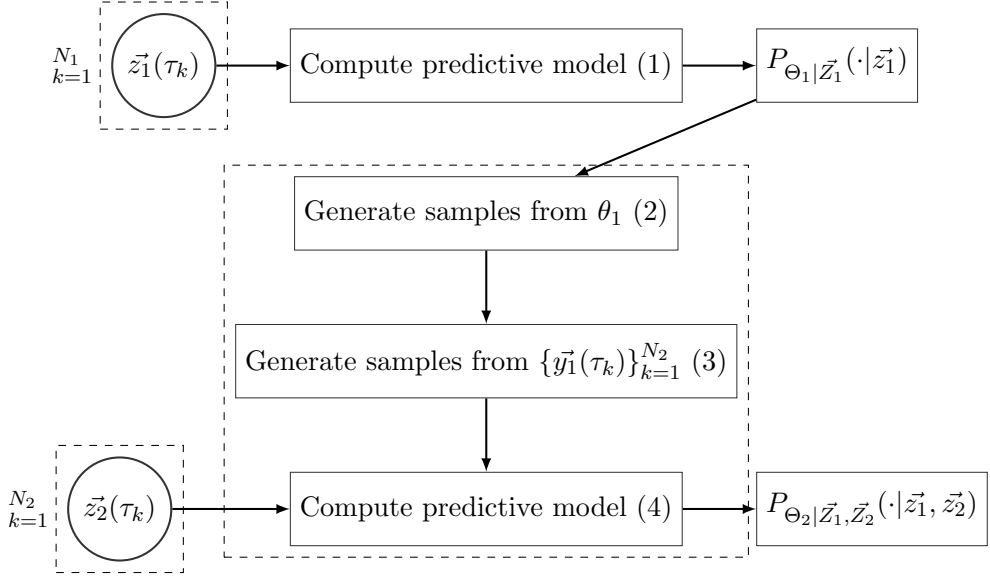
\begin{figure}[htp]
\centering
        \resizebox{0.75 \textwidth}{!}{
                \begin{tikzpicture}
        \tikzstyle{main}=[circle, minimum size = 10mm, thick, draw =black!80, node distance = 16mm]
        \tikzstyle{connect}=[-latex, thick]
        \tikzstyle{box}=[rectangle, draw=black!80, minimum size = 10mm]
          \node[main] (z1) [] { $\vec{z_1}(\tau_k)$};               
          
          \node[box] (cp1) [right = of z1] {Compute predictive model (1)};
          \node[box] (gt) [below = of cp1] {Generate samples from $\theta_1$ (2)};   
          \node[box] (gs) [below = of gt] {Generate samples from $\{\vec{y_1}(\tau_k)\}_{k=1}^{N_2}$ (3)};         
          \node[box] (cp2) [below = of gs] {Compute predictive model (4)};
          
          \node[main] (z2) [left =of cp2] { $\vec{z_2}(\tau_k)$};  
          
          \node[box] (p1) [right = of cp1] {$P_{\Theta_1 | \vec{Z_1}}(\cdot | \vec{z_1})$};
          \node[box] (p2) [right = of cp2] {$P_{\Theta_2 | \vec{Z_1}, \vec{Z_2}}(\cdot | \vec{z_1}, \vec{z_2})$};                            
          
          \path (z1) edge [connect] (cp1);
          \path (cp1) edge [connect] (p1);   
          
          \path (p1) edge [connect] (gt);   
          
          \path (gt) edge [connect] (gs);   
          \path (gs) edge [connect] (cp2);  
          \path (cp2) edge [connect] (p2); 
          
          \path (z2) edge [connect] (cp2);

		 \node[rectangle, inner sep=1mm, fit= (z2), label=left :\textbf{$_{k=1}^{N_2}$}] {};	
          \node[rectangle, inner sep=1.5mm, draw=black!100, dashed, fit = (z2) ] {};     

		 \node[rectangle, inner sep=1mm, fit= (z1), label=left :\textbf{$_{k=1}^{N_1}$}] {};	
          \node[rectangle, inner sep=1.5mm, draw=black!100, dashed, fit = (z1) ] {};    
          
          \node[rectangle, inner sep=1.5mm, draw=black!100, dashed, fit = (gs) (gt) (cp2) ] {};                           
                    
        \end{tikzpicture}
        }
    \caption{Information-flow diagram in the astrometry alone scenario.}
    \label{fig:inference_process_as}
\end{figure}

Finally, the conditional independent structures encoded in our graphical models allow us to perform the inference in a series of sequential steps, illustrated in Figure~\ref{fig:inference_process_as}, as follows:
\begin{enumerate}
\item We compute the predictive model $P_{\Theta_1 | \vec{Z_1}}(\cdot | \vec{z_1})$ using $\{\vec{z_1}(\tau_k)\}_{k=1}^{N_1}$ in a sample-based scheme.
\item We generate empirical samples of $\theta_1$ using the posterior distribution of $\Theta_1$ given $\vec{Z_1}$, i.e., using the model $P_{\Theta_1 | \vec{Z_1}}(\cdot | \vec{z_1})$.
\item We generate virtual observations $\{\vec{y_1}(\tau_k)\}_{k=1}^{N_2}$, using $P_{\Theta_1 | \vec{Z_1}}(\cdot | \vec{z_1})$, for each observation epoch from $\vec{z_2}$, in an imputations framework.
\item We compute the predictive model $P_{\Theta_2 | \vec{Z_1}, \vec{Z_2}}(\cdot | \vec{z_1}, \vec{z_2})$ using virtual observations $\{\vec{y_1}(\tau_k)\}_{k=1}^{N_2}$ and $\{\vec{y_2}(\tau_k)\}_{k=1}^{N_2}$, and observations $\{\vec{z_2}(\tau_k)\}_{k=1}^{N_2}$ in a sample-based scheme. At the end of this stage, we obtain i.i.d. samples of the posterior distributions of the whole set of parameters $\Theta_1 \cup \Theta_2$ given the observations $\{\vec{z_1}(\tau_k)\}_{k=1}^{N_1}$ and $\{\vec{z_2}(\tau_k)\}_{k=1}^{N_2}$.
\end{enumerate}
More details about each subprocess can be found in Appendix \ref{app:mcmc}. It is interesting to note that there appears an explicit dependency on $q_{A_{a}A_{b}}$ due to the wobble effect (opening thus the possibility of individual mass determination, even in the absence of RV data), but there is no dependency on $q_{AB}$ (as is the case of classical binary systems, where we can only determine the overall mass sum of the system, but not the mass ratio).

\subsection{Radial velocity alone}

In this scenario, we will consider the following notation:
\begin{align}
	&\Theta_1 = \{T_{A_{a}A_{b}}, P_{A_{a}A_{b}}, e_{A_{a}A_{b}}, a_{A_{a}A_{b}}, \omega_{A_{a}A_{b}}, i_{A_{a}A_{b}}, q_{A_{a}A_{b}}\} \\
	&\Theta_2 = \{T_{AB}, P_{AB}, e_{AB}, a_{AB}, \omega_{AB}, i_{AB}, q_{AB}, v_{CoM}\}, \nonumber
\end{align}
where the symbols are the same as in Equation~(\ref{eq:notatal}), except that now there is no dependency of the observations on $\Omega_{A_{a}A_{b}}$ nor $\Omega_{AB}$, but there appears a dependency on the overall mas ratio $ q_{AB}$, and on the systemic velocity $v_{CoM}$.

RV observations correspond to the velocity of one of the bodies involved in the triple hierarchical configuration, measured along the observer's line-of-sight. Measurements from $A_a$ are necessary; however, measurements of $A_b$ and $B$ are included if available. The RV measurements from $B$ only depends on outer parameters; but, as the $CoM$ of $A$ moves in the outer orbit, the RVs from $A_a$ and $A_b$  consider this movement and depend on inner and outer parameters. The observation model follows the general form from Equation~(\ref{eq:obsmodel}), with highly non-linear functions $f_i$ (presented in full detail in Appendix~\ref{app:rv_eq}):
\begin{align}
	{z}_3(\tau) =   &f_3(\Theta_1, \Theta_2, \tau) + \epsilon_3(\tau), \text{$A_a$'s RV measurements}, \nonumber \\
	{z}_4(\tau) =   &f_4(\Theta_1, \Theta_2, \tau) + \epsilon_4(\tau), \text{$A_b$'s RV measurements}, \nonumber \\	
	{z}_5(\tau) =   &f_5(\Theta_2, \tau) + \epsilon_5(\tau), \text{$B$'s RV measurements}. \label{eq:rv} 	
\end{align}

Given the parameters involved in the observation model, this scenario allows us to compute the individual stellar masses \footnote{The parallax of the system is required.}, but not the mutual inclination. 

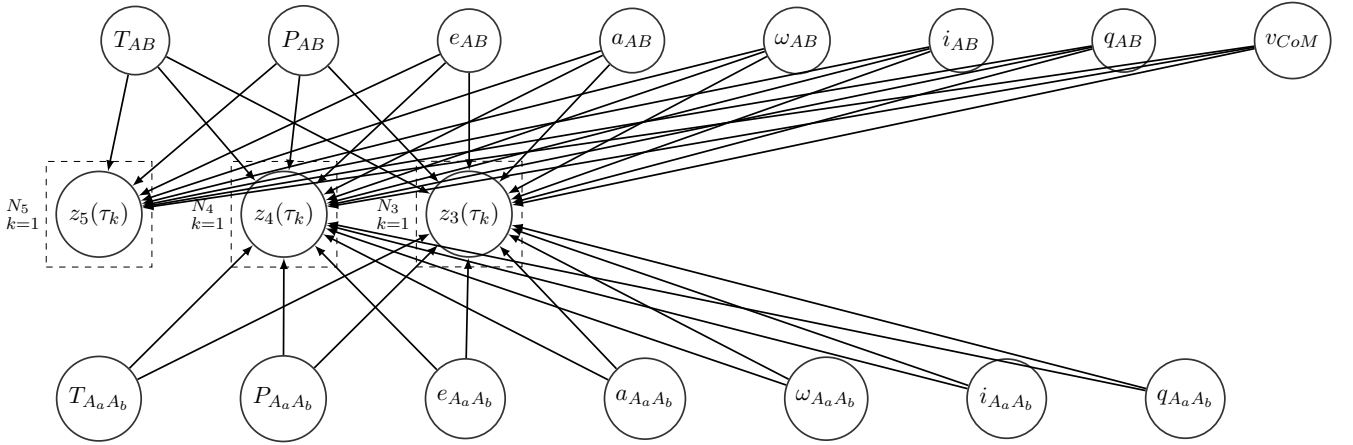
\begin{figure}[htp]
        \resizebox{\textwidth}{!}{
                \begin{tikzpicture}
        \tikzstyle{main}=[circle, minimum size = 10mm, thick, draw =black!80, node distance = 16mm]
        \tikzstyle{connect}=[-latex, thick]
        \tikzstyle{box}=[rectangle, draw=black!100]

          \node[main] (T) [] {{$T_{AB}$}};   
          \node[main] (P) [right=of T] {{$P_{AB}$}};
          \node[main] (e) [right=of P] {{$e_{AB}$}};
          \node[main] (a) [right=of e] {{$a_{AB}$}};
          \node[main] (omega) [right=of a] {{$\omega_{AB}$}};
          \node[main] (i) [right=of omega] {{$i_{AB}$}};    
          \node[main] (q) [right=of i] {{$q_{AB}$}};        
          \node[main] (vCoM) [right=of q] {{$v_{CoM}$}};                        
                
          \node[main] (zaa) [below =of e] {${z_3}(\tau_k)$};
          \node[main] (zab) [left =of zouter] { ${z_4}(\tau_k)$};  
          \node[main] (zB) [left =of zo] { ${z_5}(\tau_k)$};  
          
          \node[main] (T12) [below = of zinner] {{$T_{A_{a}A_{b}}$}};          
          \node[main] (P12) [right=of T12] {{$P_{A_{a}A_{b}}$}};
          \node[main] (e12) [right=of P12] {{$e_{A_{a}A_{b}}$}};
          \node[main] (a12) [right=of e12] {{$a_{A_{a}A_{b}}$}};
          \node[main] (omega12) [right=of a12] {{$\omega_{A_{a}A_{b}}$}};
          \node[main] (i12) [right=of omega12] {{$i_{A_{a}A_{b}}$}}; 
          \node[main] (q12) [right=of i12] {{$q_{A_{a}A_{b}}$}};       
          
          \path (T12) edge [connect] (zaa);
          \path (P12) edge [connect] (zaa);
          \path (e12) edge [connect] (zaa);
          \path (a12) edge [connect] (zaa);
          \path (omega12) edge [connect] (zaa);
          \path (i12) edge [connect] (zaa);
          \path (q12) edge [connect] (zaa);   
               
          \path (T) edge [connect] (zaa);
          \path (P) edge [connect] (zaa);
          \path (e) edge [connect] (zaa);
          \path (a) edge [connect] (zaa);
          \path (omega) edge [connect] (zaa);    
          \path (i) edge [connect] (zaa);     
          \path (q) edge [connect] (zaa);  
          \path (vCoM) edge [connect] (zaa);        
          
          \path (T12) edge [connect] (zab);
          \path (P12) edge [connect] (zab);
          \path (e12) edge [connect] (zab);
          \path (a12) edge [connect] (zab);
          \path (omega12) edge [connect] (zab);
          \path (i12) edge [connect] (zab);
          \path (q12) edge [connect] (zab);   
               
          \path (T) edge [connect] (zab);
          \path (P) edge [connect] (zab);
          \path (e) edge [connect] (zab);
          \path (a) edge [connect] (zab);
          \path (omega) edge [connect] (zab);    
          \path (i) edge [connect] (zab);     
          \path (q) edge [connect] (zab);     
          \path (vCoM) edge [connect] (zab);                 
          
          \path (T) edge [connect] (zB);
          \path (P) edge [connect] (zB);
          \path (e) edge [connect] (zB);
          \path (a) edge [connect] (zB);
          \path (omega) edge [connect] (zB);    
          \path (i) edge [connect] (zB);     
          \path (q) edge [connect] (zB);       
          \path (vCoM) edge [connect] (zB);                                     
          
		\node[rectangle, inner sep=1mm, fit= (zaa), label=left :\textbf{$_{k=1}^{N_3}$}] {};	
         \node[rectangle, inner sep=1.5mm, draw=black!100, dashed, fit = (zaa) ] {}; 
          
		\node[rectangle, inner sep=1mm, fit= (zab), label=left :\textbf{$_{k=1}^{N_4}$}] {};	
         \node[rectangle, inner sep=1.5mm, draw=black!100, dashed, fit = (zab) ] {}; 
          
		\node[rectangle, inner sep=1mm, fit= (zB), label=left :\textbf{$_{k=1}^{N_5}$}] {};	
         \node[rectangle, inner sep=1.5mm, draw=black!100, dashed, fit = (zB) ] {};                         
                    
        \end{tikzpicture}
        }
    \caption{Graphical model representation of the RV alone scenario when inner and outer observations are disconnected.}
    \label{fig:graphModelRV1}
\end{figure}

Based on the dependencies observed in Equation~(\ref{eq:rv}), a graphical model was designed expressing the relationship between the parameters and the observations $z_3$, $z_4$ and $z_5$. This model is shown in Figure~\ref{fig:graphModelRV1}, where $N_3$, $N_4$ and $N_5$ are the number of observations of $z_3$, $z_4$ and $z_5$, respectively. 

Even though the measurements related to the outer body ($z_5$) are disconnected from the inner parameters (as can be seen in Equation~(\ref{eq:rv}) and Figure~\ref{fig:graphModelRV1}), mainly due to the long outer periods, it is not common to have plenty of observations. Then a preliminary \textit{outer body} stage would involve just a few or none observations. On the other hand, measurements from the inner secondary ($z_4$) are also uncommon, due mostly to being too faint for precise RV measurements.

\begin{figure}[htp]
        \resizebox{\textwidth}{!}{
                \begin{tikzpicture}
			\usetikzlibrary{shapes}
        \tikzstyle{main}=[circle, minimum size = 10mm, thick, draw =black!80, node distance = 16mm]
        \tikzstyle{connect}=[-latex, thick]
        \tikzstyle{box}=[rectangle, draw=black!80]
        \tikzstyle{ell}=[ellipse, draw=black!80]        

          \node[main] (T) [] {{$T_{AB}$}};   
          \node[main] (P) [right=of T] {{$P_{AB}$}};
          \node[main] (e) [right=of P] {{$e_{AB}$}};
          \node[main] (a) [right=of e] {{$a_{AB}$}};
          \node[main] (omega) [right=of a] {{$\omega_{AB}$}};
          \node[main] (i) [right=of omega] {{$i_{AB}$}};    
          \node[main] (q) [right=of i] {{$q_{AB}$}};        
          \node[main] (vCoM) [right=of q] {{$v_{CoM}$}};                        
                
          \node[ell] (obs) [below =of e] {$\{{z_3}(\tau_k)_{k=1}^{N_3}, {z_4}(\tau_k)_{k=1}^{N_4},  {z_5}(\tau_k)_{k=1}^{N_5}\}$};
          
          \node[main] (T12) [below = of zinner] {{$T_{A_{a}A_{b}}$}};          
          \node[main] (P12) [right=of T12] {{$P_{A_{a}A_{b}}$}};
          \node[main] (e12) [right=of P12] {{$e_{A_{a}A_{b}}$}};
          \node[main] (a12) [right=of e12] {{$a_{A_{a}A_{b}}$}};
          \node[main] (omega12) [right=of a12] {{$\omega_{A_{a}A_{b}}$}};
          \node[main] (i12) [right=of omega12] {{$i_{A_{a}A_{b}}$}}; 
          \node[main] (q12) [right=of i12] {{$q_{A_{a}A_{b}}$}};       
          
          \path (T12) edge [connect] (obs);
          \path (P12) edge [connect] (obs);
          \path (e12) edge [connect] (obs);
          \path (a12) edge [connect] (obs);
          \path (omega12) edge [connect] (obs);
          \path (i12) edge [connect] (obs);
          \path (q12) edge [connect] (obs);   
               
          \path (T) edge [connect] (obs);
          \path (P) edge [connect] (obs);
          \path (e) edge [connect] (obs);
          \path (a) edge [connect] (obs);
          \path (omega) edge [connect] (obs);    
          \path (i) edge [connect] (obs);     
          \path (q) edge [connect] (obs);  
          \path (vCoM) edge [connect] (obs);        
          
        \end{tikzpicture}
        }
    \caption{Graphical model representation of the RV alone scenario considering the complete joint distribution between parameters and observations.}
    \label{fig:graphModelRV2}
\end{figure}
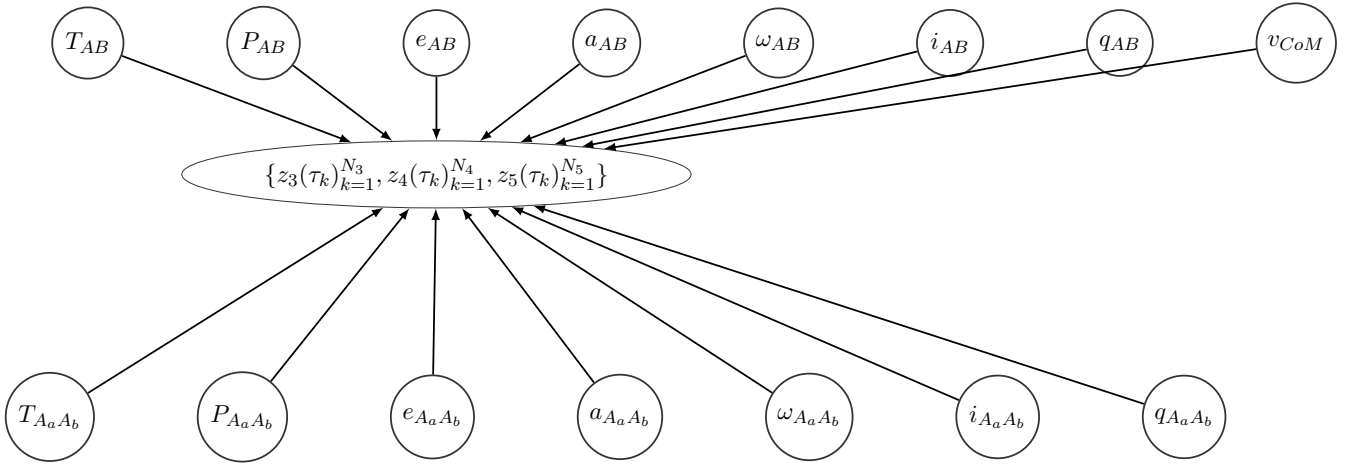

\begin{figure}[htp]
\centering
        \resizebox{0.8 \textwidth}{!}{
                \begin{tikzpicture}
        \tikzstyle{main}=[circle, minimum size = 10mm, thick, draw =black!80, node distance = 16mm]
        \tikzstyle{connect}=[-latex, thick]
        \tikzstyle{box}=[rectangle, draw=black!80, minimum size = 10mm]
        
          \node[main] (z) [] { $Z_{RV}$};                         
          \node[box] (cp) [right = of z] {Compute predictive model (1)};
		\node[box] (p1) [right = of cp] {$P_{\Theta_1 \cup \Theta_2 | Z_3, Z_4, Z_5}(\cdot | z_3, z_4, z_5)$};          
		
          \path (z) edge [connect] (cp);
          \path (cp) edge [connect] (p1);          		         
                  
        \end{tikzpicture}
        }
    \caption{Information-flow diagram in the RV alone scenario.}
    \label{fig:inference_process_rv}
\end{figure}
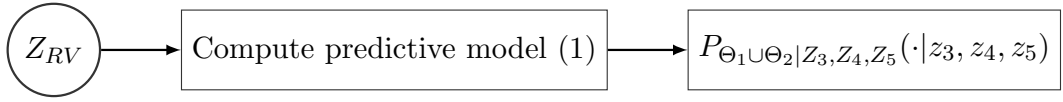

Therefore, in this case, a sequential factorization of the joint in terms of partial posteriors is not simple to develop and we prefer to model the complete joint distribution between parameters and observations (see Figure~\ref{fig:graphModelRV2}). This network allows us to factorize the joint distribution in conditional components and, therefore, arrange the inference process for this scenario. In this context, the factorization is expressed as in Equation~(\ref{eq:gm_rv}), and the inference is conducted in one step using the complete joint distribution of the problem (see Figure~{\ref{fig:inference_process_rv}}).

\begin{equation}
    p(Z_3, Z_4, Z_5, \Theta_1, \Theta_2) =  p(Z_3, Z_4, Z_5 | \Theta_1, \Theta_2) \cdot \prod_{\theta \in \Theta_1 \cup \Theta_2} p(\theta).  \label{eq:gm_rv}
\end{equation}

Concerning the predictive model $P_{\Theta_1 \cup \Theta_2 | Z_3, Z_4, Z_5}(\cdot | z_3, z_4, z_5)$, this is approximated with samples using a sample-based scheme. More details about this estimation approach can be found in Appendix \ref{app:mcmc}.

\subsection{Combined scenario} \label{sec:comb}

\begin{figure}[H]
    \centering
    \gridline{
        \resizebox{\textwidth}{!}{
        \begin{tikzpicture}
        \tikzstyle{main}=[circle, minimum size = 10mm, thick, draw =black!80, fill=white!100, node distance = 16mm]
        \tikzstyle{connect}=[-latex, thick]
        \tikzstyle{box}=[rectangle, draw=black!100]
        \tikzstyle{ell}=[ellipse, draw=black!80]   
        \node[main] (Omega12) [] {$\Omega_{A_{a}A_{b}}$};
        \node[main] (a12) [right=of Omega12] {$a_{A_{a}A_{b}}$};
        \node[main] (zi) [below  = 3cm of a12] { $\Vec{z}_{1}(\tau_k)$}; 
        \node[main] (i12) [right=of a12] {$i_{A_{a}A_{b}}$}; 
        \node[main] (q12) [right=of i12] {$q_{A_{a}A_{b}}$}; 
        \node[main] (T12) [right=of q12] {$T_{A_{a}A_{b}}$};
        \node[main] (P12) [right=of T12] {$P_{A_{a}A_{b}}$};
        \node[main] (e12) [right=of P12] {$e_{A_{a}A_{b}}$};
        \node[main] (omega12) [right=of e12] {$\omega_{A_{a}A_{b}}$};
        \node[main] (VCoM) [right=of omega12] {$V_{CoM}$}; 
        \node[main] (Omega) [below = 3cm of zi] {$\Omega_{AB}$};
        \node[main] (a) [right=of Omega] {$a_{AB}$};
        \node[main] (zo) [below  = 3cm of i12] {$\Vec{z}_{2}(\tau_k)$};
        \node[main] (i) [right=of a] {$i_{AB}$};  
        \node[main] (T) [right= 3cm of i] {$T_{AB}$};
        \node[main] (P) [right=of T] {$P_{AB}$};
        \node[main] (e) [right=of P] {$e_{AB}$};
        \node[main] (omega) [right=of e] {$\omega_{AB}$};
        \node[main] (q) [right=of omega] {$q_{AB}$}; 
        \node[ell] (v_obs) [below =3cm of e12] {$\{{z_3}(\tau_k)_{k=1}^{N_3}, {z_4}(\tau_k)_{k=1}^{N_4},  {z_5}(\tau_k)_{k=1}^{N_5}\}$}; 
          \path (T12) edge [left]  (zi);
          \path (P12) edge [left]  (zi);
          \path (e12) edge [left]  (zi);
          \path (a12) edge [left]  (zi);
          \path (omega12) edge [left]  (zi);
          \path (Omega12) edge [connect]  (zi);  
          \path (i12) edge [connect]  (zi);
          \path (T) edge [connect]  (zo);
          \path (P) edge [connect]  (zo);
          \path (e) edge [connect]  (zo);
          \path (a) edge [connect]  (zo);
          \path (omega) edge [connect]  (zo);  
          \path (Omega) edge [connect]  (zo);    
          \path (i) edge [connect]  (zo);    
          \path (q12) edge [connect]  (zo);    
          \path (omega) edge [connect]  (zo);    
          \path (zi) edge [connect] (zo);
          \path (T12) edge [connect] (v_obs);
          \path (P12) edge [connect] (v_obs);
          \path (e12) edge [connect] (v_obs);
          \path (a12) edge [connect] (v_obs);
          \path (i12) edge [connect] (v_obs);
          \path (omega12) edge [connect] (v_obs);
          \path (q12) edge [connect] (v_obs);
          \path (T) edge [connect] (v_obs);
          \path (P) edge [connect] (v_obs);
          \path (e) edge [connect] (v_obs);
          \path (a) edge [connect] (v_obs);
          \path (i) edge [connect] (v_obs);
          \path (omega) edge [connect] (v_obs);
          \path (q) edge [connect] (v_obs);
          \path (VCoM) edge [connect] (v_obs);  
		 \node[rectangle, inner sep=1mm, fit= (zo), label=left :\textbf{$_{k=1}^{N_2}$}] {};	
          \node[rectangle, inner sep=1.5mm, draw=black!100, dashed, fit = (zo) ] {};     
		 \node[rectangle, inner sep=1mm, fit= (zi), label=left :\textbf{$_{k=1}^{N_1}$}] {};	
          \node[rectangle, inner sep=1.5mm, draw=black!100, dashed, fit = (zi) ] {};              
          \node[rectangle, inner sep=2mm, fit= (zo) (zi) (a12) (omega12) (Omega12) (a) (omega), label=above left :\textbf{\textcolor{red}{Astrometric scenario}}] {};	
          \node[rectangle, inner sep=1.5mm, draw=red!100, fill=red!50, opacity=0.2, fit = (zo) (zi) (a12) (omega12) (Omega12) (a) (omega)] {};   
        \end{tikzpicture}
        }
    }{}
    \gridline{
        \resizebox{\textwidth}{!}{
        \begin{tikzpicture}
        \tikzstyle{main}=[circle, minimum size = 10mm, thick, draw =black!80, fill=white!100, node distance = 16mm]
        \tikzstyle{connect}=[-latex, thick]
        \tikzstyle{box}=[rectangle, draw=black!100]
        \tikzstyle{ell}=[ellipse, draw=black!80]   
        \node[main] (Omega12) [] {$\Omega_{A_{a}A_{b}}$};
        \node[main] (a12) [right=of Omega12] {$a_{A_{a}A_{b}}$};
        \node[main] (zi) [below  = 3cm of a12] { $\Vec{z}_{1}(\tau_k)$}; 
        \node[main] (i12) [right=of a12] {$i_{A_{a}A_{b}}$}; 
        \node[main] (q12) [right=of i12] {$q_{A_{a}A_{b}}$}; 
        \node[main] (T12) [right=of q12] {$T_{A_{a}A_{b}}$};
        \node[main] (P12) [right=of T12] {$P_{A_{a}A_{b}}$};
        \node[main] (e12) [right=of P12] {$e_{A_{a}A_{b}}$};
        \node[main] (omega12) [right=of e12] {$\omega_{A_{a}A_{b}}$};
        \node[main] (VCoM) [right=of omega12] {$V_{CoM}$}; 
        \node[main] (Omega) [below = 3cm of zi] {$\Omega_{AB}$};
        \node[main] (a) [right=of Omega] {$a_{AB}$};
        \node[main] (zo) [below  = 3cm of i12] {$\Vec{z}_{2}(\tau_k)$};
        \node[main] (i) [right=of a] {$i_{AB}$};  
        \node[main] (T) [right= 3cm of i] {$T_{AB}$};
        \node[main] (P) [right=of T] {$P_{AB}$};
        \node[main] (e) [right=of P] {$e_{AB}$};
        \node[main] (omega) [right=of e] {$\omega_{AB}$};
        \node[main] (q) [right=of omega] {$q_{AB}$}; 
        \node[ell] (v_obs) [below =3cm of e12] {$\{{z_3}(\tau_k)_{k=1}^{N_3}, {z_4}(\tau_k)_{k=1}^{N_4},  {z_5}(\tau_k)_{k=1}^{N_5}\}$}; 
          \path (T12) edge [left]  (zi);
          \path (P12) edge [left]  (zi);
          \path (e12) edge [left]  (zi);
          \path (a12) edge [left]  (zi);
          \path (omega12) edge [left]  (zi);
          \path (Omega12) edge [connect]  (zi);  
          \path (i12) edge [connect]  (zi);
          \path (T) edge [connect]  (zo);
          \path (P) edge [connect]  (zo);
          \path (e) edge [connect]  (zo);
          \path (a) edge [connect]  (zo);
          \path (omega) edge [connect]  (zo);  
          \path (Omega) edge [connect]  (zo);    
          \path (i) edge [connect]  (zo);    
          \path (q12) edge [connect]  (zo);    
          \path (omega) edge [connect]  (zo);    
          \path (zi) edge [connect] (zo);
          \path (T12) edge [connect] (v_obs);
          \path (P12) edge [connect] (v_obs);
          \path (e12) edge [connect] (v_obs);
          \path (a12) edge [connect] (v_obs);
          \path (i12) edge [connect] (v_obs);
          \path (omega12) edge [connect] (v_obs);
          \path (q12) edge [connect] (v_obs);
          \path (T) edge [connect] (v_obs);
          \path (P) edge [connect] (v_obs);
          \path (e) edge [connect] (v_obs);
          \path (a) edge [connect] (v_obs);
          \path (i) edge [connect] (v_obs);
          \path (omega) edge [connect] (v_obs);
          \path (q) edge [connect] (v_obs);
          \path (VCoM) edge [connect] (v_obs);  
		 \node[rectangle, inner sep=1mm, fit= (zo), label=left :\textbf{$_{k=1}^{N_2}$}] {};	
          \node[rectangle, inner sep=1.5mm, draw=black!100, dashed, fit = (zo) ] {};     
		 \node[rectangle, inner sep=1mm, fit= (zi), label=left :\textbf{$_{k=1}^{N_1}$}] {};	
          \node[rectangle, inner sep=1.5mm, draw=black!100, dashed, fit = (zi) ] {};              
          \node[rectangle, inner sep=2mm, fit= (a12) (i12) (a) (i) (q12) (VCoM) (v_obs) (q),label=above right :\textbf{\textcolor{blue}{RV scenario}}] {};	
          \node[rectangle, inner sep=1.5mm, draw=blue!100, fill=blue!50, opacity=0.2, fit = (a12) (i12) (a) (i) (q12) (VCoM) (v_obs) (q)] {};   
        \end{tikzpicture}
        }
    }{}
    \caption{Graphical model representation considering astrometric and RV measurements. This representation can be further split into the upper (red) box that shows the influence and dependencies of the astrometric data, while the lower (blue) box shows the equivalent for the RV data (see also Figure~\ref{fig:graphModelRV01}).}
    \label{fig:graphModelCombined}
\end{figure}
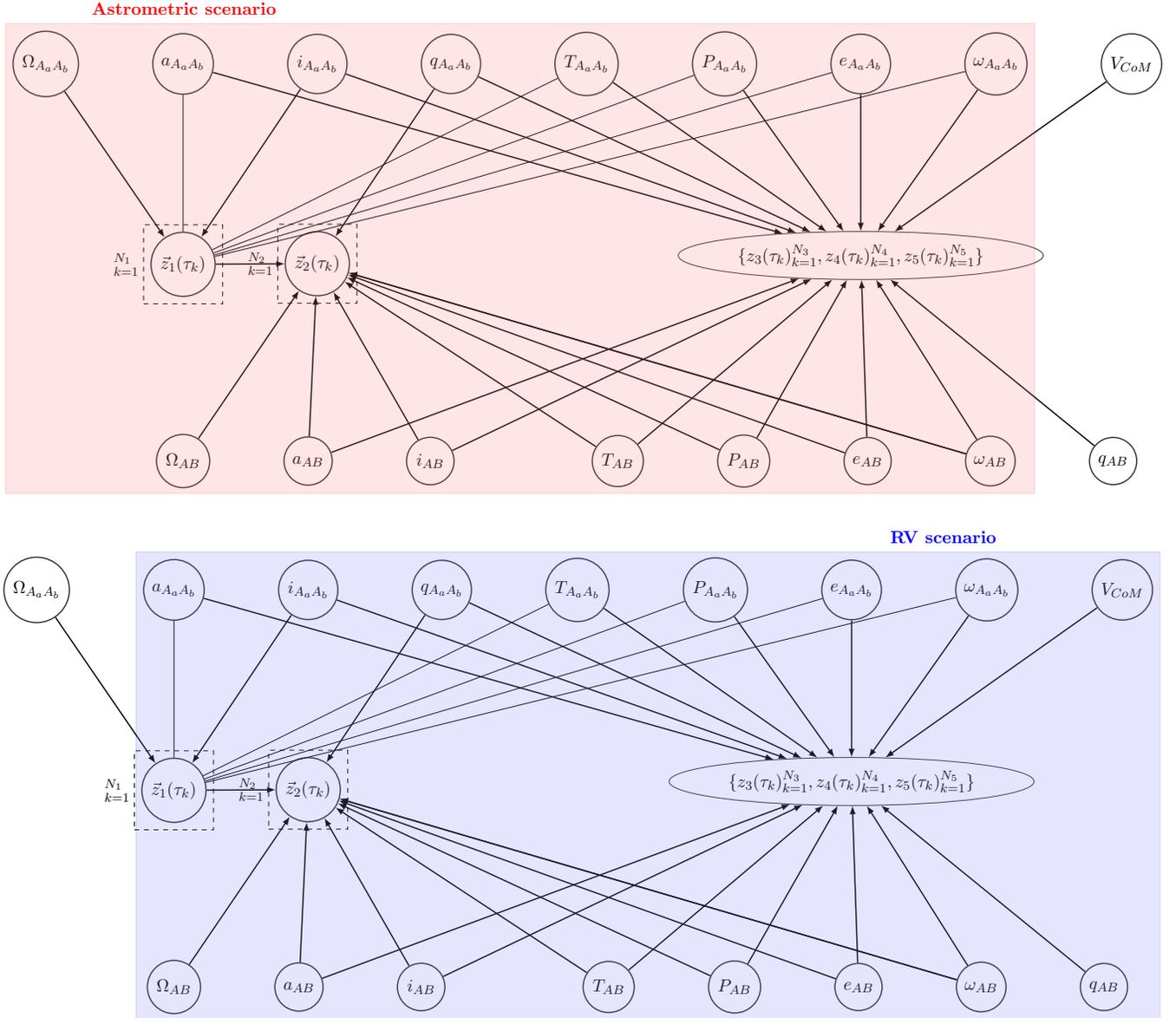

Finally, we consider the scenario where both astrometric and RV measurements are available. {This rich scenario allows us to compute the relative orbital orientation, the stellar masses and, in principle, also, the so-called orbital parallax (derived from the ratio of the semi-major axes), which is independent of the trigonometric parallax. Due to the high-dimensionality of this setting, the estimation is complex both computationally and analytically, and in this particular application, we have left the parallax as a fixed external parameter, not to be fitted in our algorithm\footnote{In future applications, if enough high-quality data is available for a given system, one could certainly determine the orbital parallax self-consistently from our overall fit as done e.g., in \citet{mendez2017orbits, Mendet2021}.}.} On the modeling side, there are several interdependencies between the parameters and the observations. A major effort was made in this work to encode this relationship by the graphical model presented in Figure~\ref{fig:graphModelCombined}. Consequently, a factorization as the one presented on our previous simpler (unimodal) models is difficult to illustrate in a simple diagram.

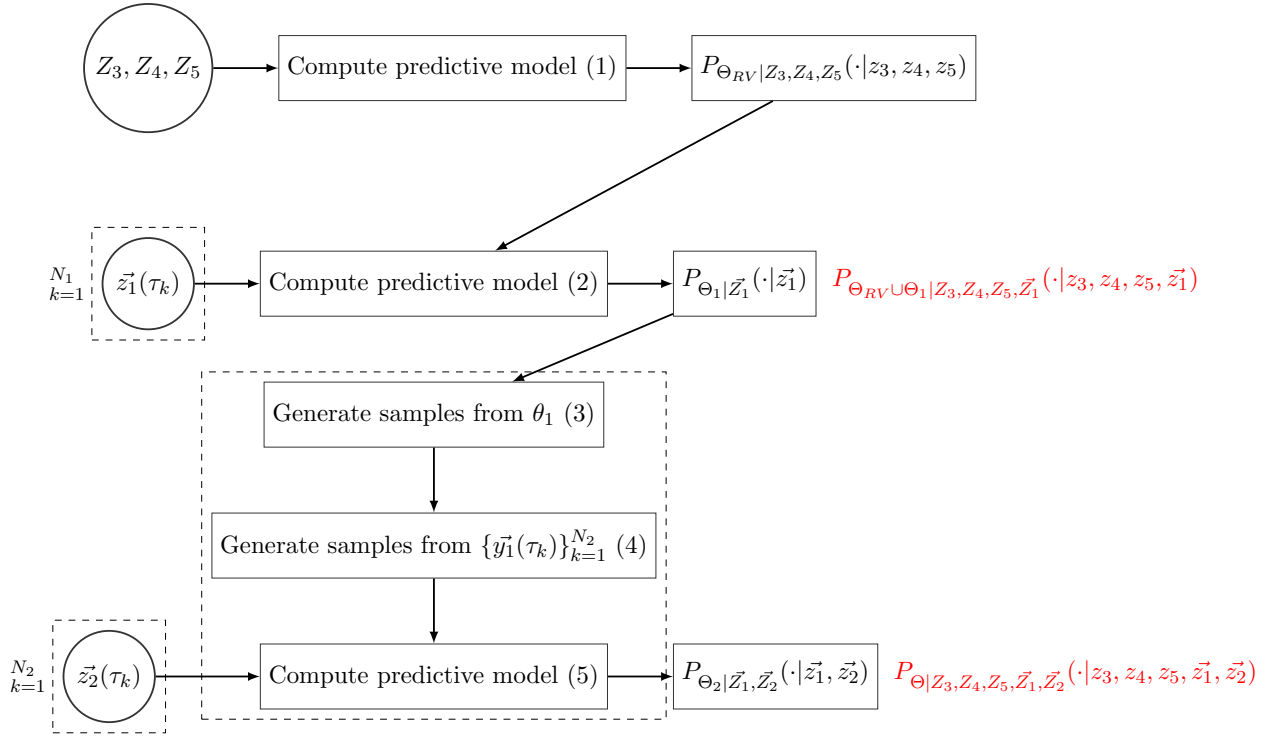
\begin{figure}[htp]
\centering
        \resizebox{0.95 \textwidth}{!}{
        \begin{tikzpicture}
        \tikzstyle{main}=[circle, minimum size = 10mm, thick, draw =black!80, node distance = 16mm]
        \tikzstyle{connect}=[-latex, thick]
        \tikzstyle{box}=[rectangle, draw=black!80, minimum size = 10mm]
        
        \node[main] (z_rv) [] { $Z_3, Z_4, Z_5$};  
        \node[box] (cp_rv) [right = of z_rv] {Compute predictive model (1)};  
        \node[box] (p_rv) [right = of cp_rv] {$P_{\Theta_{RV} | Z_3, Z_4, Z_5}(\cdot | z_3, z_4, z_5)$};  
          
        \node[main] (z1) [below = of z_rv] { $\vec{z_1}(\tau_k)$};      
        \node[box] (cp1) [right = of z1] {Compute predictive model (2)};
        \node[box] (p1) [right = of cp1] {$P_{\Theta_1 | \vec{Z_1}}(\cdot | \vec{z_1})$};
        
        \node[box] (gt) [below = of cp1] {Generate samples from $\theta_1$ (3)};   
        \node[box] (gs) [below = of gt] {Generate samples from $\{\vec{y_1}(\tau_k)\}_{k=1}^{N_2}$ (4)};         
        \node[box] (cp2) [below = of gs] {Compute predictive model (5)};
          
        \node[main] (z2) [left =of cp2] { $\vec{z_2}(\tau_k)$};  
        \node[box] (p2) [right = of cp2] {$P_{\Theta_2 | \vec{Z_1}, \vec{Z_2}}(\cdot | \vec{z_1}, \vec{z_2})$};  
        
        \path (z_rv) edge [connect] (cp_rv);
        \path (cp_rv) edge [connect] (p_rv);     
        
        \path (p_rv) edge [connect] (cp1);  
        
        \path (z1) edge [connect] (cp1);
        \path (cp1) edge [connect] (p1);   
        
        \path (p1) edge [connect] (gt);   
        \path (gt) edge [connect] (gs);   
        \path (gs) edge [connect] (cp2); 
        \path (z2) edge [connect] (cp2);
        \path (cp2) edge [connect] (p2); 

		\node[rectangle, inner sep=1mm, fit= (z2), label=left :\textbf{$_{k=1}^{N_2}$}] {};	
        \node[rectangle, inner sep=1.5mm, draw=black!100, dashed, fit = (z2) ] {};     

		\node[rectangle, inner sep=1mm, fit= (z1), label=left :\textbf{$_{k=1}^{N_1}$}] {};	
        \node[rectangle, inner sep=1.5mm, draw=black!100, dashed, fit = (z1) ] {};    
          
        \node[rectangle, inner sep=1.5mm, draw=black!100, dashed, fit = (gs) (gt) (cp2) ] {};    
        
        \node[rectangle, inner sep=1mm, fit= (p1), label=right :\textbf{\color{red}$P_{\Theta_{RV} \cup \Theta_1 |Z_3, Z_4, Z_5, \vec{Z_1}}(\cdot | z_3, z_4, z_5, \vec{z_1})$}] {};	
        
        \node[rectangle, inner sep=1mm, fit= (p2), label=right :\textbf{\color{red}$P_{\Theta | Z_3, Z_4, Z_5, \vec{Z_1}, \vec{Z_2}}(\cdot | z_3, z_4, z_5, \vec{z_1}, \vec{z_2})$}] {};	
                   
        \end{tikzpicture}
        }
    \caption{Information-flow diagram in the combined scenario.}
    \label{fig:inference_process_combined}
\end{figure}

For the inference, we propose an approach built upon the aforementioned scenarios, which comes from the mathematical formulation of the problem. Interestingly, the resulting steps resemble the ones presented by \citet{tokovinin2017}. The procedure is shown in Figure~\ref{fig:inference_process_combined} and considers the following steps:
\begin{enumerate}
\item We compute the predictive model $P_{\Theta_{RV} | Z_3, Z_4, Z_5}(\cdot | z_3, z_4, z_5)$ using $\{\vec{z_3}(\tau_k)\}_{k=1}^{N_3}$, $\{\vec{z_4}(\tau_k)\}_{k=1}^{N_4}$ and $\{\vec{z_5}(\tau_k)\}_{k=1}^{N_5}$ in a sample-based scheme (Stage~1 of Figure~\ref{fig:inference_process_combined}).
\item We compute the predictive model $P_{\Theta_1 | \vec{Z_1}}(\cdot | \vec{z_1})$ in a sample-based scheme, using the observations $\{\vec{z_1}(\tau_k)\}_{k=1}^{N_1}$ and the posteriors from Stage~1 as priors (Stage~2 of Figure~\ref{fig:inference_process_combined}).
\item We generate empirical samples of $\theta_1$ using the posterior distribution of $\Theta_1$ given $\vec{Z_1}$, $P_{\Theta_1 | \vec{Z_1}}(\cdot | \vec{z_1})$ (Stage~3 of Figure~\ref{fig:inference_process_combined}).
\item We generate virtual observations $\{\vec{y_1}(\tau_k)\}_{k=1}^{N_2}$, using $P_{\Theta_1 | \vec{Z_1}}(\cdot | \vec{z_1})$, for each observation epoch from $\vec{z_2}$, in an imputations framework (Stage~4 of Figure~\ref{fig:inference_process_combined}).
\item We compute the predictive model $P_{\Theta_2 | \vec{Z_1}, \vec{Z_2}}(\cdot | \vec{z_1}, \vec{z_2})$ using $\{\vec{y_1}(\tau_k)\}_{k=1}^{N_2}$ and $\{\vec{z_2}(\tau_k)\}_{k=1}^{N_2}$ in a sample-based scheme (Stage~5 of Figure~\ref{fig:inference_process_combined}).
\item We return to (1) until a stopping criterion or the maximum amount of iterations are reached.
\end{enumerate}

At the end of each stages, we obtain i.i.d. samples of the posterior distribution of the parameters, partially conditioned on the set of observations associated to that stage. Besides, with each new stage we reach the target distribution, given \textit{all} the available observations. More details about each subprocess can be found in Appendix \ref{app:mcmc}.

\section{Benchmark Results} \label{sec:benchmarks}

In order to test the methodology presented in Section~\ref{sec:orbitscalc}, and to uncover its advantages and possible limitations, we have selected three well-studied triple hierarchical stellar systems published by Tokovinin and collaborators \citep{tokovinin2017, tokovinin2018dancing}, and kindly provided by them upon request. The astrometric and spectroscopic data has been supplemented with more recent SOAR HRCam observations and data from the 9th Catalogue of Spectroscopic Binary Orbits (\cite{Pouret2004})\footnote{Updated regularly, and available at \url{https://sb9.astro.ulb.ac.be/}}, whenever available. We then regard these objects as ``benchmark'' systems from the point of view of our estimation algorithm. Tokovinin et al. provide an in-depth description of the orbital architecture for each of these systems, including a report on the available measurements and previous studies on them. Given that no further analysis has been reported in the literature for these objects since then, we refer the reader to those papers for further details regarding the specific properties of the selected objects.

It has been mentioned previously that the parallax is needed to obtain relevant physical parameters of the system, such as the sum of the masses (astrometry observations alone) or the individual ones (combined astrometry plus RV observations). Our adopted parallaxes are shown on the second column of Table~\ref{tab:derived}, indicating the source.

These benchmark systems allow us to evaluate the astrometry-only and the combined scenarios described before, and in what follows we provide a brief discussion regarding the estimation processes performed in each of these cases.

\subsection{WDS00247-2653 (LHS 1070)}

The triple system (identified as LP 881-64 in the SIMBAD database \citep{SIMBAD}, but also known as GJ~2005 and LHS~1070), consists of a secondary on a tight binary ($B_a, B_b$) accompanying a distant primary ($A$). In the Washington Double Star Catalogue (\cite{WDSCat2001}, hereafter WDS), the components are identified as LEI~A (the single primary,``outer'') and LEI~B, C (the double secondary, ``inner''). There are astrometric measurements available for the inner and outer subsystems \citep{kohler2012orbits, tokovinin2018dancing}, so the first scenario from the methodology was applied. The priors were chosen as follows:

\paragraph{Inner system:} The period is known to be $\sim 17$~years, so a uniform prior between 10 and 100~years was chosen, and the same boundaries were determined to limit the exploration of the state space. The (normalized) periastron passage $T$ and the eccentricity $e$ were restricted only between their physical boundaries, $[0,1]$ and $[0,0.99]$, respectively. After some trials, a number of $250,000$ iterations and a burn-in of $25,000$ iterations of our method were adopted, which allows the convergence of the algorithm and gives us a stable solution (based on the steadiness of the $\chi^2$ value).

\paragraph{Outer system:} \citet{tokovinin2018dancing} estimated a period $\sim 77$~years, so a uniform prior between 70 and 90~years was selected. Regarding the eccentricity, the uniform prior was bounded in $[0.001,0.05]$, based on the value estimated by \citet{tokovinin2018dancing} of $\sim 0.039$. The lower bound was necessary because the algorithm had some preference towards very small eccentricities. The angles $\Omega$ and $i$ were known $\sim 13.9$ and $\sim 62.5$~degrees, respectively, so both uniform priors were set between $[0, 100]$. On the other hand, the prior for $\omega$ was chosen in the interval $[100, 360]$ degrees. A number of $3,000,000$ iterations and a burn-in of $300,000$ iterations were adopted in this case.

\begin{figure}[htp]
    \centering
    \gridline{
    \fig{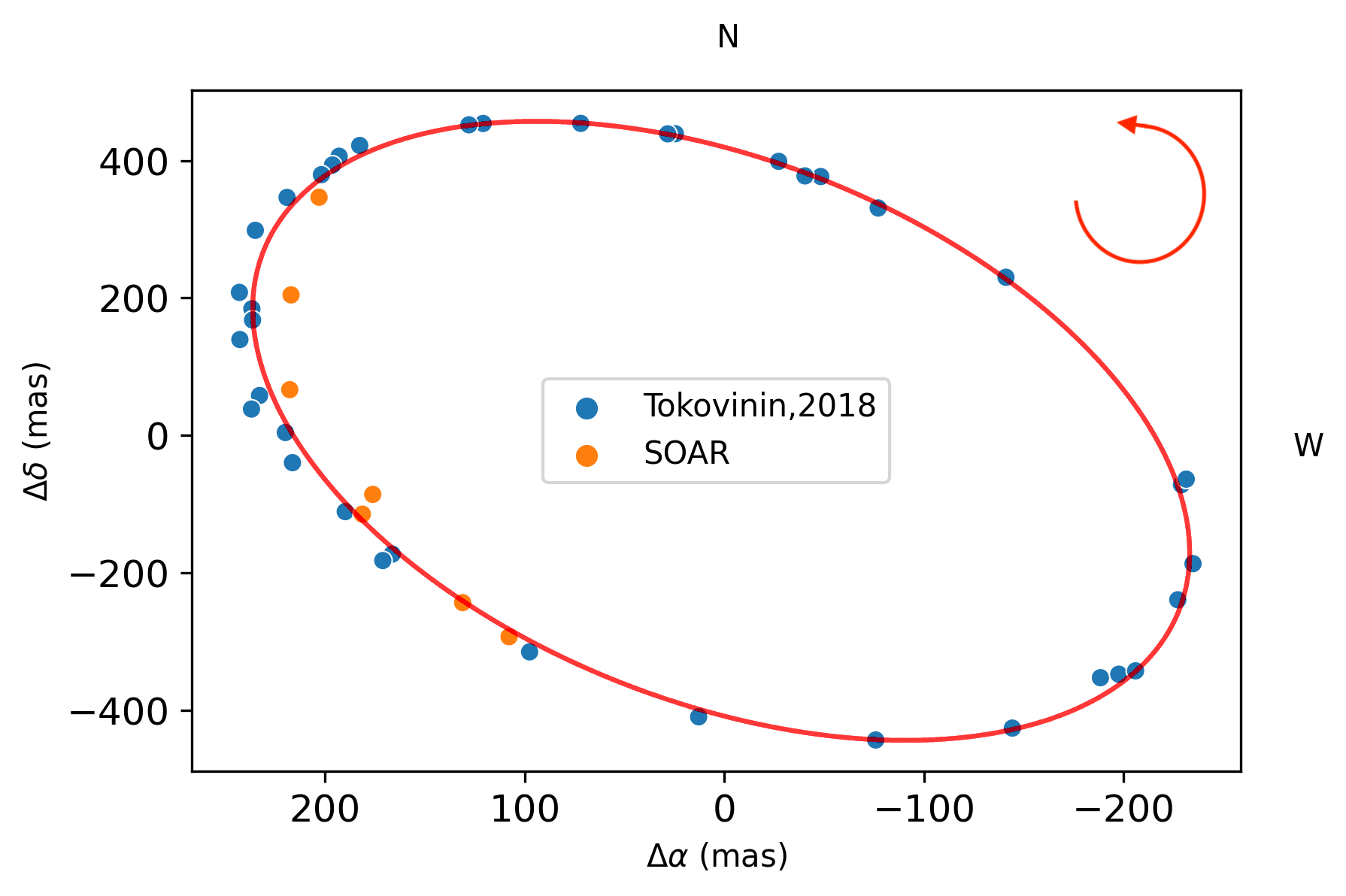}{0.5\textwidth}{}
    }
    \gridline{
    \fig{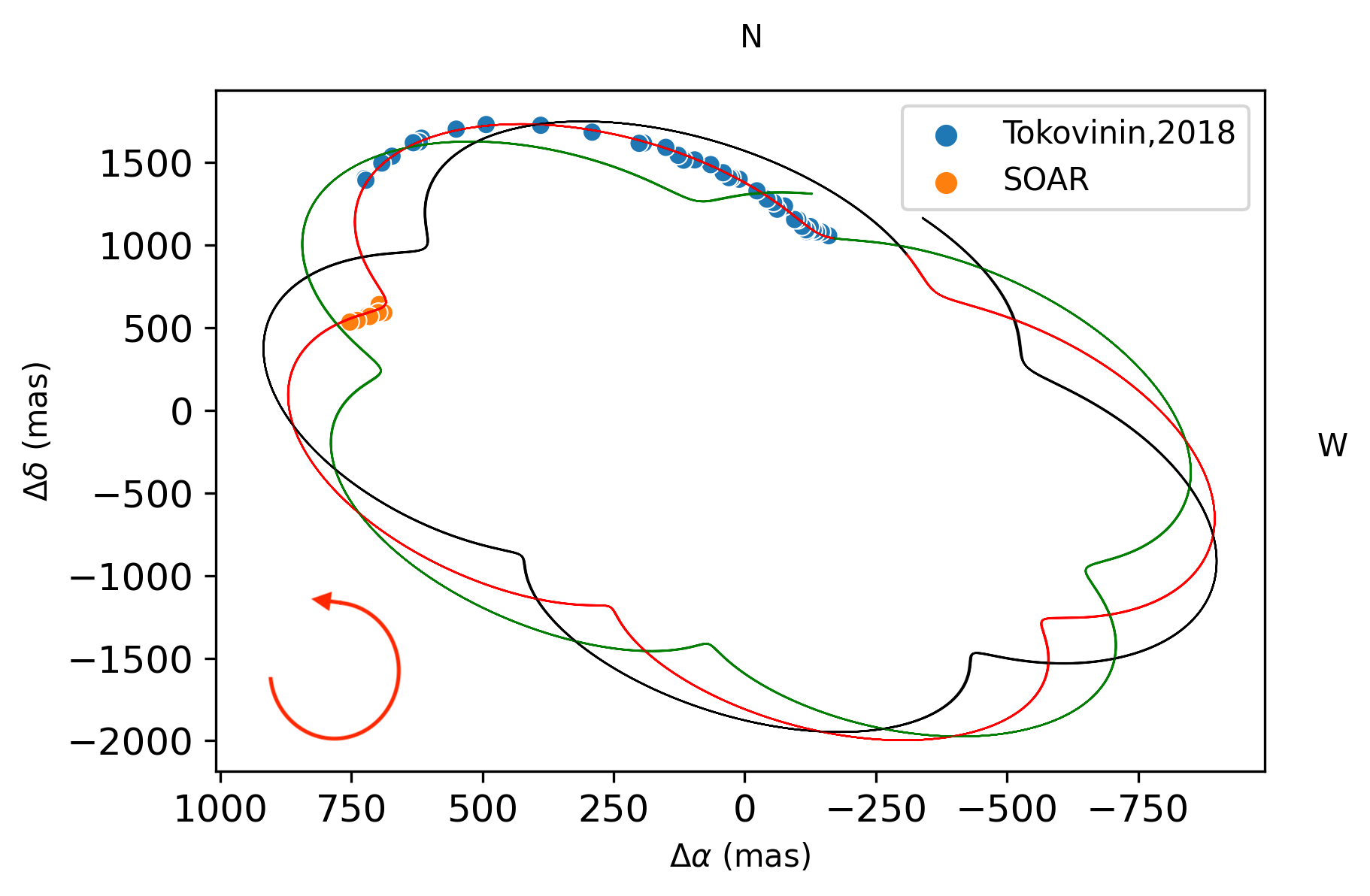}{0.5\textwidth}{}    
    \fig{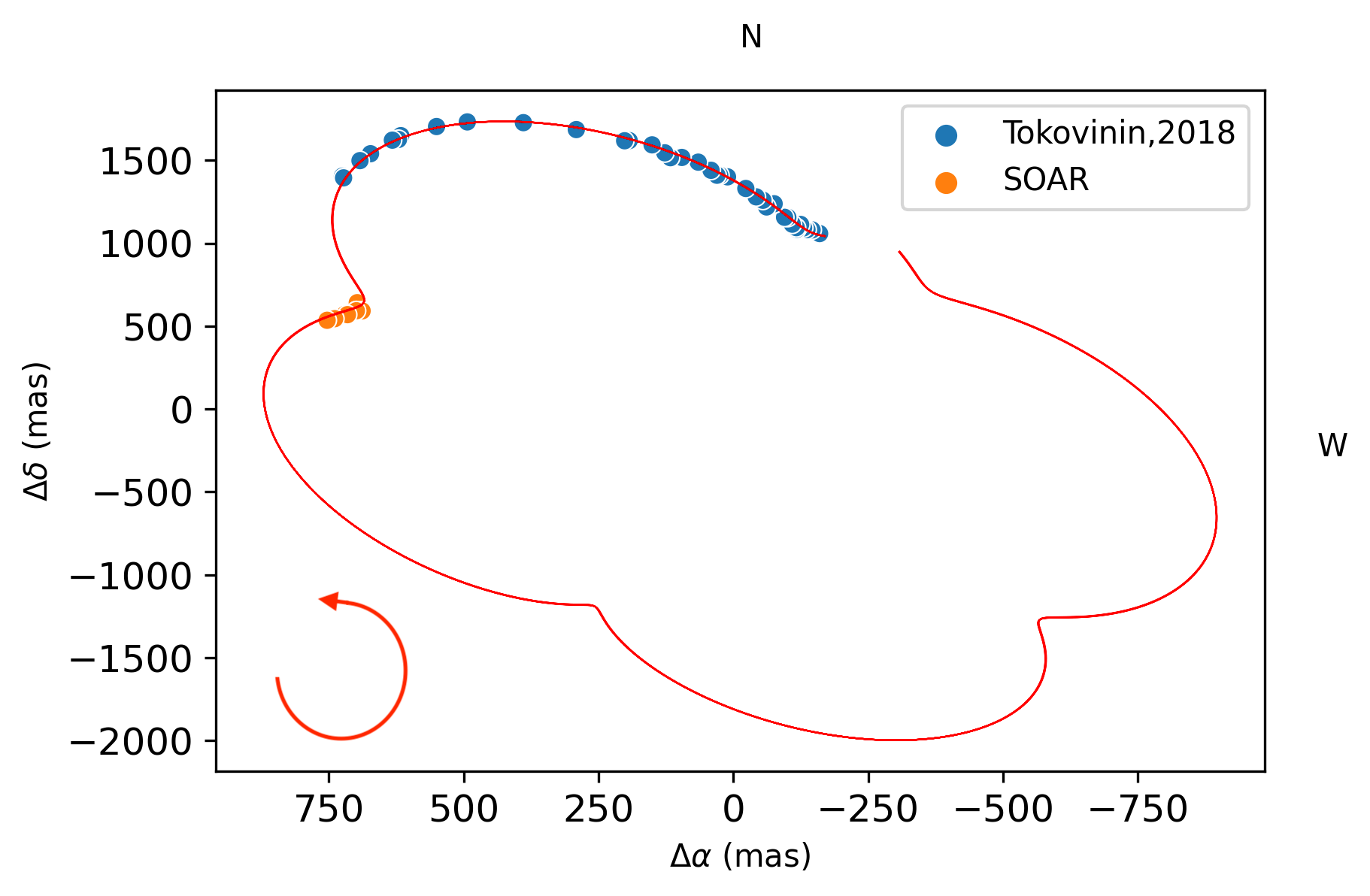}{0.5\textwidth}{}
    }  
\caption{Inner (upper) and outer (lower) orbits for LHS~1070. The size of the dots is proportional to their weight in the solution (which, in turn, is equal to the inverse square of its uncertainty). The most recent SOAR HRCam, data, not included in the original study by Tokovinin, are shown in these (and in the subsequent) plots, with a red dot. In all figures we overplot the 1000 most probable orbits from our solution. The fine width of the line is an indication that the orbits are quite well determined. As explained in the text, in the bottom panels, the orbits are not closed because the origin of the reference system in all three panels (at $(0,0)$) is the Aa component. \label{fig:LHS1070}}
\end{figure}

The resulting most likely 1000 orbits can be seen in Figure~\ref{fig:LHS1070} (inner system - left panel, outer system - center and right panels). It is important to note a subtle issue regarding the orbit of the outer system: the ``wobble'' (due to the presence of the inner system) implies that the orbit, as measured from the primary, is not necessarily a closed orbit (see Appendix \ref{app:general_dyn}, Eqs.~(\ref{eq12},~\ref{eq13})). For this reason, when plotting the outer orbit we set the time $t$ as the independent variable and not the (outer) \textit{true anomaly} $\nu$ (as commonly done in binary star calculations). For clarity, in the center panel of Figure~\ref{fig:LHS1070} we therefore consecutively plot the curves obtained for the following range of epochs:
\begin{itemize}
\item $[t_0 - P, t_0]$ in green,
\item $[t_0, t_0 + P]$ in red: this is the curve that minimizes the O-C residuals, 
\item $[t_0 + P, t_0 + 2P]$ in black,
\end{itemize}
where $t_0$ is the epoch of the first astrometric data available. The right panel shows only the curve that minimizes the O-C residuals. The same is done for all the subsequent orbital plots.

\begin{figure}[htp]    
    \centering
    \gridline{
    \fig{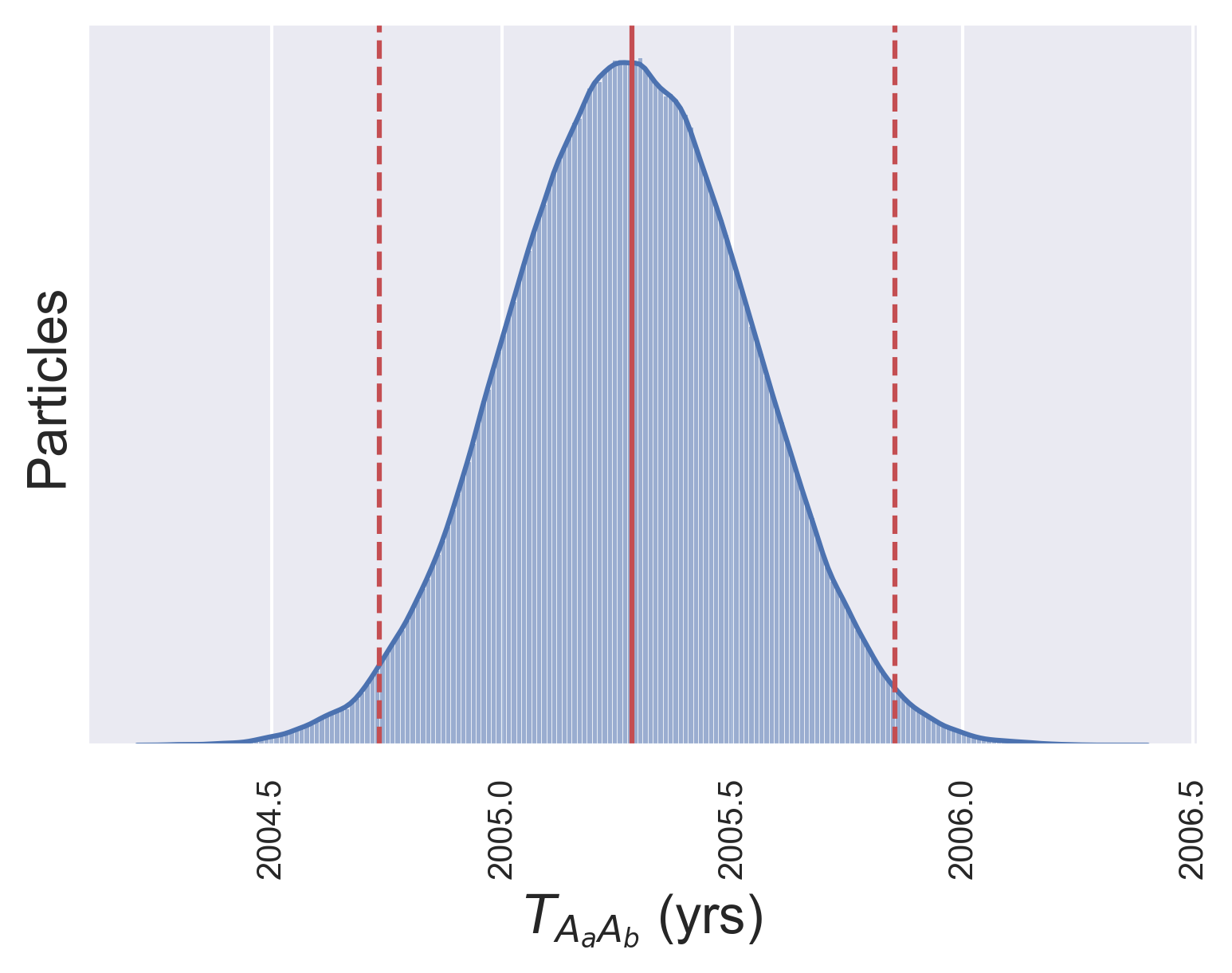}{0.33\textwidth}{}
    \fig{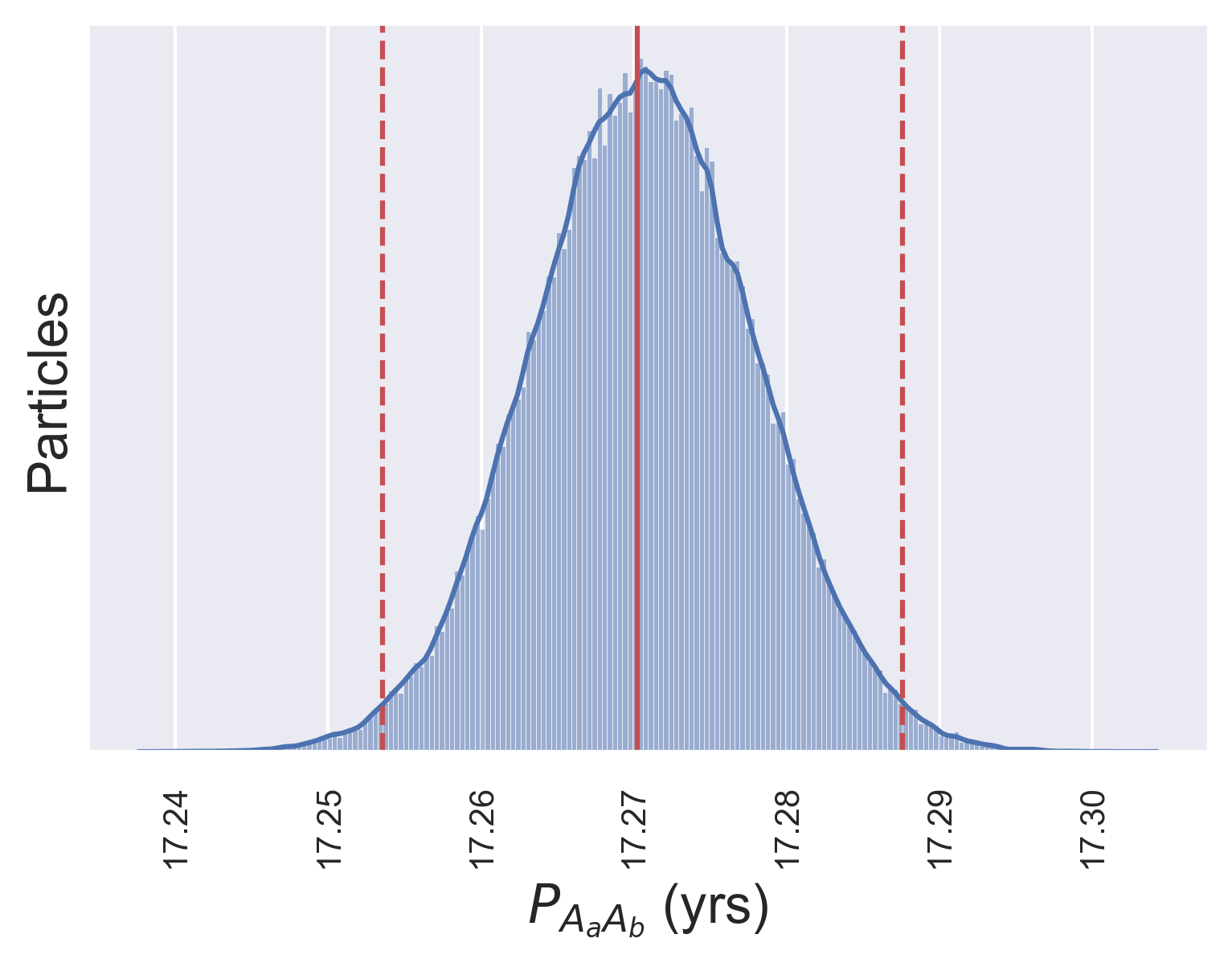}{0.33\textwidth}{}
    \fig{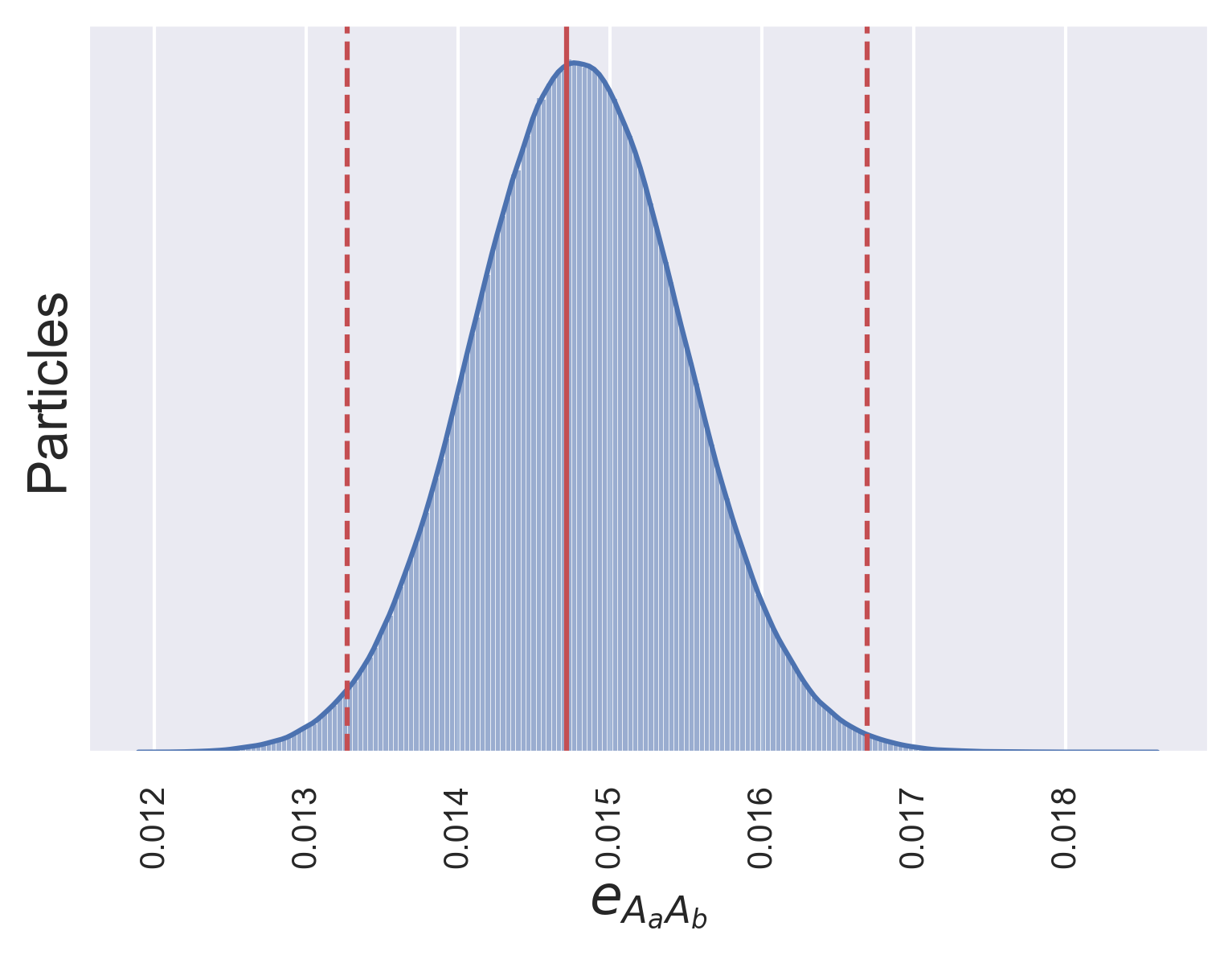}{0.33\textwidth}{}
    }
    \gridline{    
    \fig{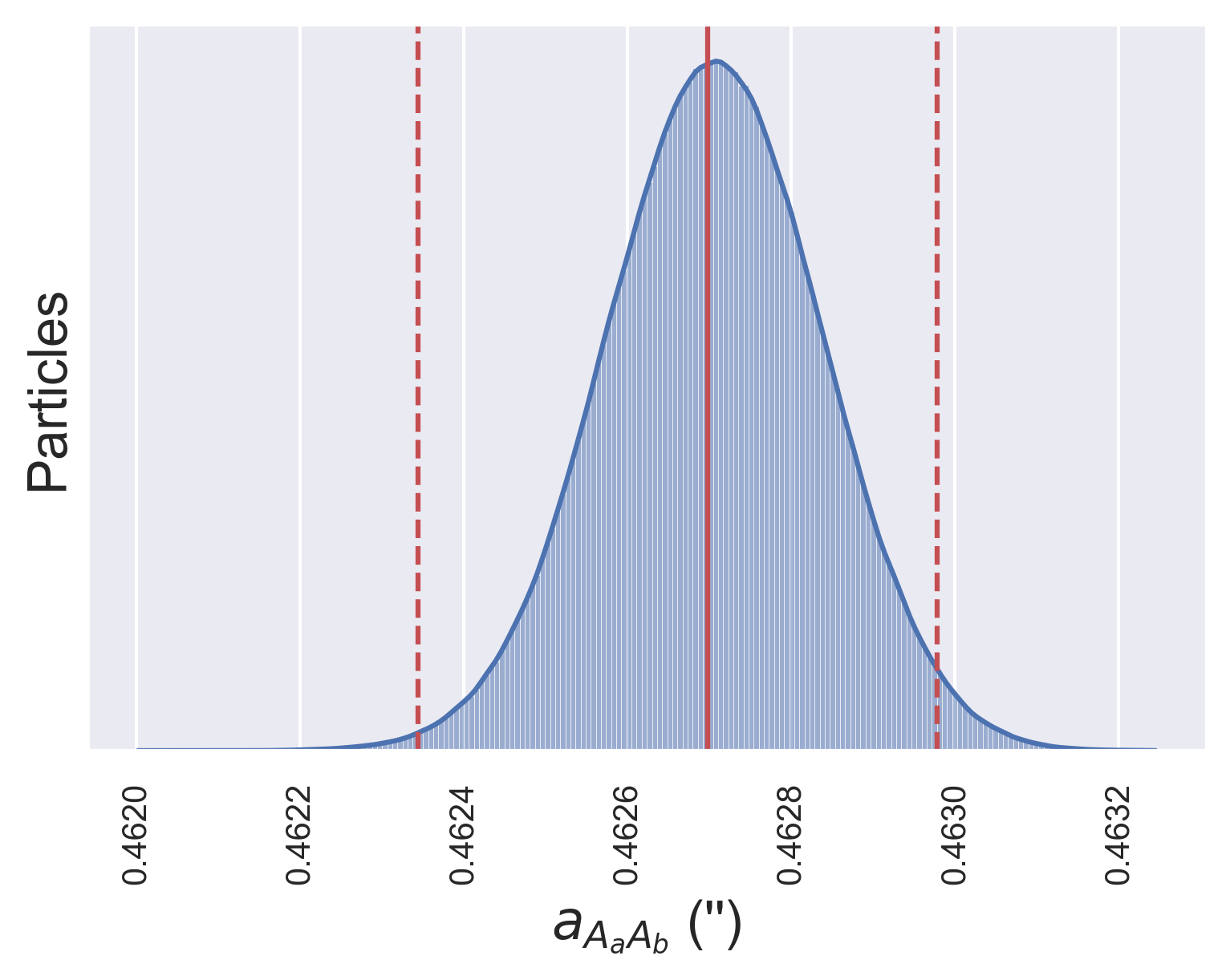}{0.33\textwidth}{}
    \fig{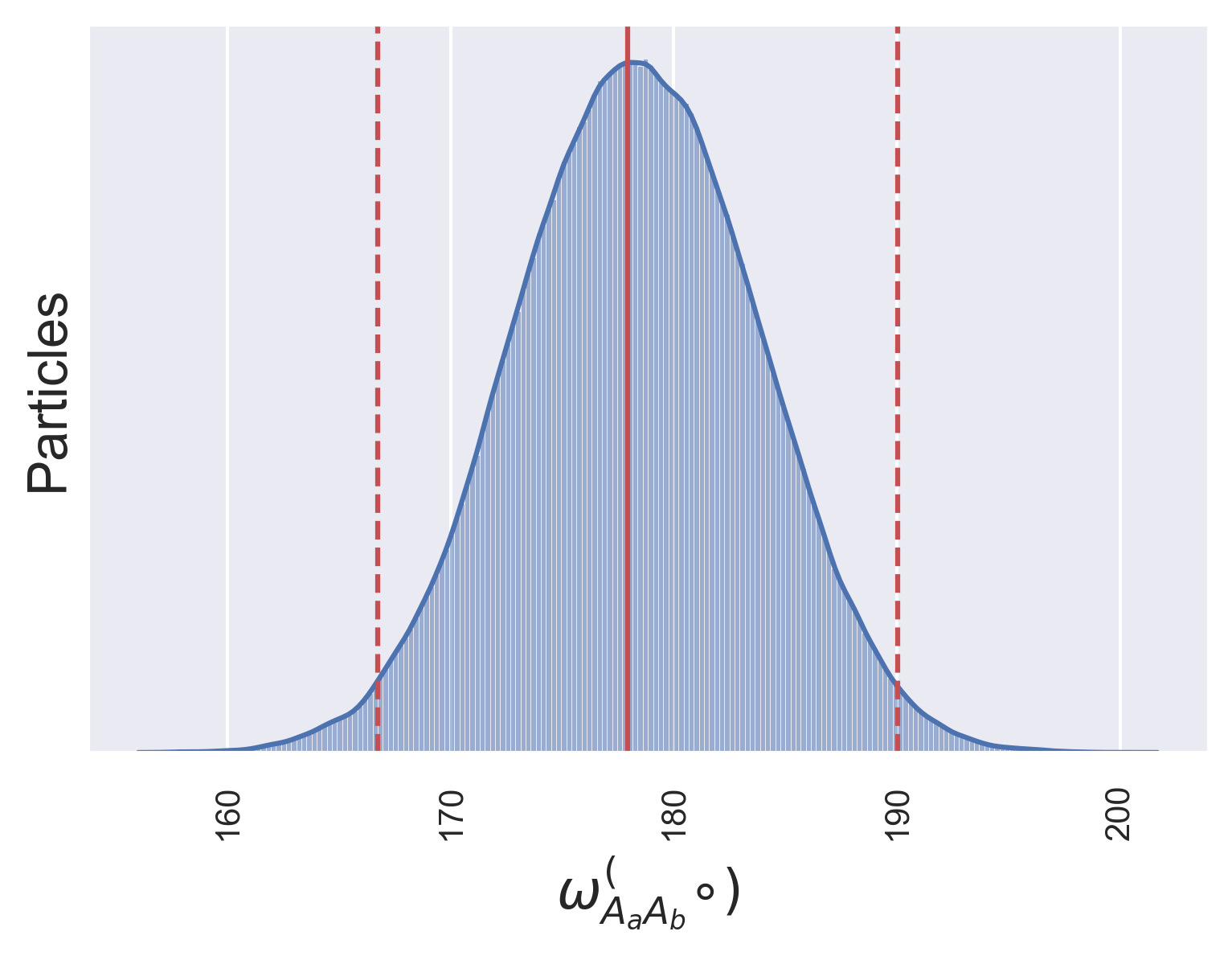}{0.33\textwidth}{}
    \fig{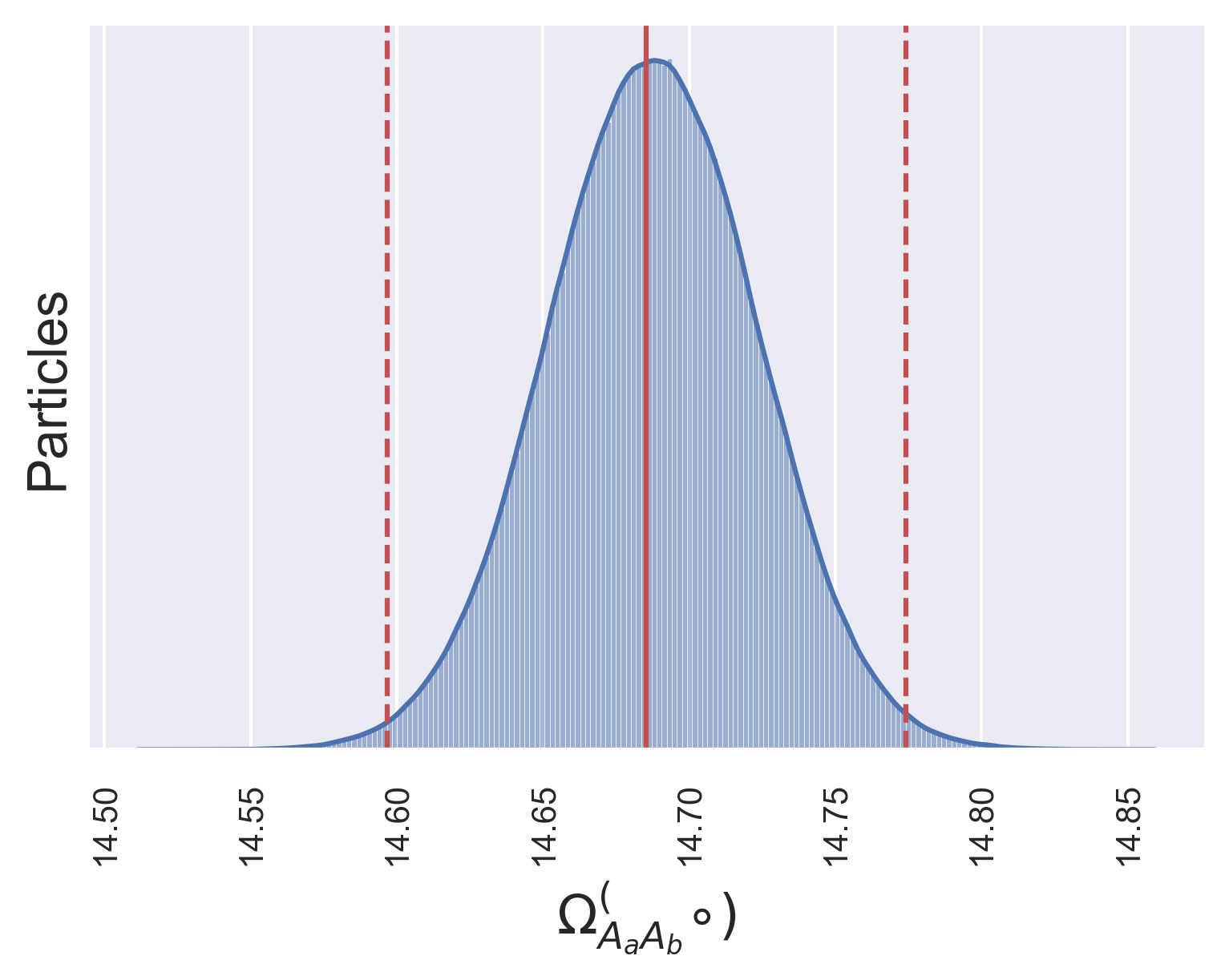}{0.33\textwidth}{}
    }
    \gridline{    
    \fig{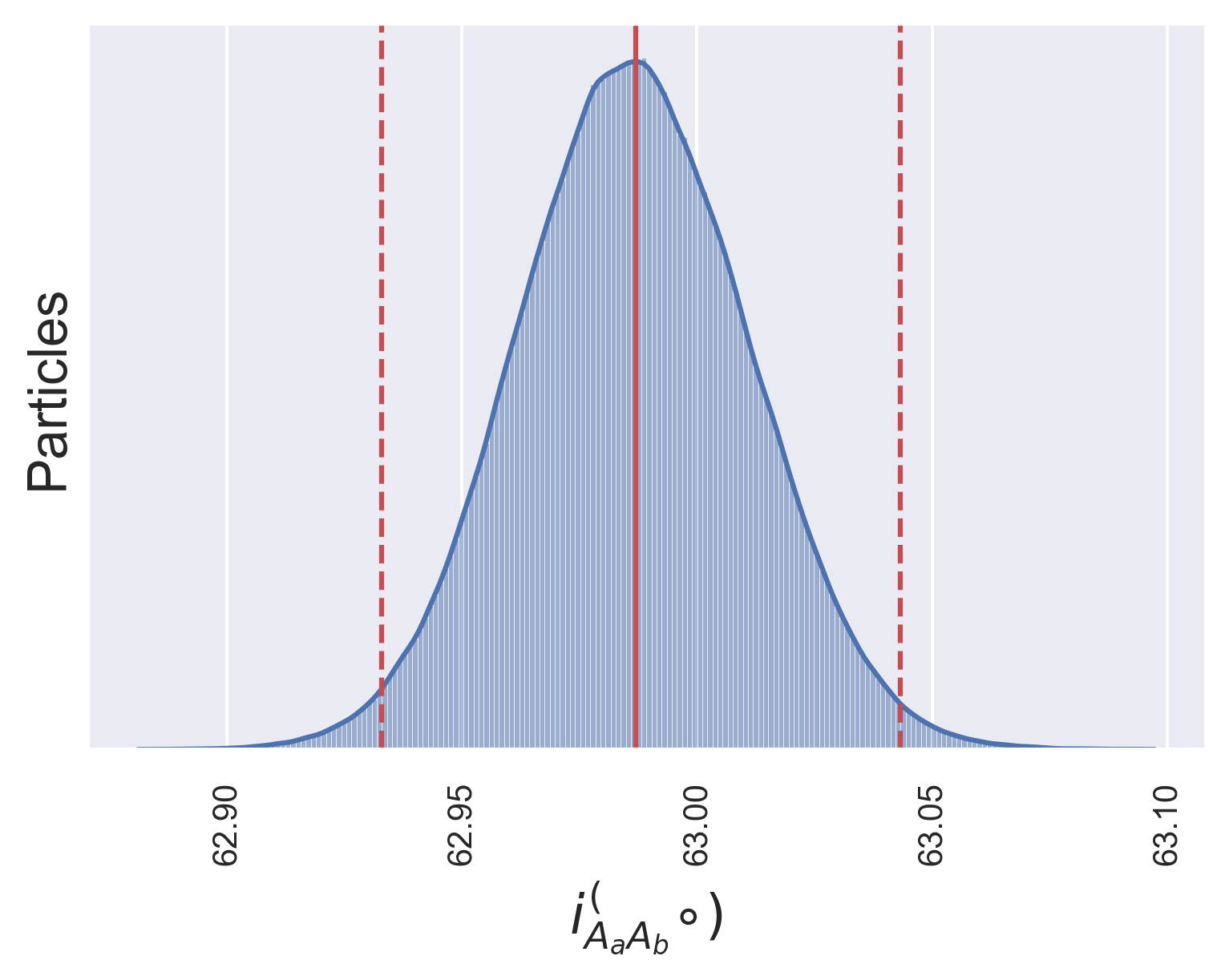}{0.33\textwidth}{}
    }      
\caption{Marginal empirical PDF distributions of the orbital elements obtained after performing the first stage for the inner orbit for LHS~1070. The MAP estimator is indicated by the red vertical line, while the lower and upper quartiles are shown in blue. The uni-modal Gaussian-like distributions indicate a robust fit, meaning that the corresponding parameters are well constrained.}
\label{fig:LHS1070inner}
\end{figure}

\begin{figure}[htp]
    \centering
    \gridline{
    \fig{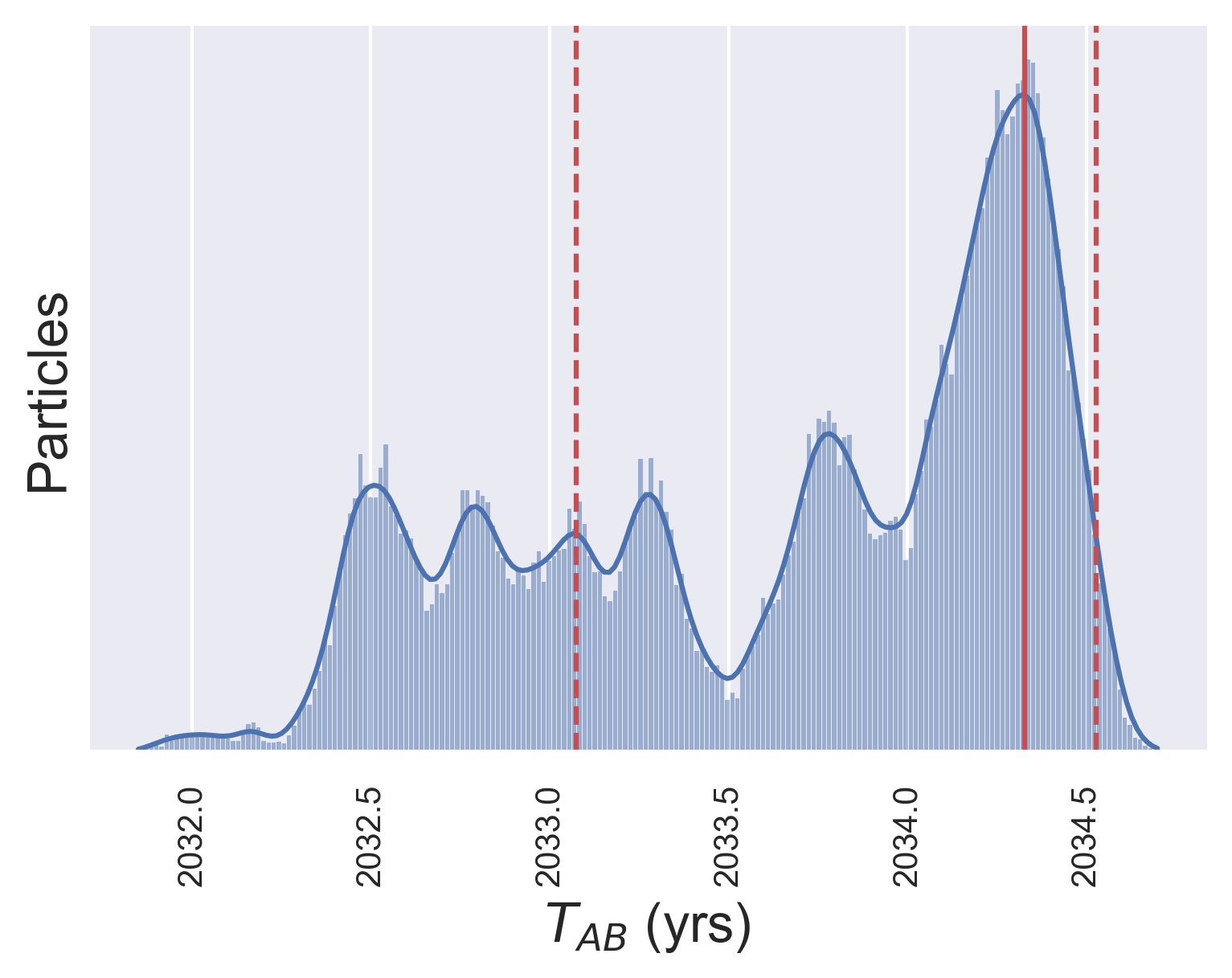}{0.33\textwidth}{}
    \fig{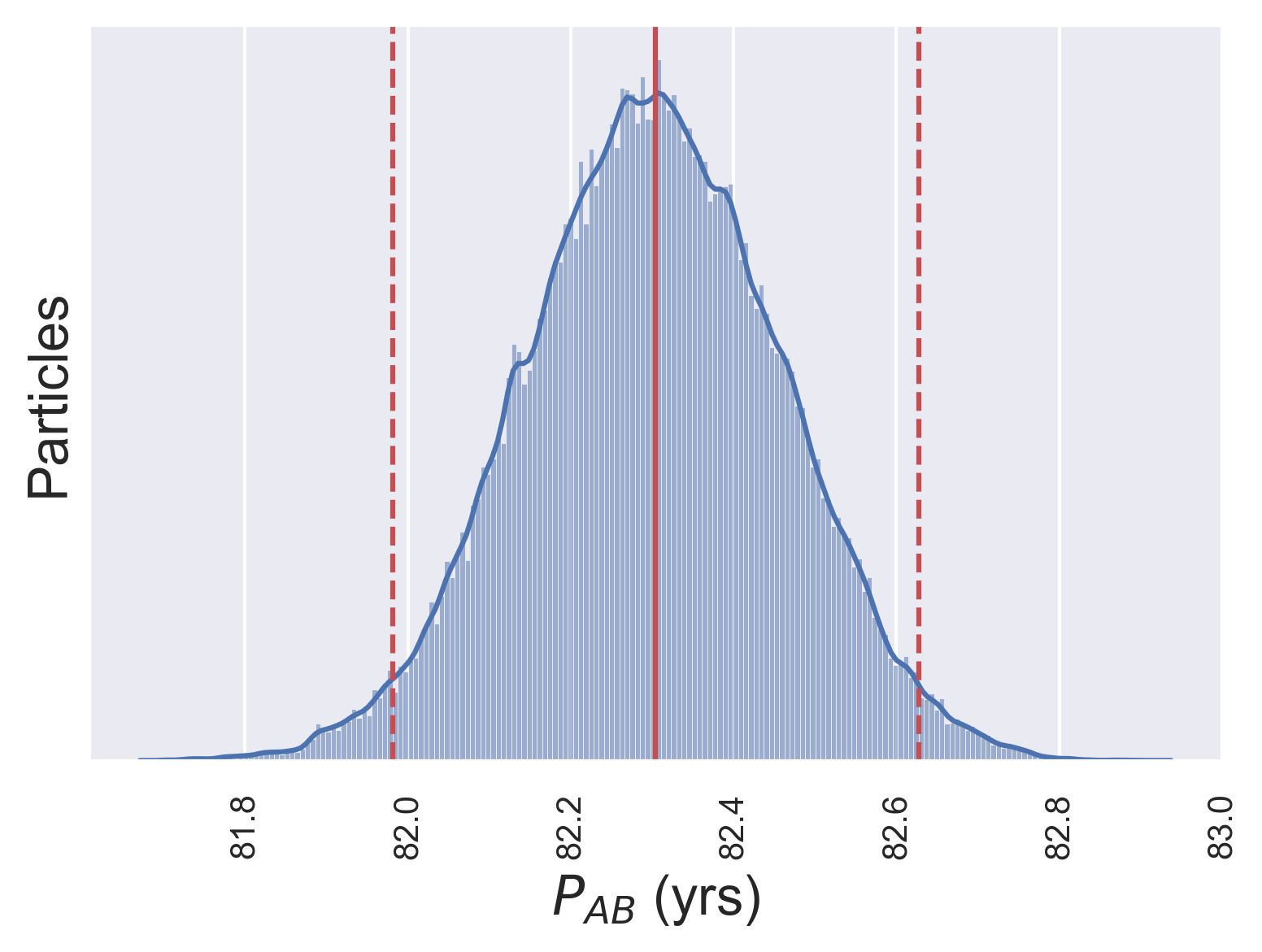}{0.33\textwidth}{}
    \fig{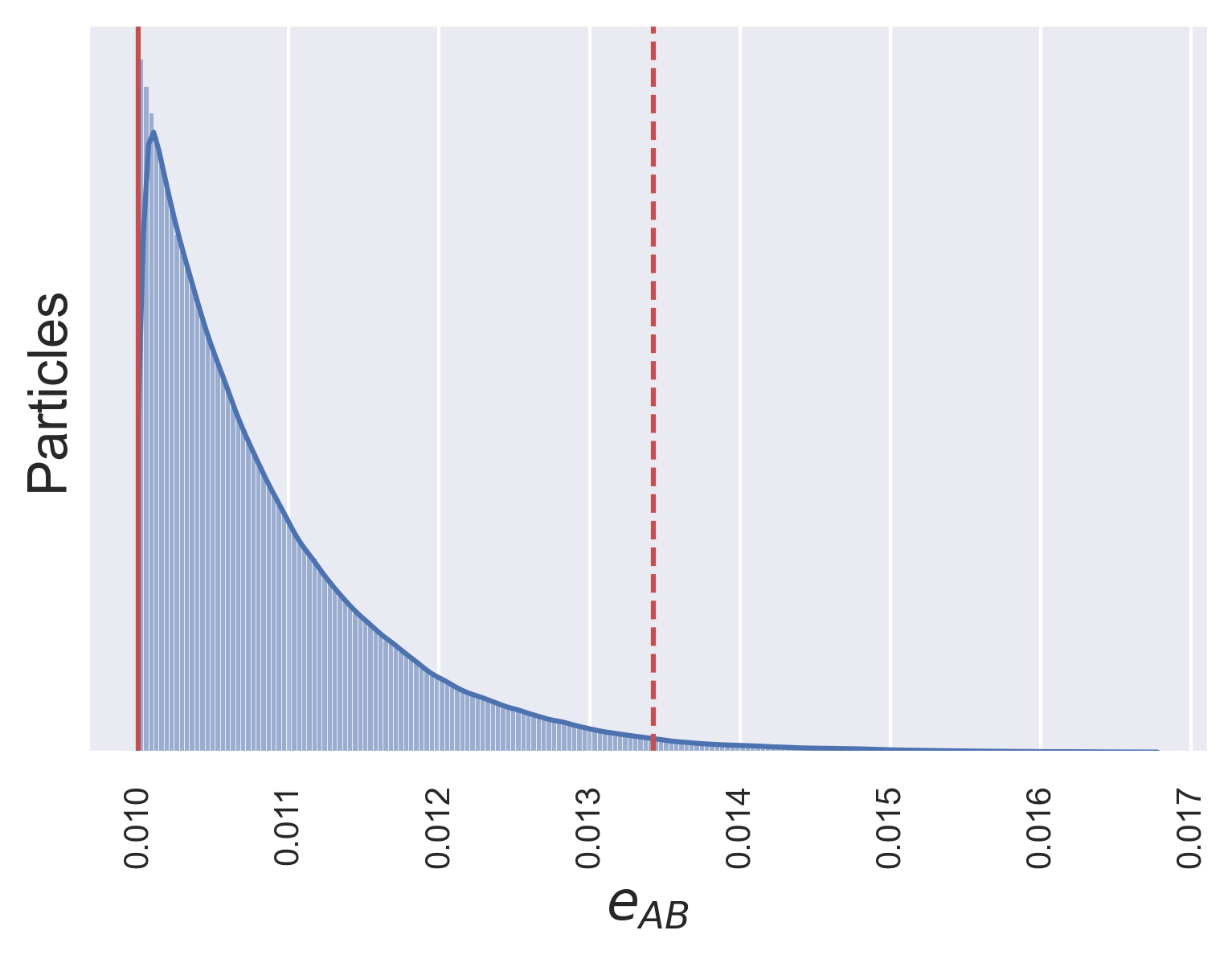}{0.33\textwidth}{}
    }
    \gridline{    
    \fig{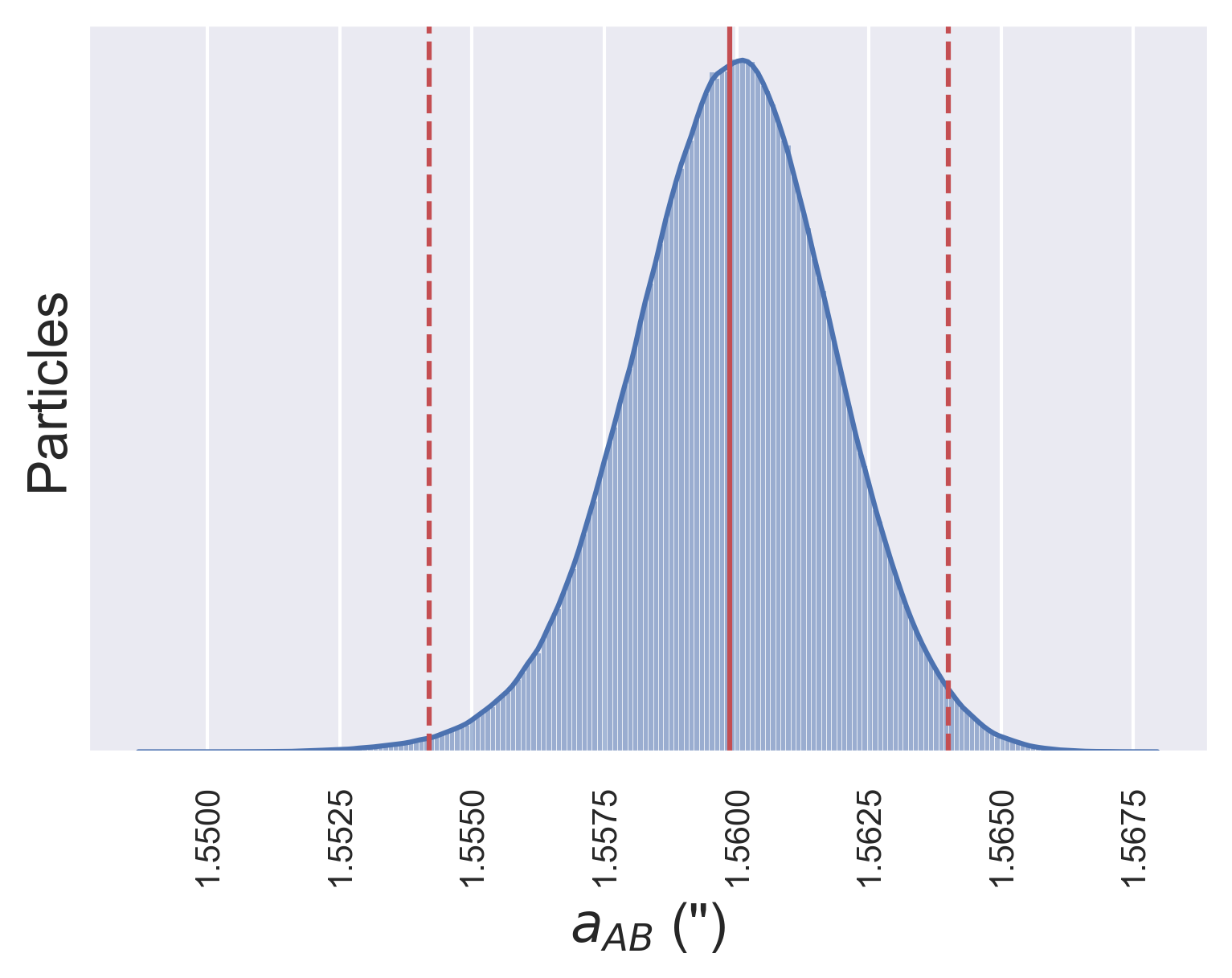}{0.33\textwidth}{}
    \fig{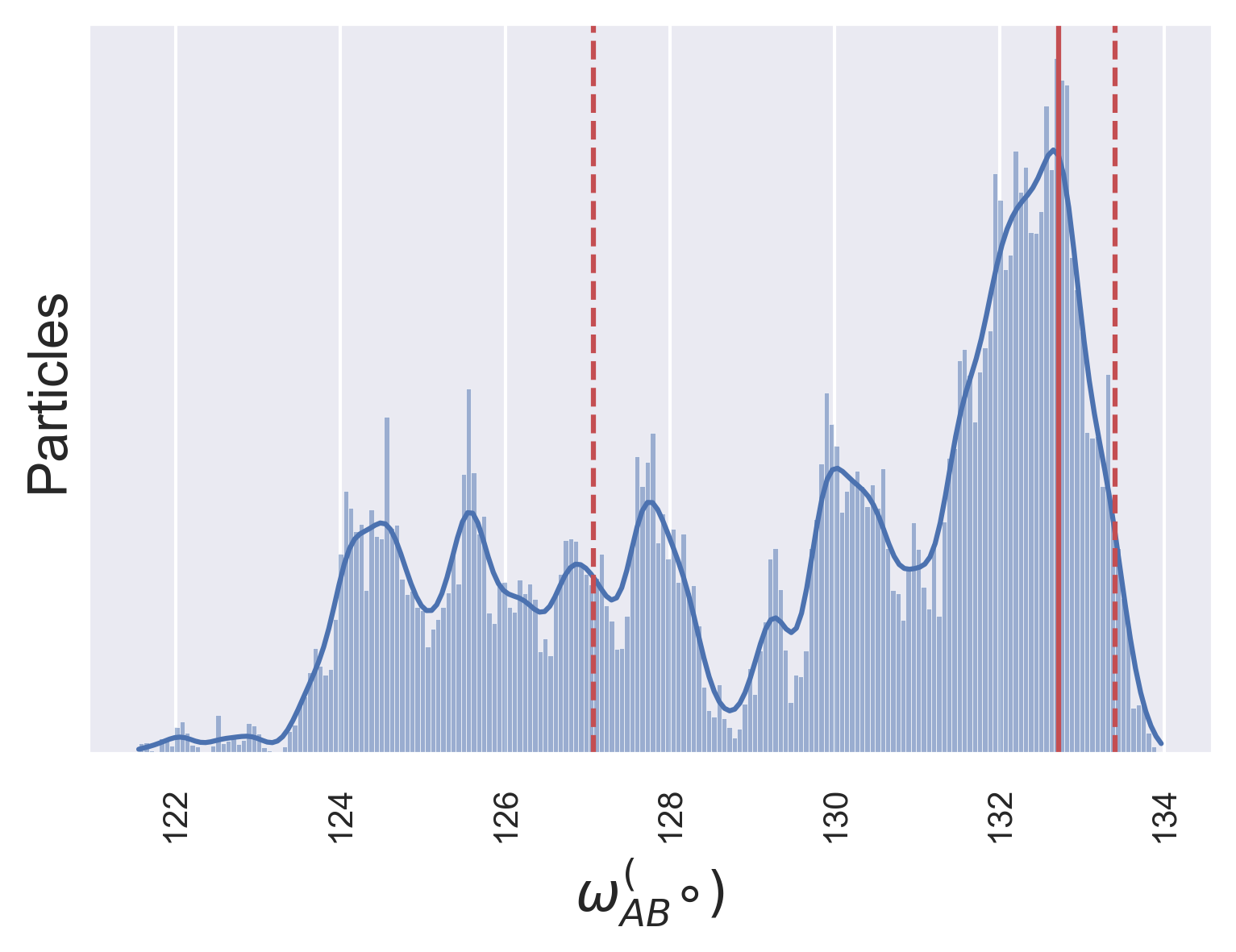}{0.33\textwidth}{}
    \fig{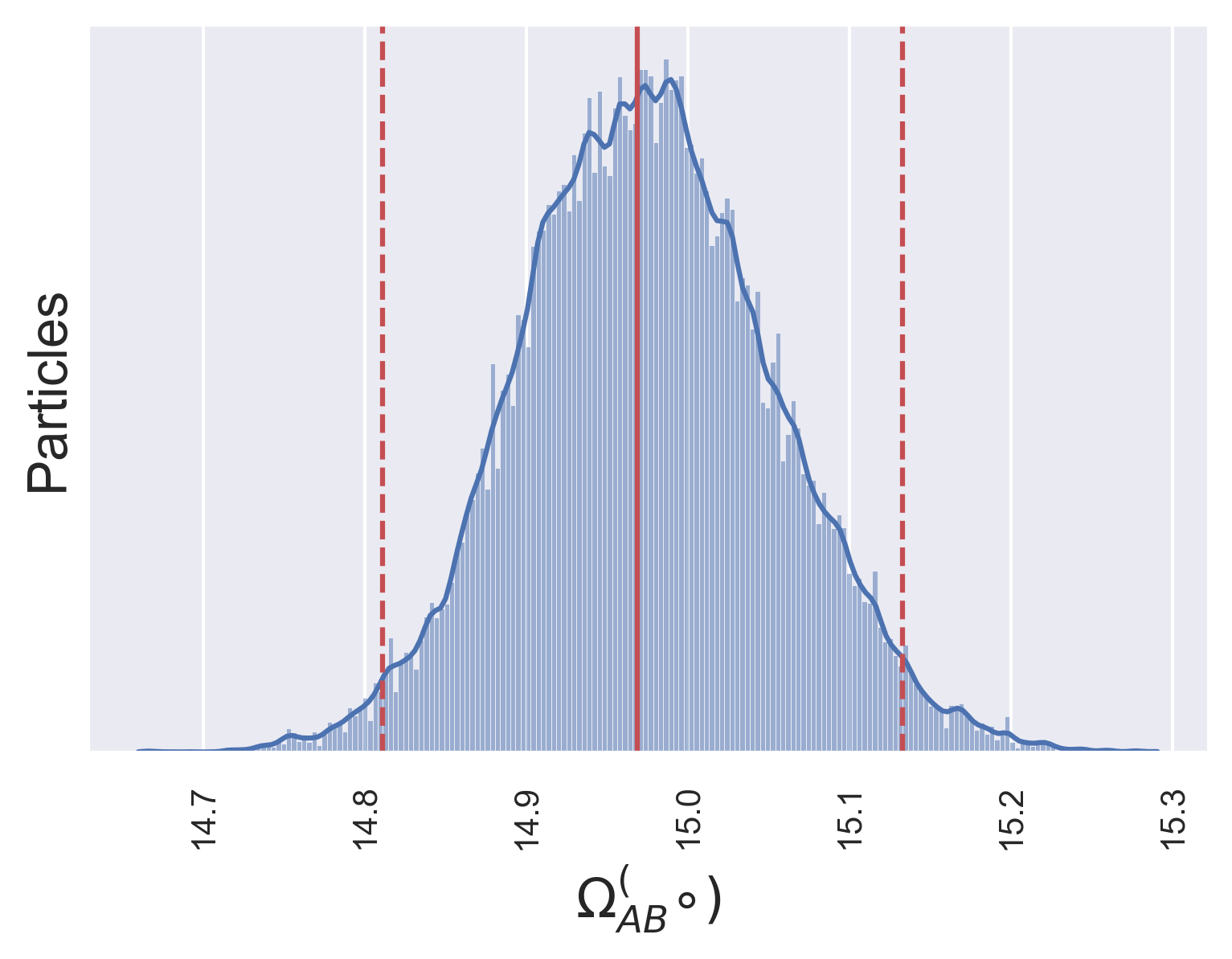}{0.33\textwidth}{}
    }
    \gridline{    
    \fig{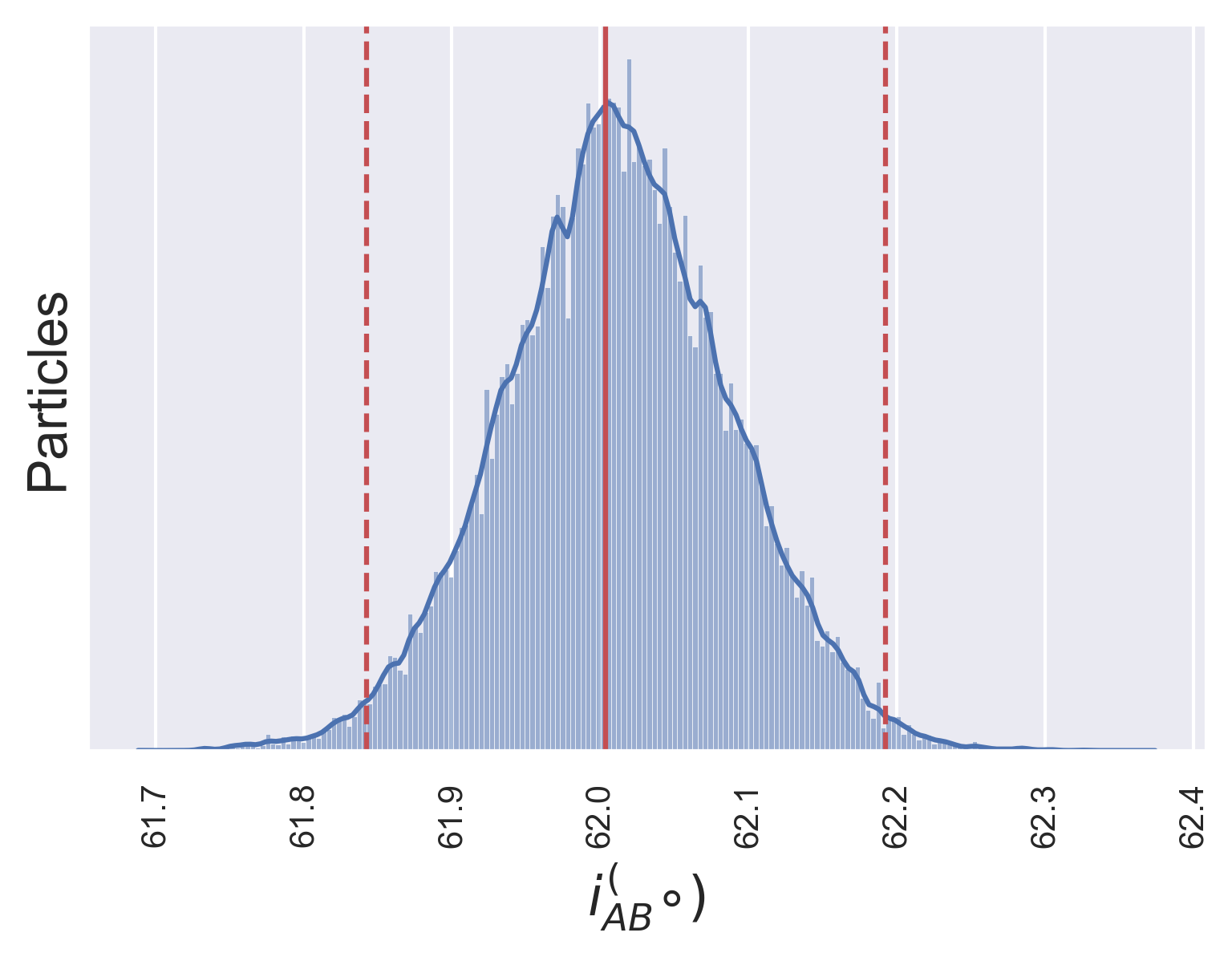}{0.33\textwidth}{}
    \fig{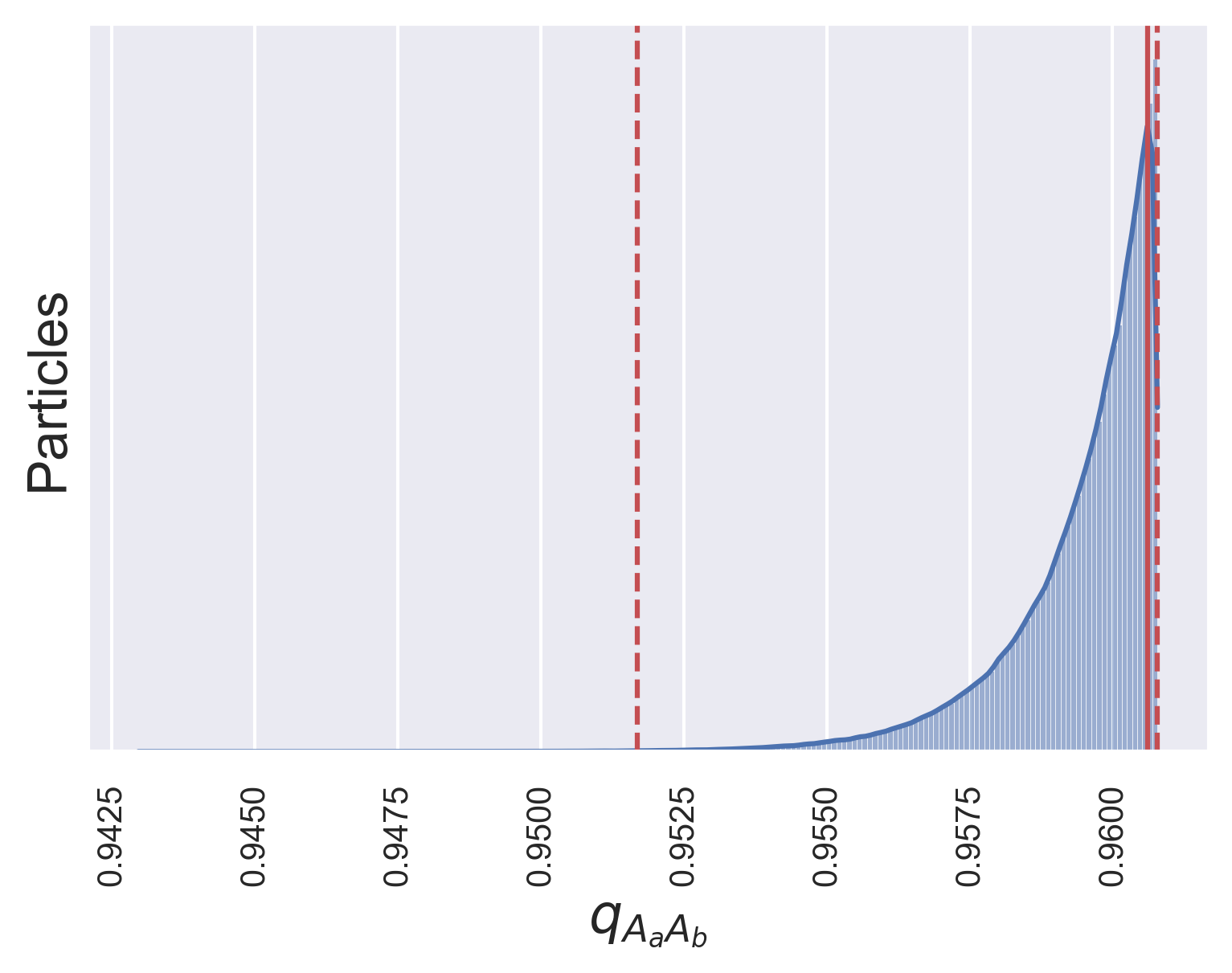}{0.33\textwidth}{}
    }
\caption{Marginal empirical PDF distributions of the orbital elements obtained
  after performing the second stage for the outer orbit of LHS~1070. The MAP
  estimator is indicated in red vertical line, while the lower and upper
  quartiles are shown in blue. It is clear that, in this case, the parameters
  exhibit a larger uncertainty than those of the inner system, shown in
  Figure~\ref{fig:LHS1070inner}. It may seem surprising at first that, in some
  cases, the MAP value does not necessarily coincide with the maximum of the
  marginal PDFs, but we must recall that MAP is the maximum of the joint
  (multi-dimensional) PDF, while the quartiles are defined in terms of the
  marginal PDF itself. We must also note that on these and other PDFs, the distributions look a bit ragged despite the fact that our solution is stable (see text), because they are based directly on the MCMC particles and the histograms have not been smoothed.}
\label{fig:LHS1070outer}
\end{figure}

The marginal PDF of individual parameters are built upon the histograms  shown in Figures~\ref{fig:LHS1070inner} and~\ref{fig:LHS1070outer} for the inner and outer systems, respectively. In these PDF plots, we indicate the best solution and the confidence intervals around it. In particular, we adopt as the ``best solution'' the one obtained from the MAP estimator, derived in turn from the posterior joint PDF distributions (see Section~\ref{sec:intro}). On the other hand, the confidence interval is obtained using \textit{modified} quartiles over the marginal distributions: the lower ($Q_1$) and upper ($Q_3$) quartiles were computed considering the MAP estimator as the median value ($Q_2$).

\subsection{WDS20396+0458 (HIP 101955)} \label{sec:case2}

This system, recognized as HD~196795 and GJ~795AB by SIMBAD, consists of an inner subsystem $A_a, A_b$ as the primary, and a distant object $B$, forming the outer subsystem known as KUI~99~AB in the WDs catalogue. Astrometry is available for the inner and outer orbits, and there are also RV measurements for $A_a$,  $A_b$ and $B$; then, the combined scenario described in Section~\ref{sec:comb} is performed. 
\begin{figure}[htp]
        \resizebox{\textwidth}{!}{
        \begin{tikzpicture}
        \tikzstyle{main}=[circle, minimum size = 10mm, thick, draw =black!80, fill=white!100, node distance = 16mm]
        \tikzstyle{connect}=[-latex, thick]
        \tikzstyle{box}=[rectangle, draw=black!100]
        \tikzstyle{ell}=[ellipse, draw=black!80]   

        \node[main] (Omega12) [] {$\Omega_{A_{a}A_{b}}$};
        \node[main] (a12) [right=of Omega12] {$a_{A_{a}A_{b}}$};
        \node[main] (zi) [below  = 6cm of a12] { $\Vec{z}_{1}(\tau_k)$}; 
        \node[main] (i12) [right=of a12] {$i_{A_{a}A_{b}}$}; 
        \node[main] (q12) [right=of i12] {$q_{A_{a}A_{b}}$}; 
        \node[main] (T12) [right=of q12] {$T_{A_{a}A_{b}}$};
        \node[main] (P12) [right=of T12] {$P_{A_{a}A_{b}}$};
        \node[main] (e12) [right=of P12] {$e_{A_{a}A_{b}}$};
        \node[main] (omega12) [right=of e12] {$\omega_{A_{a}A_{b}}$};
        \node[main] (VCoM) [right=of omega12] {$V_{CoM}$}; 
        \node[main] (Omega) [below = 6cm of zi] {$\Omega_{AB}$};
        \node[main] (a) [right=of Omega] {$a_{AB}$};
        \node[main] (zo) [below  = 6cm of i12] {$\Vec{z}_{2}(\tau_k)$};
        \node[main] (i) [right=of a] {$i_{AB}$};  
        \node[main] (T) [right= 6cm of i] {$T_{AB}$};
        \node[main] (P) [right=of T] {$P_{AB}$};
        \node[main] (e) [right=of P] {$e_{AB}$};
        \node[main] (omega) [right=of e] {$\omega_{AB}$};
        \node[main] (q) [right=of omega] {$q_{AB}$}; 
        \node[main] (K1) [below=3cm of VCoM] {$K_1$};
        \node[main] (K2) [right=of K1] {$K_2$};
        \node[main] (K3) [above=3cm of omega] {$K_3$};
        \node[main] (K4) [right=of K3] {$K_4$};        
        \node[ell] (v_obs) [below =6cm of e12] {$\{{z_3}(\tau_k)_{k=1}^{N_3}, {z_4}(\tau_k)_{k=1}^{N_4},  {z_5}(\tau_k)_{k=1}^{N_5}\}$}; 

          \path (T12) edge [left]  (zi);
          \path (P12) edge [left]  (zi);
          \path (e12) edge [left]  (zi);
          \path (a12) edge [left]  (zi);
          \path (omega12) edge [left]  (zi);
          \path (Omega12) edge [connect]  (zi);  
          \path (i12) edge [connect]  (zi);
          \path (T) edge [connect]  (zo);
          \path (P) edge [connect]  (zo);
          \path (e) edge [connect]  (zo);
          \path (a) edge [connect]  (zo);
          \path (omega) edge [connect]  (zo);  
          \path (Omega) edge [connect]  (zo);    
          \path (i) edge [connect]  (zo);    
          \path (q12) edge [connect]  (zo);    
          \path (omega) edge [connect]  (zo);    
          \path (zi) edge [connect] (zo);
          \path (T12) edge [connect] (v_obs);
          \path (P12) edge [connect] (v_obs);
          \path (e12) edge [connect] (v_obs);
          \path (T12) edge [connect, draw=blue!100] (K1);
          \path (P12) edge [connect, draw=blue!100] (K1);
          \path (e12) edge [connect, draw=blue!100] (K1);
          \path (T12) edge [connect, draw=blue!100] (K2);
          \path (P12) edge [connect, draw=blue!100] (K2);
          \path (e12) edge [connect, draw=blue!100] (K2);          
          \path (a12) edge [connect, draw=blue!100] (K1);
          \path (i12) edge [connect, draw=blue!100] (K1);
          \path (a12) edge [connect, draw=blue!100] (K2);
          \path (i12) edge [connect, draw=blue!100] (K2);          
          \path (omega12) edge [connect] (v_obs);
          \path (q12) edge [connect, draw=blue!100] (K1);
          \path (q12) edge [connect, draw=blue!100] (K2);
          \path (T) edge [connect] (v_obs);
          \path (P) edge [connect] (v_obs);
          \path (e) edge [connect] (v_obs);
          \path (T) edge [connect, draw=blue!100] (K3);
          \path (P) edge [connect, draw=blue!100] (K3);
          \path (e) edge [connect, draw=blue!100] (K3);       
          \path (T) edge [connect, draw=blue!100] (K4);
          \path (P) edge [connect, draw=blue!100] (K4);
          \path (e) edge [connect, draw=blue!100] (K4);              
          \path (a) edge [connect, draw=blue!100] (K3);
          \path (i) edge [connect, draw=blue!100] (K3);
          \path (a) edge [connect, draw=blue!100] (K4);
          \path (i) edge [connect, draw=blue!100] (K4);          
          \path (omega) edge [connect] (v_obs);
          \path (q) edge [connect, draw=blue!100] (K3);
          \path (q) edge [connect, draw=blue!100] (K4);
          \path (VCoM) edge [connect] (v_obs);  
          \path (K1) edge [connect, draw=red!100] (v_obs);  
          \path (K2) edge [connect, draw=red!100] (v_obs);  
          \path (K3) edge [connect, draw=red!100] (v_obs);  
          \path (K4) edge [connect, draw=red!100] (v_obs);  
		 \node[rectangle, inner sep=1mm, fit= (zo), label=left :\textbf{$_{k=1}^{N_2}$}] {};	
          \node[rectangle, inner sep=1.5mm, draw=black!100, dashed, fit = (zo) ] {};     
		 \node[rectangle, inner sep=1mm, fit= (zi), label=left :\textbf{$_{k=1}^{N_1}$}] {};	
          \node[rectangle, inner sep=1.5mm, draw=black!100, dashed, fit = (zi) ] {};              
          \node[rectangle, inner sep=2mm, fit= (a12) (i12) (a) (i) (q12) (VCoM) (v_obs) (q),label=above right :\textbf{\textcolor{blue}{RV scenario}}] {};	
          \node[rectangle, inner sep=1.5mm, draw=blue!100, fill=blue!50, opacity=0.2, fit = (a12) (i12) (a) (i) (q12) (VCoM) (v_obs) (q)] {};   
        \end{tikzpicture}
        }
    \caption{Graphical model representation for the RV component of the fit in the combined (self-consistent) astrometric plus RV scenario as described in Section~\ref{sec:comb} (see the blue panel on Figure~\ref{fig:graphModelCombined}). Blue arrows represent relationships between parameters. Magenta arrows show the relationship between RV amplitudes and observations. Some relationships may not be so obvious, e.g., the time of passage by periastron defines the phase zero-point of the radial velocity curve and, as such, it determines the instant (in phase) where the velocity semi-amplitude reaches its maximum (and minimum), but of course not the value of the semi-amplitude itself. Additionally, this parameter enters into the calculation of the true anomaly (see Section~\ref{app:trueanomaly}), which is in turn used to compute the radial velocities (see Section~\ref{app:rv_eq}.).}

    \label{fig:graphModelRV0}
\end{figure}
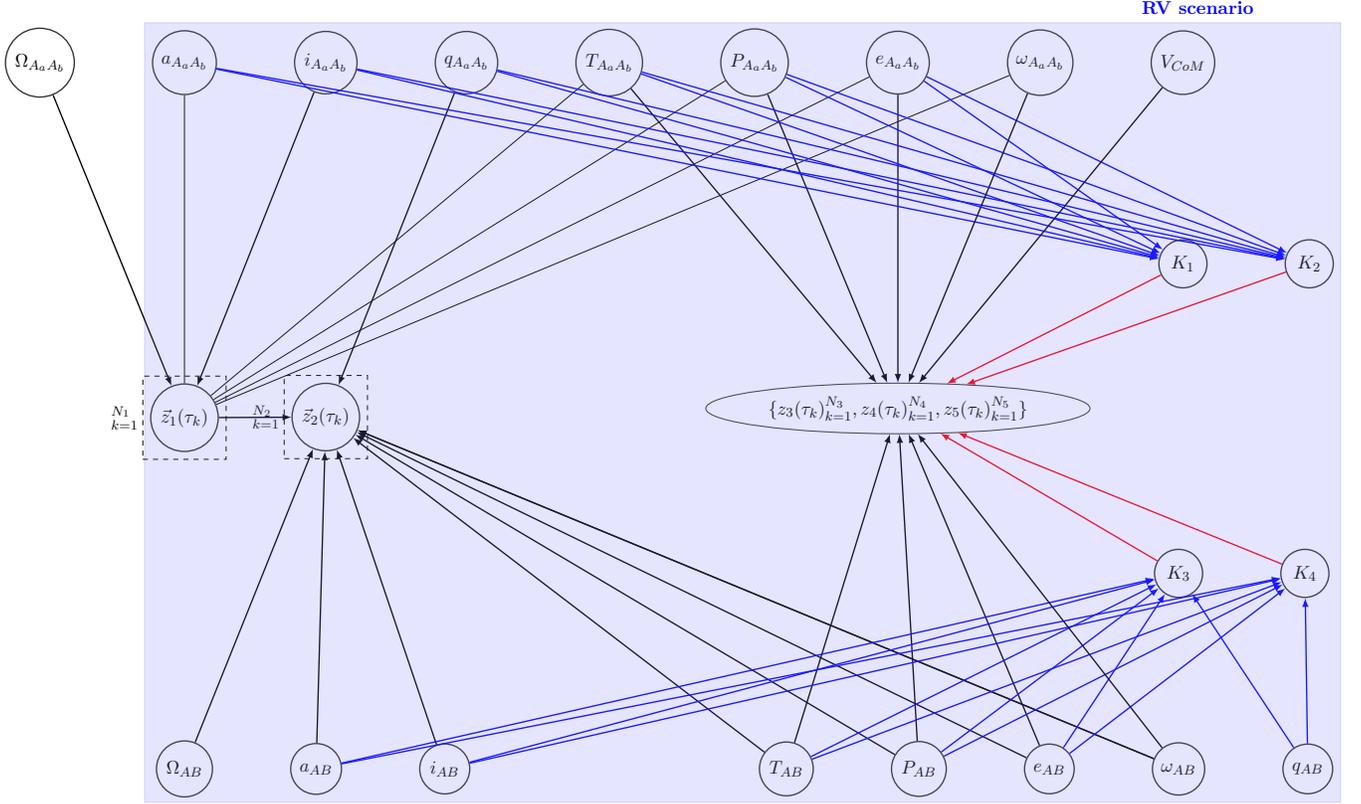

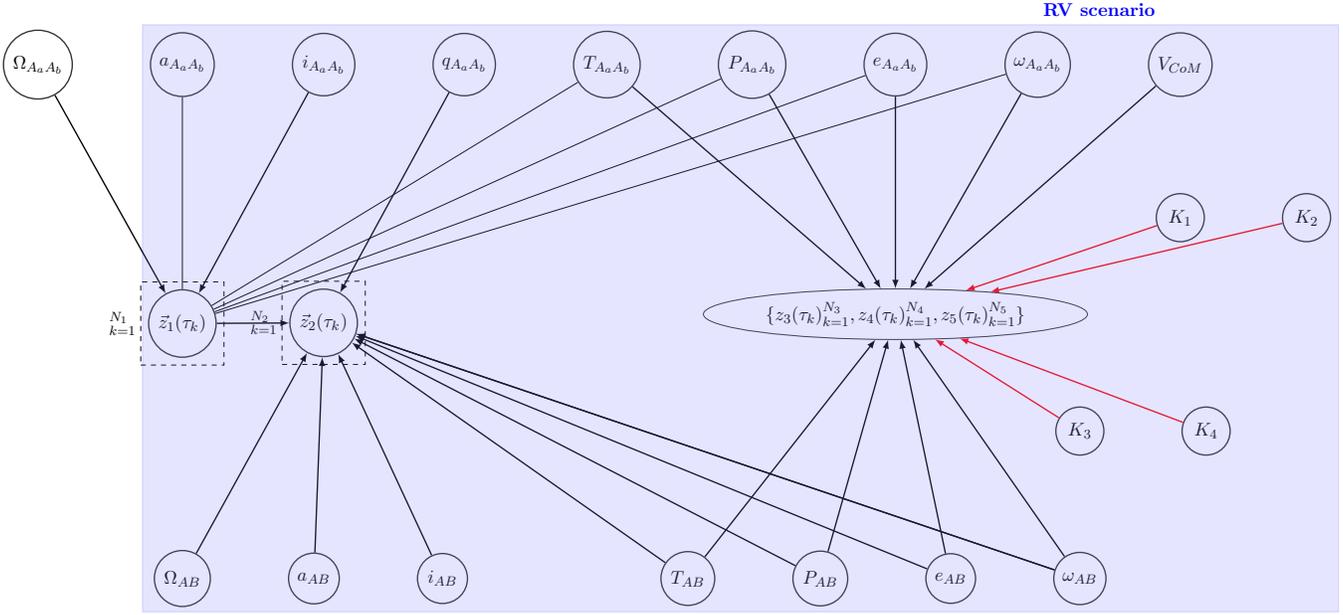
\begin{figure}[htp]
        \resizebox{\textwidth}{!}{
        \begin{tikzpicture}
        \tikzstyle{main}=[circle, minimum size = 10mm, thick, draw =black!80, fill=white!100, node distance = 16mm]
        \tikzstyle{connect}=[-latex, thick]
        \tikzstyle{box}=[rectangle, draw=black!100]
        \tikzstyle{ell}=[ellipse, draw=black!80]   

        \node[main] (Omega12) [] {$\Omega_{A_{a}A_{b}}$};
        \node[main] (a12) [right=of Omega12] {$a_{A_{a}A_{b}}$};
        \node[main] (zi) [below  = 4cm of a12] { $\Vec{z}_{1}(\tau_k)$}; 
        \node[main] (i12) [right=of a12] {$i_{A_{a}A_{b}}$}; 
        \node[main] (q12) [right=of i12] {$q_{A_{a}A_{b}}$}; 
        \node[main] (T12) [right=of q12] {$T_{A_{a}A_{b}}$};
        \node[main] (P12) [right=of T12] {$P_{A_{a}A_{b}}$};
        \node[main] (e12) [right=of P12] {$e_{A_{a}A_{b}}$};
        \node[main] (omega12) [right=of e12] {$\omega_{A_{a}A_{b}}$};
        \node[main] (VCoM) [right=of omega12] {$V_{CoM}$}; 
        \node[main] (Omega) [below = 4cm of zi] {$\Omega_{AB}$};
        \node[main] (a) [right=of Omega] {$a_{AB}$};
        \node[main] (zo) [below  = 4cm of i12] {$\Vec{z}_{2}(\tau_k)$};
        \node[main] (i) [right=of a] {$i_{AB}$};  
        \node[main] (T) [right= 4cm of i] {$T_{AB}$};
        \node[main] (P) [right=of T] {$P_{AB}$};
        \node[main] (e) [right=of P] {$e_{AB}$};
        \node[main] (omega) [right=of e] {$\omega_{AB}$};
        \node[main] (K1) [below=2cm of VCoM] {$K_1$};
        \node[main] (K2) [right=of K1] {$K_2$};
        \node[main] (K3) [above=2cm of omega] {$K_3$};
        \node[main] (K4) [right=of K3] {$K_4$};        
        \node[ell] (v_obs) [below =4cm of e12] {$\{{z_3}(\tau_k)_{k=1}^{N_3}, {z_4}(\tau_k)_{k=1}^{N_4},  {z_5}(\tau_k)_{k=1}^{N_5}\}$}; 

          \path (T12) edge [left]  (zi);
          \path (P12) edge [left]  (zi);
          \path (e12) edge [left]  (zi);
          \path (a12) edge [left]  (zi);
          \path (omega12) edge [left]  (zi);
          \path (Omega12) edge [connect]  (zi);  
          \path (i12) edge [connect]  (zi);
          \path (T) edge [connect]  (zo);
          \path (P) edge [connect]  (zo);
          \path (e) edge [connect]  (zo);
          \path (a) edge [connect]  (zo);
          \path (omega) edge [connect]  (zo);  
          \path (Omega) edge [connect]  (zo);    
          \path (i) edge [connect]  (zo);    
          \path (q12) edge [connect]  (zo);    
          \path (omega) edge [connect]  (zo);    
          \path (zi) edge [connect] (zo);
          \path (T12) edge [connect] (v_obs);
          \path (P12) edge [connect] (v_obs);
          \path (e12) edge [connect] (v_obs);
          \path (omega12) edge [connect] (v_obs);
          \path (T) edge [connect] (v_obs);
          \path (P) edge [connect] (v_obs);
          \path (e) edge [connect] (v_obs);
          \path (omega) edge [connect] (v_obs);
          \path (VCoM) edge [connect] (v_obs);  
          \path (K1) edge [connect, draw=red!100] (v_obs);  
          \path (K2) edge [connect, draw=red!100] (v_obs);  
          \path (K3) edge [connect, draw=red!100] (v_obs);  
          \path (K4) edge [connect, draw=red!100] (v_obs);  
		 \node[rectangle, inner sep=1mm, fit= (zo), label=left :\textbf{$_{k=1}^{N_2}$}] {};	
          \node[rectangle, inner sep=1.5mm, draw=black!100, dashed, fit = (zo) ] {};     
		 \node[rectangle, inner sep=1mm, fit= (zi), label=left :\textbf{$_{k=1}^{N_1}$}] {};	
          \node[rectangle, inner sep=1.5mm, draw=black!100, dashed, fit = (zi) ] {};              
          \node[rectangle, inner sep=2mm, fit= (a12) (i12) (a) (i) (q12) (VCoM) (v_obs) (K1) (K2) (K3) (K4),label=above right :\textbf{\textcolor{blue}{RV scenario}}] {};	
          \node[rectangle, inner sep=1.5mm, draw=blue!100, fill=blue!50, opacity=0.2, fit = (a12) (i12) (a) (i) (q12) (VCoM) (v_obs)  (K1) (K2) (K3) (K4)] {};   
        \end{tikzpicture}
        }
    \caption{Graphical model representation for the RV component of the the fit in the simplified (not-self-consistent) astrometric plus RV scenario. Red arrows show the relationship between RV amplitudes and observations. This is the scheme finally adopted for the HIP101955 and HIP111805 systems for the reasons explained in the text.}
    \label{fig:graphModelRV01}
\end{figure}

\paragraph{Inner system:} The inner system is known to be highly eccentric \citep{malogolovets2007nearby, tokovinin2017}, therefore the prior for $ e \in [0.5,0.8]$. On the other hand, \citet{duquennoy1987study} reports that the period is $\sim 2.5$~yr, so the prior for $P$ is chosen uniform in the interval $[1.5, 3.5]$~yr. 

Due to apparent inconsistencies in the data (see below for further details), \citet{tokovinin2017} found different values for the mass ratio from the astrometric wobble ($q=0.84$) and from spectroscopy ($q=0.45$). As our method does not consider the amplitudes $K_1$ and $K_2$ as independent variables (instead, they are derived in a dynamically self-consistent way from the astrometric orbit and the RV curve, see Figure~\ref{fig:graphModelCombined}), we established a search range for $q$ in the interval $[0.7, 0.95]$, which is in agreement with the fractional mass $f \in [0.42, 0.48]$ and the value of $q=0.8$ from \citet{malogolovets2007nearby}. For our solution, all the speckle measurements before 1981 were considered, but their associated error was increased considerably with respect of the more recent measurements. 

\begin{figure}[htp]
    \centering
    \gridline{
    \fig{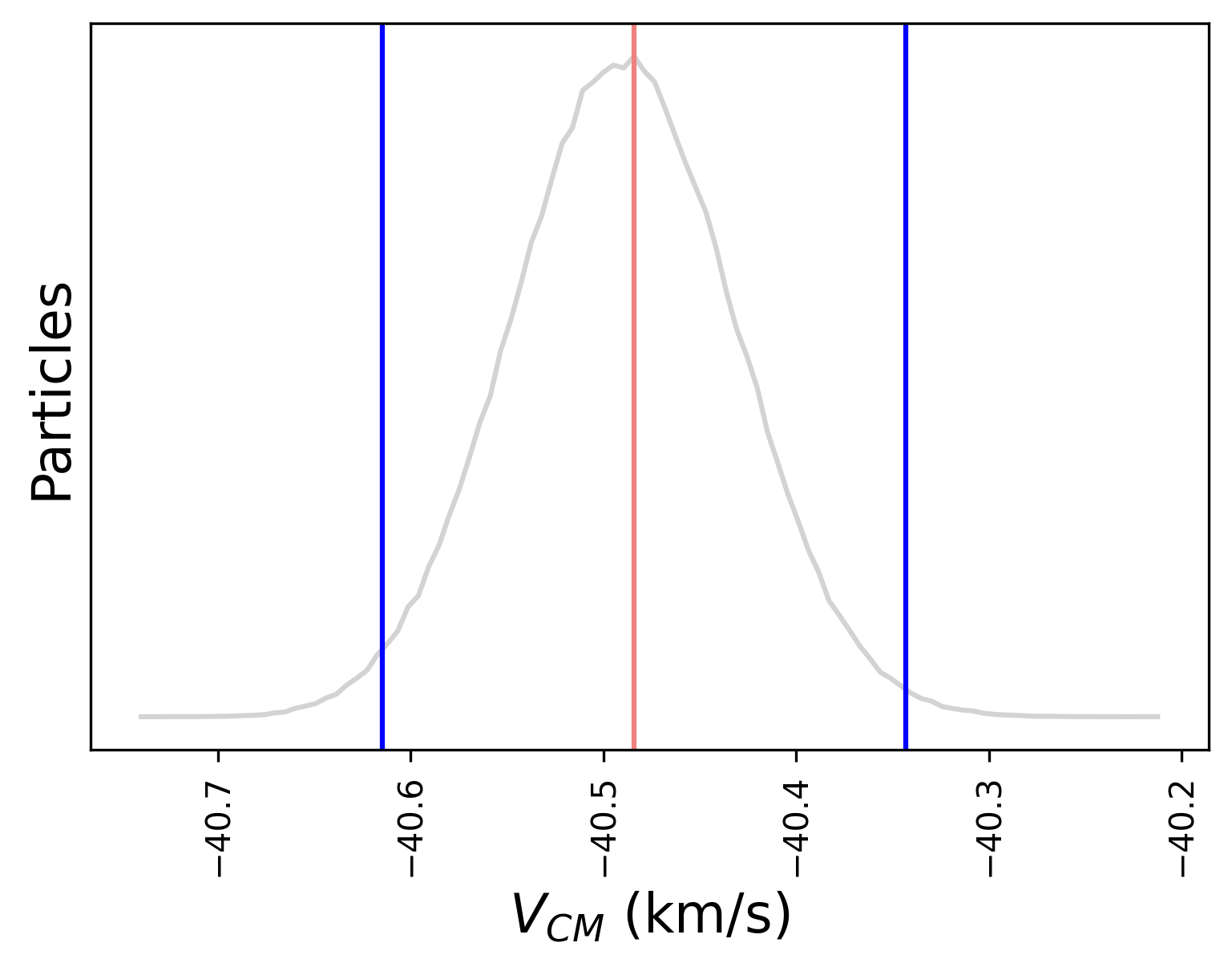}{0.4\textwidth}{}
    \fig{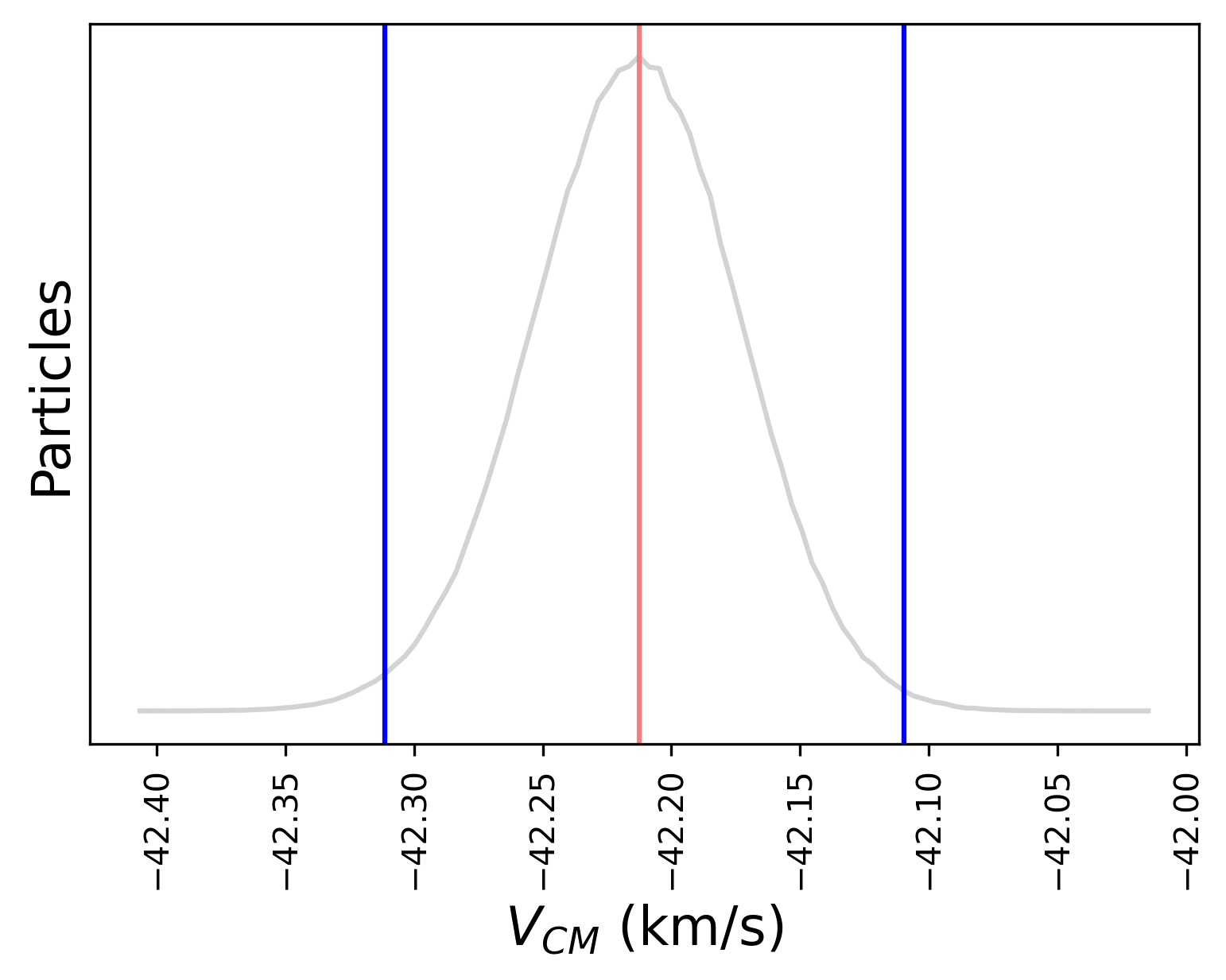}{0.4\textwidth}{}
    }  
\caption{$V_{COM}$ histograms using \citep{tokovinin2017} (left) and SB9 (right) data sets for HIP101955. \label{fig:KUI99_correction}}
\end{figure}

Due to a systematic offset between the RV data from \citep{tokovinin2017} and the SB9 data (measured by CORAVEL), reflected in the corresponding $V_{COM}$ (see Figure \ref{fig:KUI99_correction}), we computed the best parameters using these data separately. We then applied an offset of $\sim 2 \frac{km}{s}$ to the CORAVEL measurements, and a new full joint solution was performed using both datasets simultaneously.

\paragraph{Outer system:} The period is known to be $\sim 40$~years, with a small eccentricity and a highly inclined orbit \citep{baize1981orbital, heintz1984orbits, malogolovets2007nearby}. Then, the priors for these parameters were set in the range $[35, 45]$~year, $[0, 0.25]$, and $[0, 180]$~deg, respectively. 

As reported by \citet{tokovinin2017}, the RV amplitudes for, both, the $A_aA_b$ pair as well as for the $B$ component should be taken with caution since they were blended in the spectra, and difficult to measure. This poses a problem for our methodology, since our imposed self-consistency between the amplitudes $K_1$ and $K_2$ and the astrometric orbit (as shown in Figure~\ref{fig:graphModelCombined}) is not guaranteed by the data. In Figure~\ref{fig:graphModelRV0} we show an expanded version of the blue box of Figure~\ref{fig:graphModelCombined} that makes explicit the dependency of the radial velocity amplitudes on the orbital parameters in the self-consistent combined scenario, in graphical form. To circumvent this observational issue with the spectroscopic data (which affects the amplitudes, but not the periods), we adopt a simplified scheme, where we fit the RV amplitudes {\it independently} of the rest of the parameters. This is graphically shown in Figure~\ref{fig:graphModelRV01}. By doing this, we find that we are able to obtain a better match to the RV curves of both the inner and outer components. We note that this is the same approach adopted by \citet{tokovinin2017} (see the (b) panels on their  Figures~ 5 and 6). As a corollary of this, we are in a position to make a fair comparison between our results and theirs, which is presented in Section~\ref{sec:anaben}.

\begin{figure}[htp]
    \centering
    \gridline{
    \fig{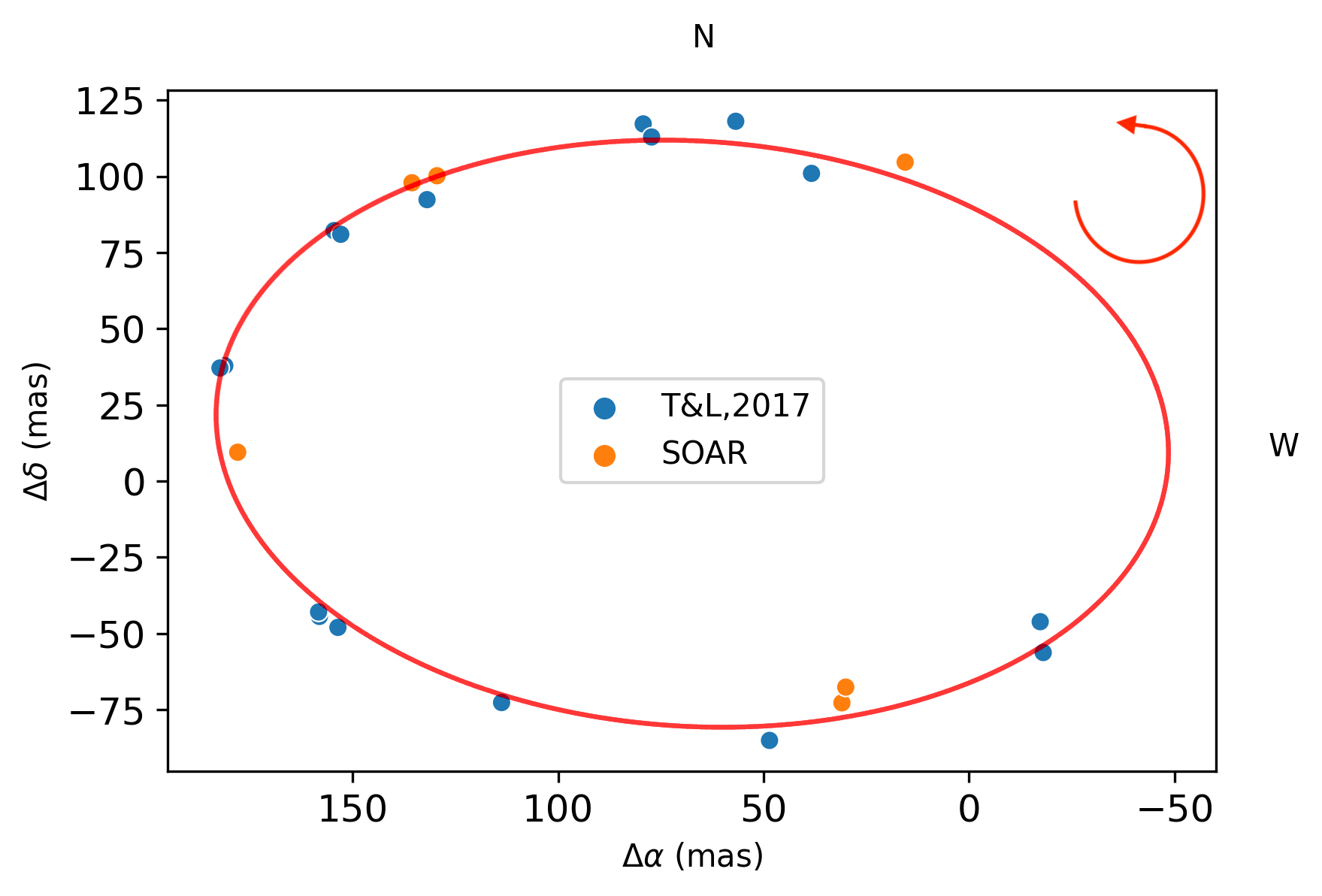}{0.5\textwidth}{}
    }
    \gridline{    
    \fig{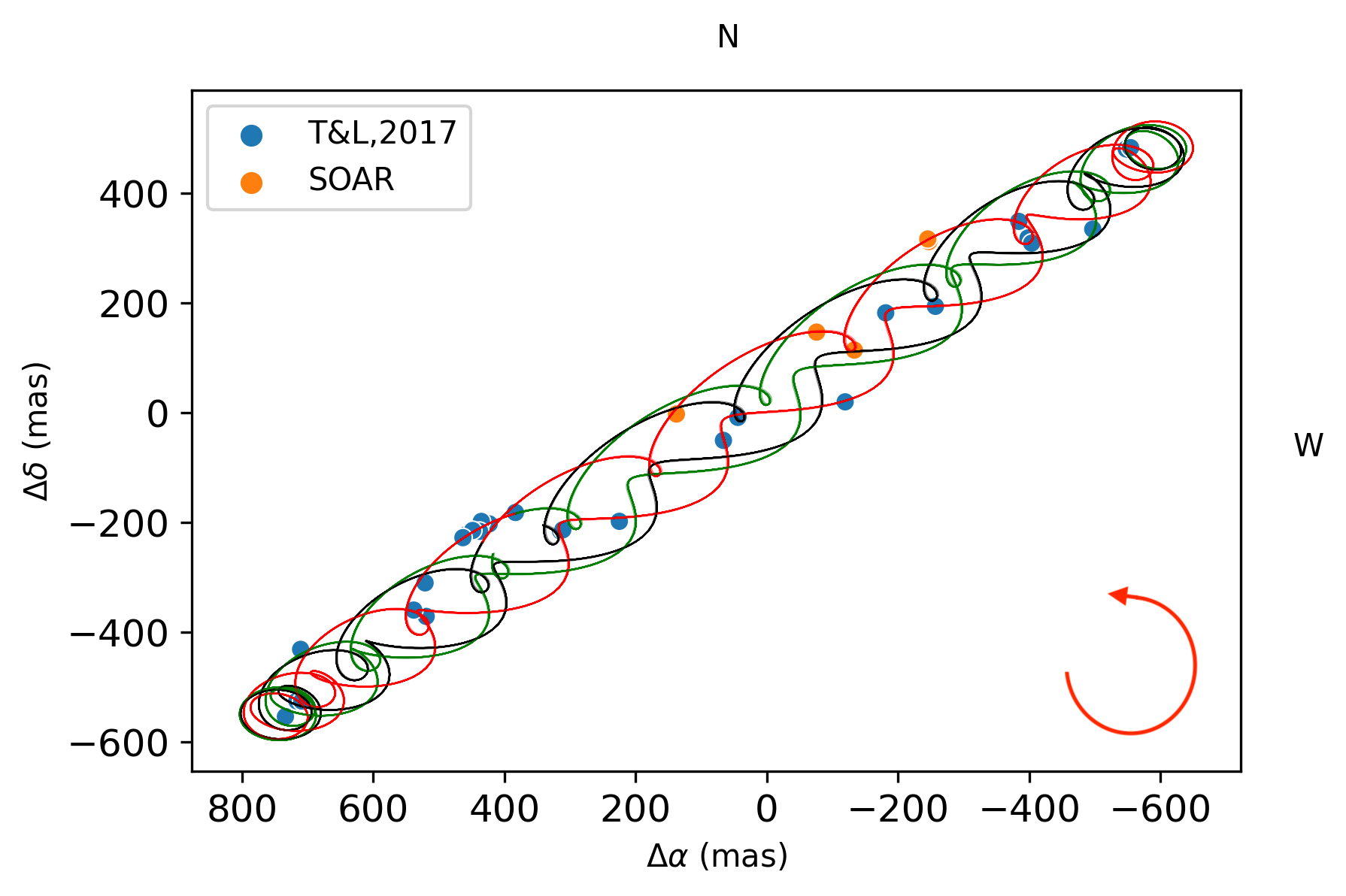}{0.5\textwidth}{}
    \fig{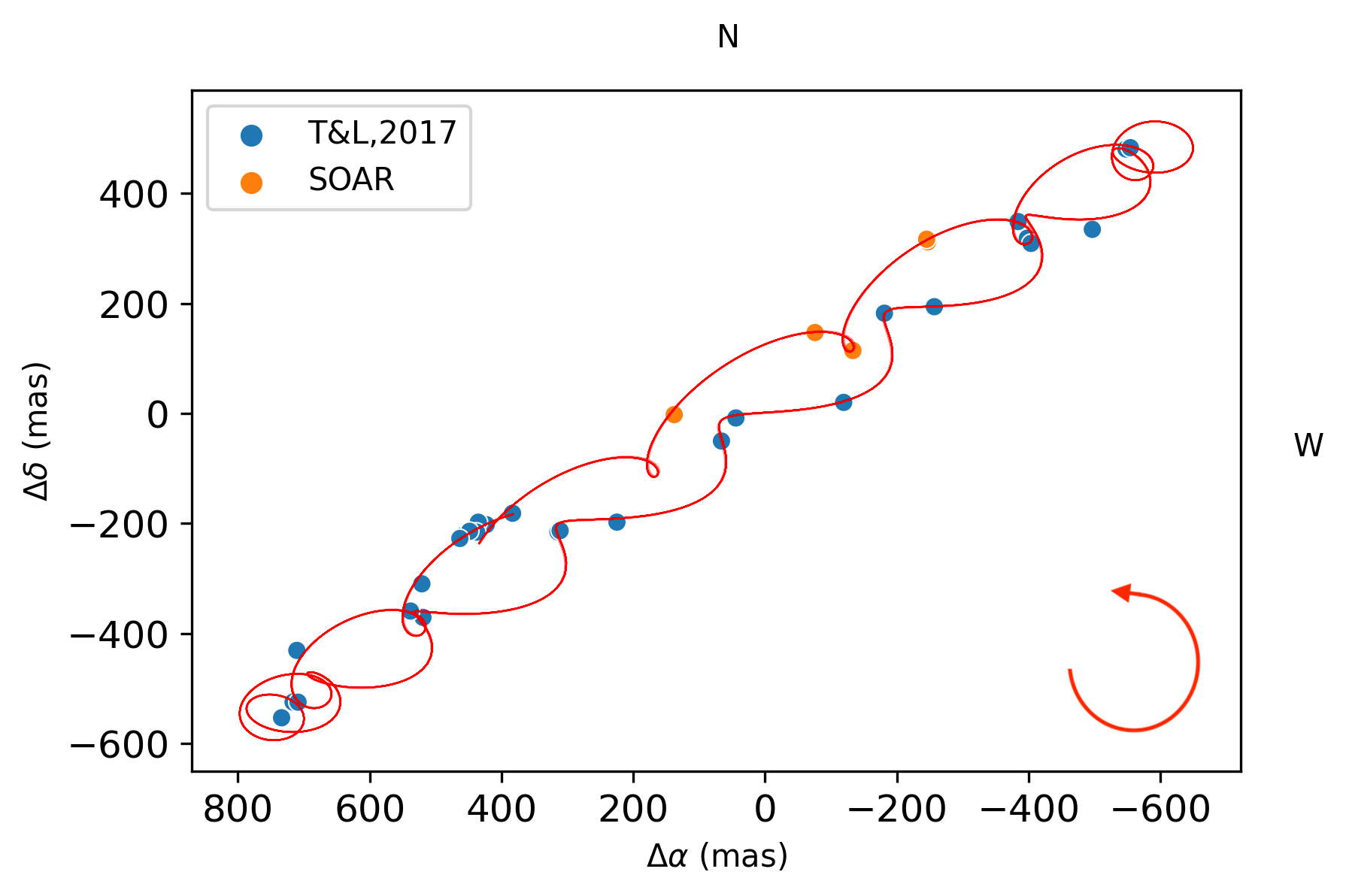}{0.5\textwidth}{}
    }  
\caption{Same as Figure~\ref{fig:LHS1070} but for HIP101955. \label{fig:KUI99as}}
\end{figure}

\begin{figure}[htp]
    \centering
    \gridline{\fig{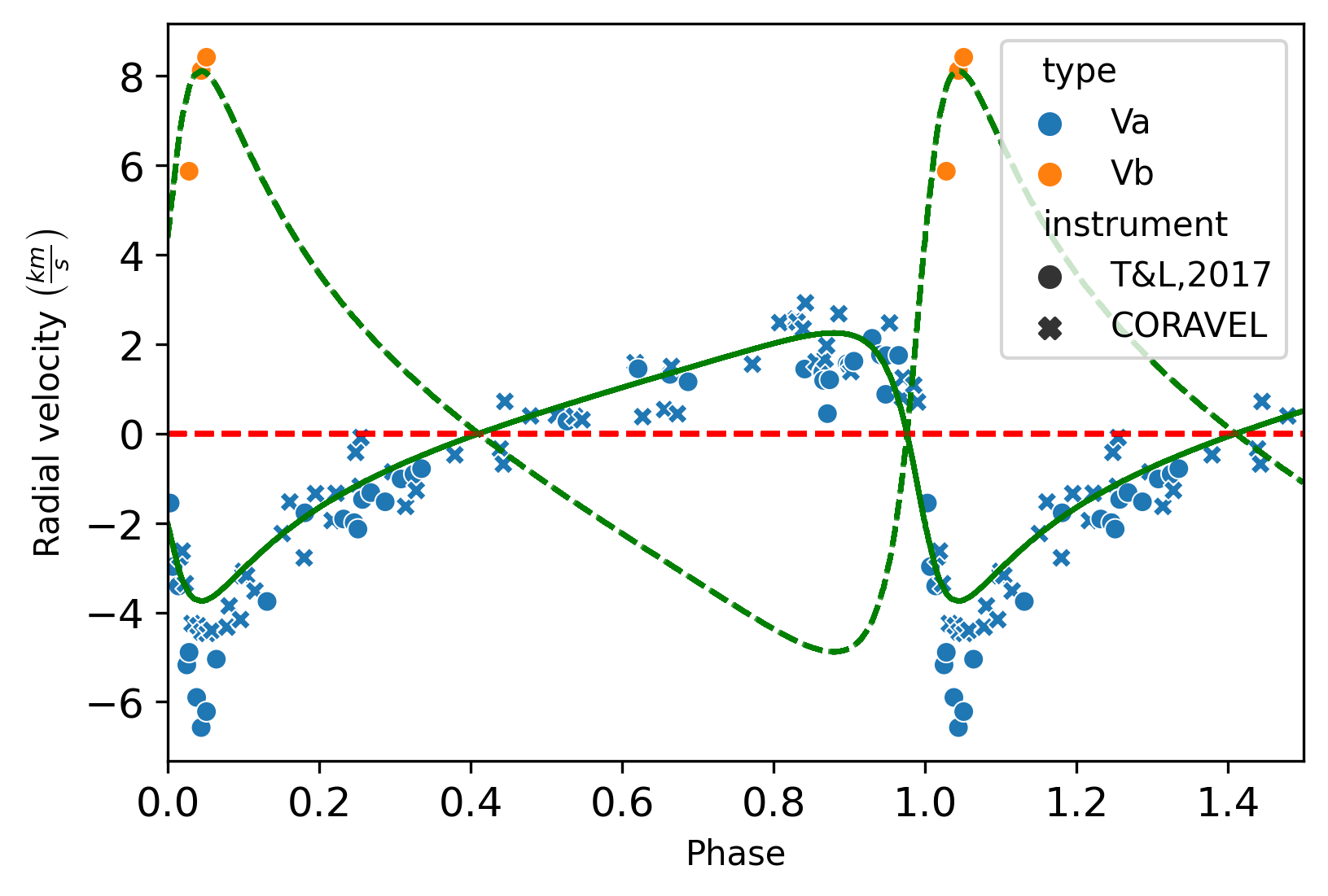}{0.45\textwidth}{}
    \fig{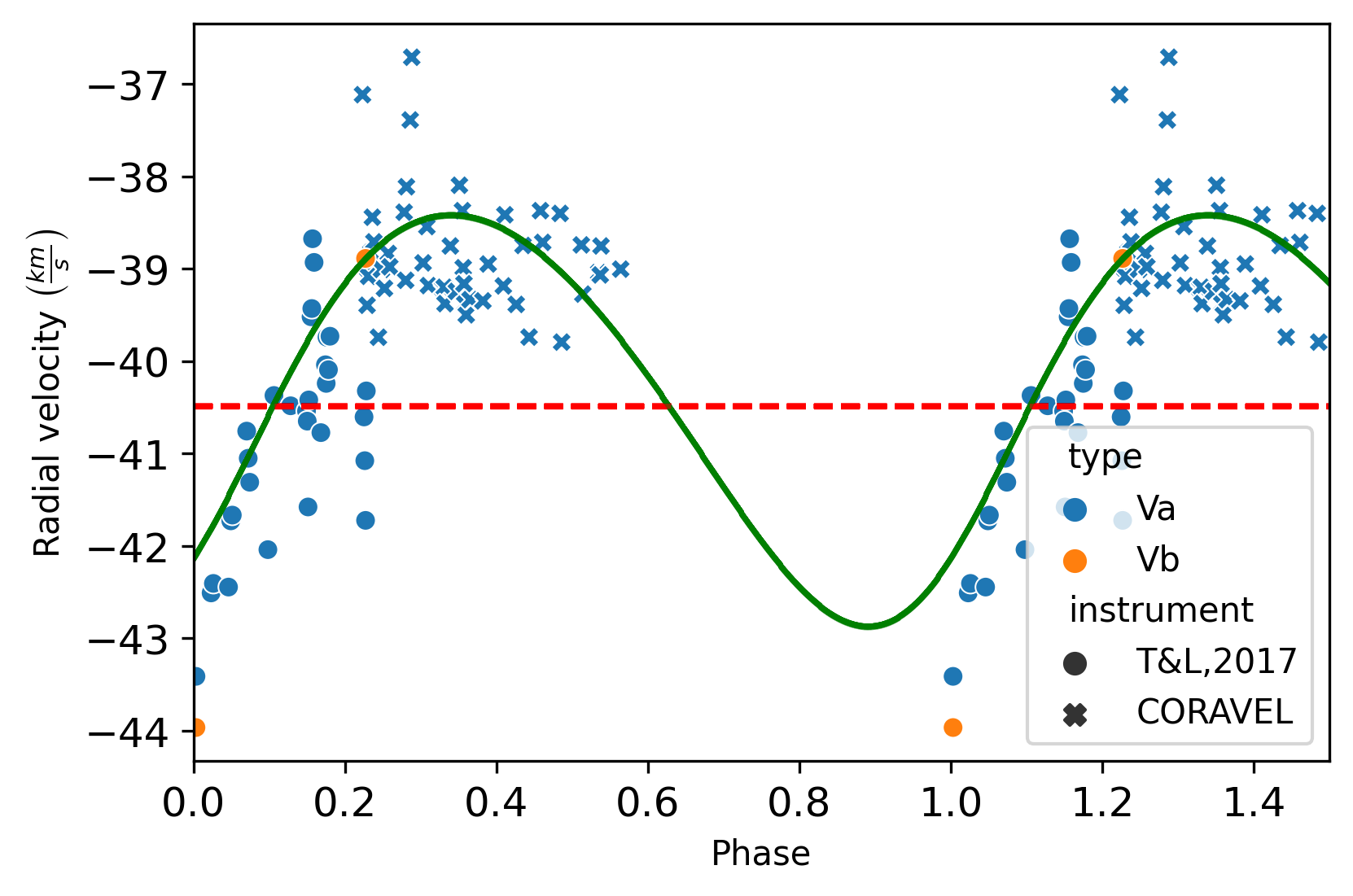}{0.45\textwidth}{}
    }  
\caption{Inner (left) and outer (right) RV curves of HIP101955. In the left panel the motion due to the presence of the external system has been subtracted, the solid and dashed lines for the primary and secondary components, respectively. The right panel shows the motion of the $CoM$ of the inner $AaAb$ pair. \label{fig:KUI99rv}}
\end{figure}

Adopting the simplified scheme described in the previous paragraph, we obtain the best 1000 inner (outer) orbits shown in the left (center and right) panel of Figure~\ref{fig:KUI99as}, while the MAP's RV curve for the inner (left panel) and outer (right panel) systems are shown in Figure~\ref{fig:KUI99rv}.

\subsection{WDS22388+4419 (HIP 111805)}

This system, also known as HD~214608 in SIMBAD, consists of a secondary with the inner subsystem $B_a, B_b$, and a distant primary $A$. In the WDS catalogue the discoverer designation for the outer is HO~295, while the inner is known as BAG~15. Astrometry is available for the inner and outer orbits, and there are also RV measurements for $B_a$, $B_b$ and $A$; then, the combined scenario is performed. Unfortunately, just as in the case of HIP101955, \citet{tokovinin2017} indicate that the RV measurements present line blending and denoted them as ``noisy''. We thus are forced to to adopt the same procedure regarding the fitting of $K_1$ and $K_2$ as done for HIP101955.

\paragraph{Inner system:} The inner system is also known as HO 265 and it was recognized as highly inclined and with a small eccentricity \citep{duquennoy1987study, balega2002speckle, tokovinin2017}, so the priors were set in the range $[80, 100]$~deg and $[0.01, 0.08]$, respectively. The period is known to be around 551 days, so the starting value was chosen to be 1.5~yr, but set free in the range $[0.1, 10]$~yr. Due to the issues with the measured RV mentioned before, \citet{tokovinin2017} find different mass ratios $q$ when derived from the visual or spectroscopic data. To account for this, and also based on previous studies, we set the prior for $q$ in the interval $[0.5, 0.9]$.

\paragraph{Outer system:} The outer system is also known as HDO 295 or ADS16138, with a period of $\sim 30$~yr \citep{hough1890catalogue, duquennoy1987study, tokovinin2017}. Therefore, the starting value for $P$ was chosen to be 30~yr, but set free in the range $[10, 100]$~yr, and the prior for $i \in [80, 100]$~deg. The eccentricity was known to be $\sim 0.3$, so the prior was allowed to explore the range $[0.2, 0.5]$. Finally, the prior for the mass ratio $q$ was selected in the interval $[0.6, 0.8]$.

\begin{figure}[htp]
    \centering
    \gridline{
    \fig{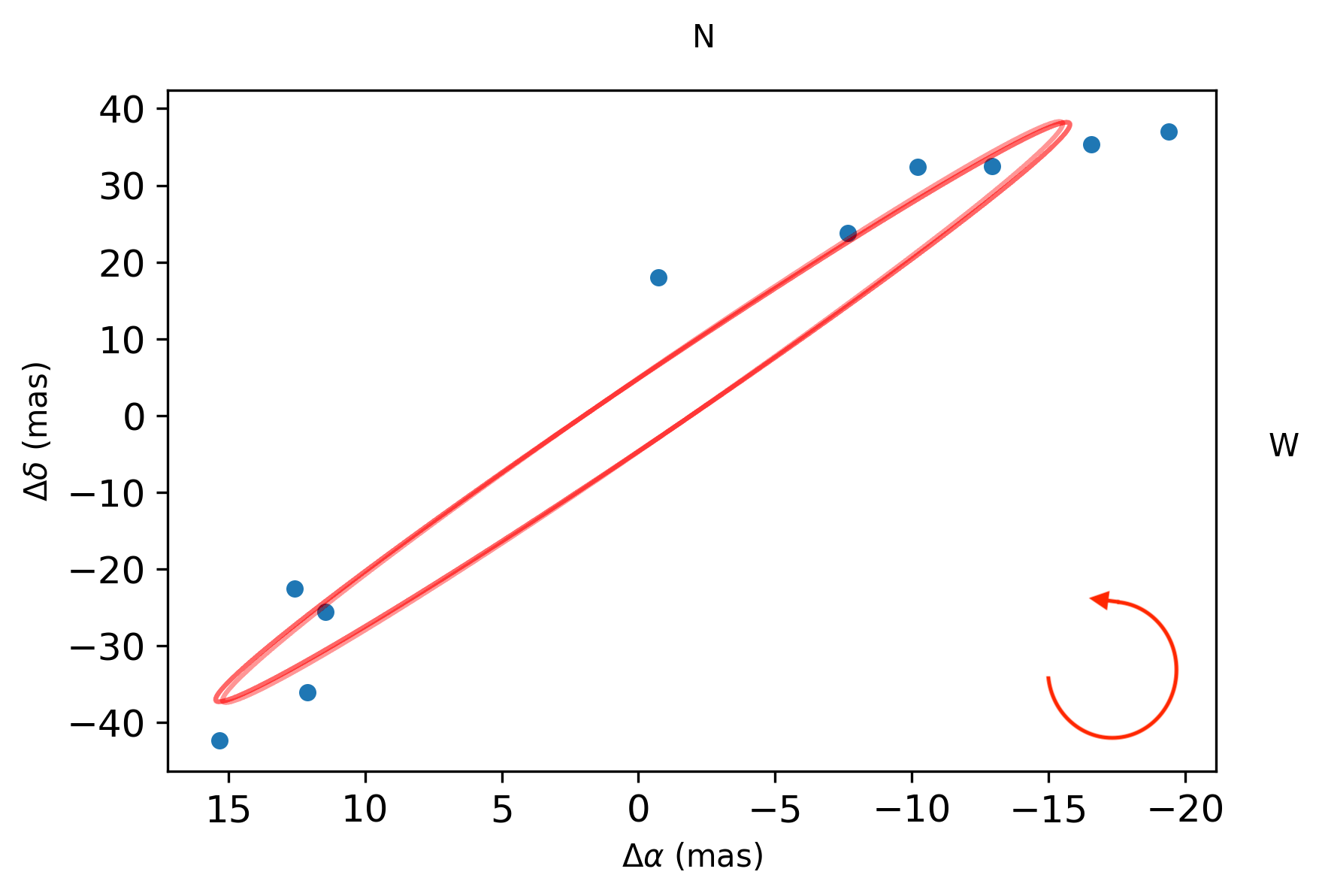}{0.5\textwidth}{}
    }
    \gridline{    
    \fig{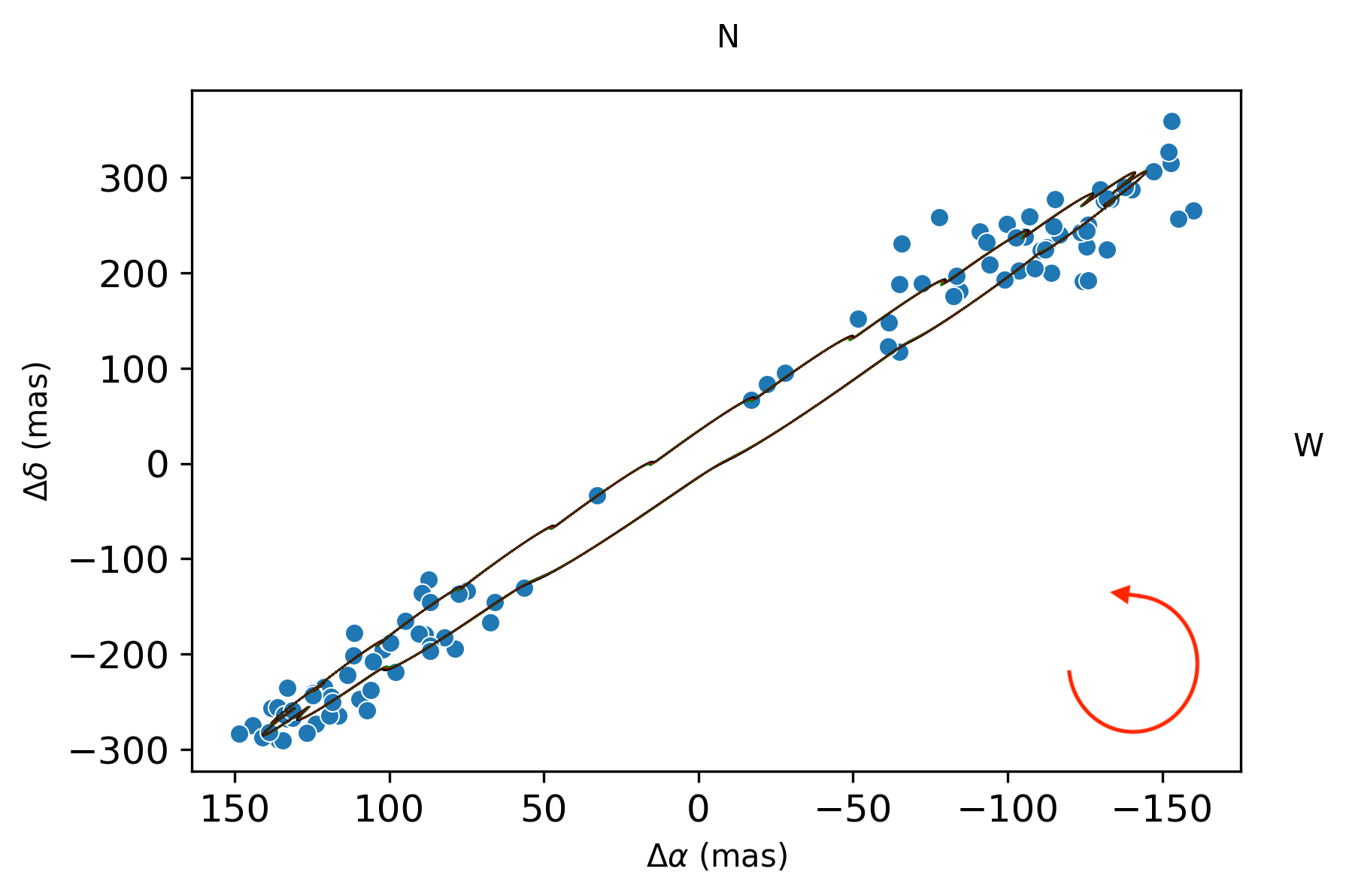}{0.5\textwidth}{}
    \fig{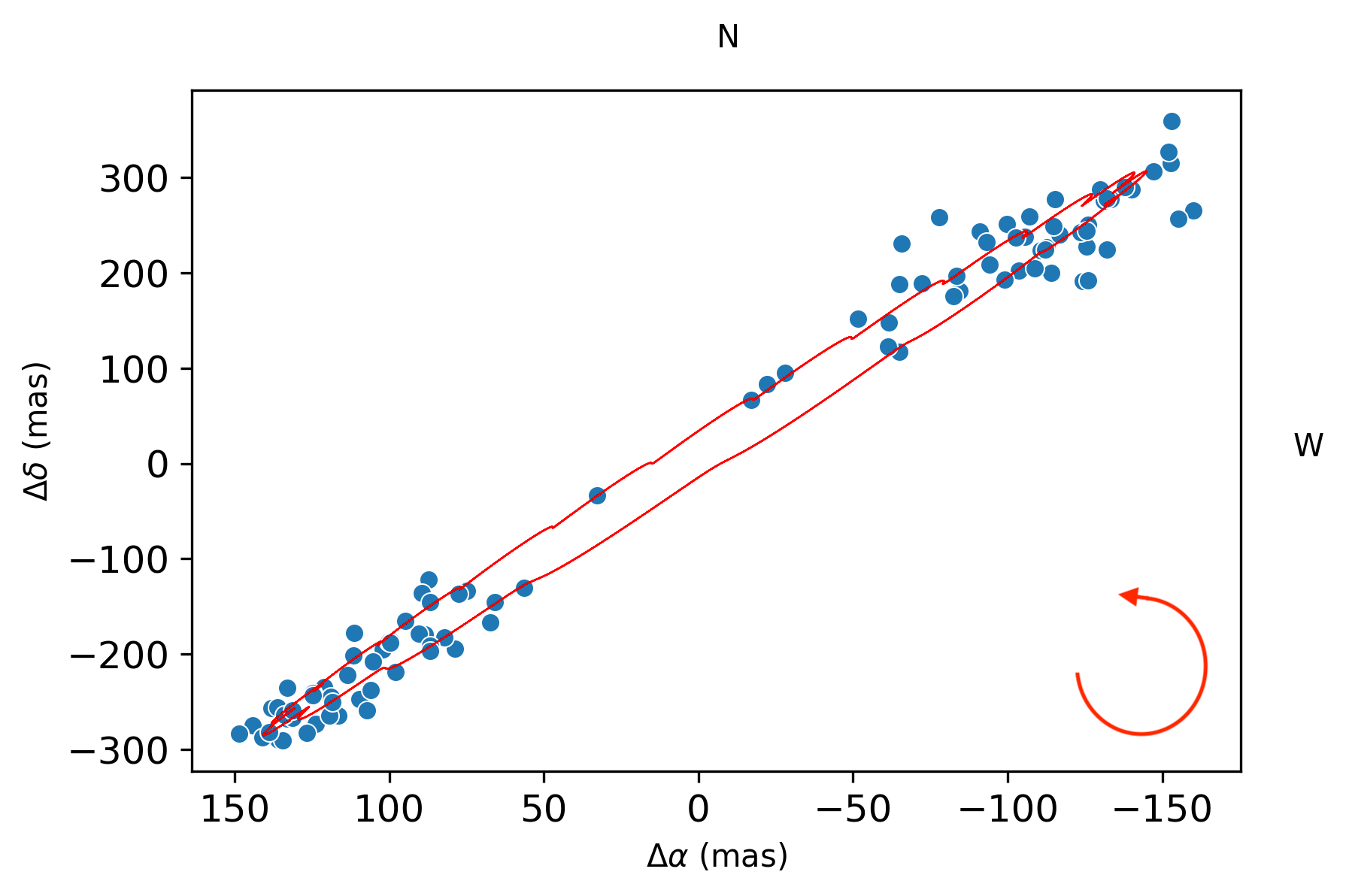}{0.5\textwidth}{}
    }  
\caption{Same as Figure~\ref{fig:LHS1070} but for HIP111805. \label{fig:HIP111805as}}
\end{figure}

\begin{figure}[htp]
    \centering
    \gridline{\fig{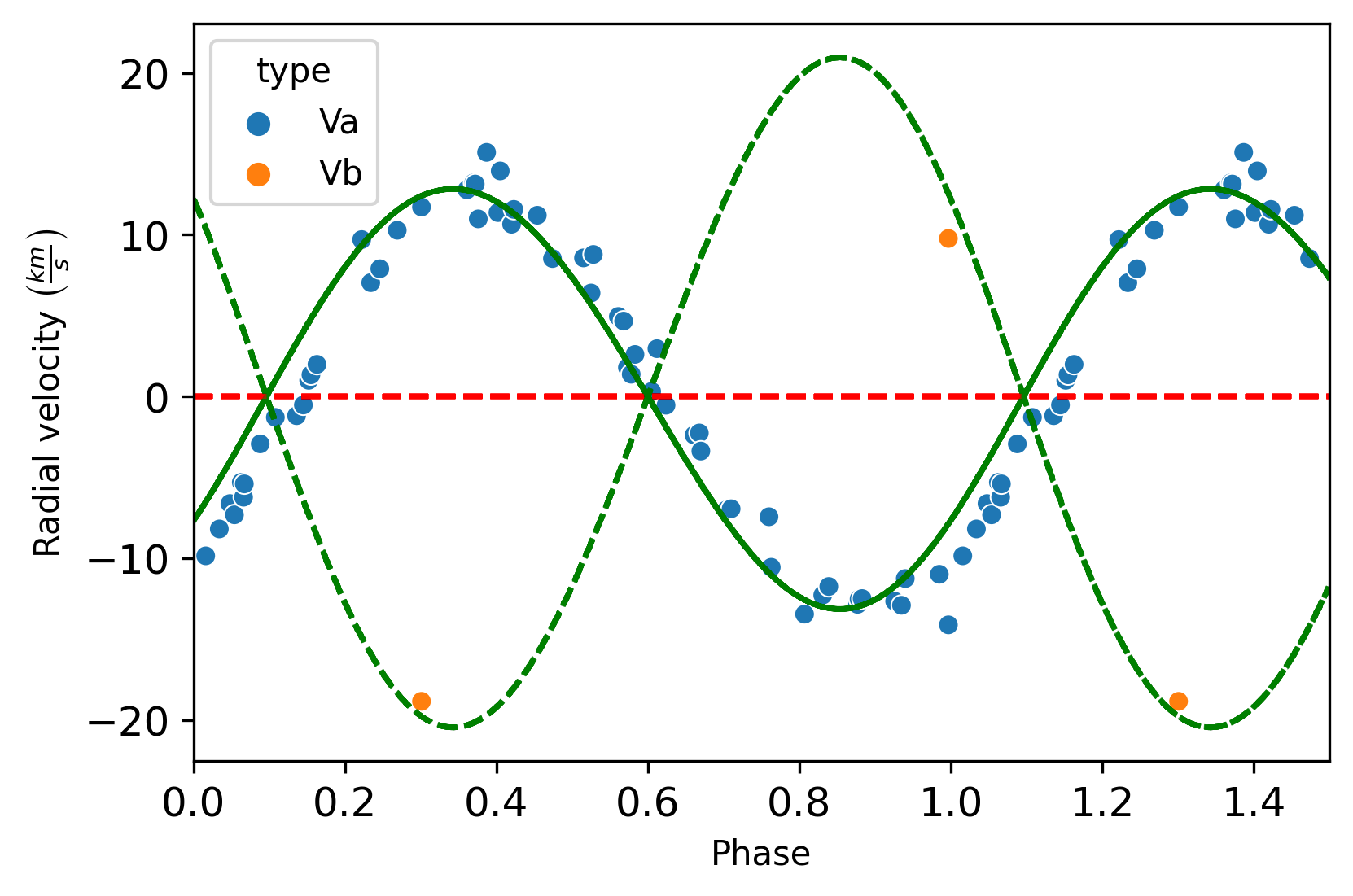}{0.45\textwidth}{}
    \fig{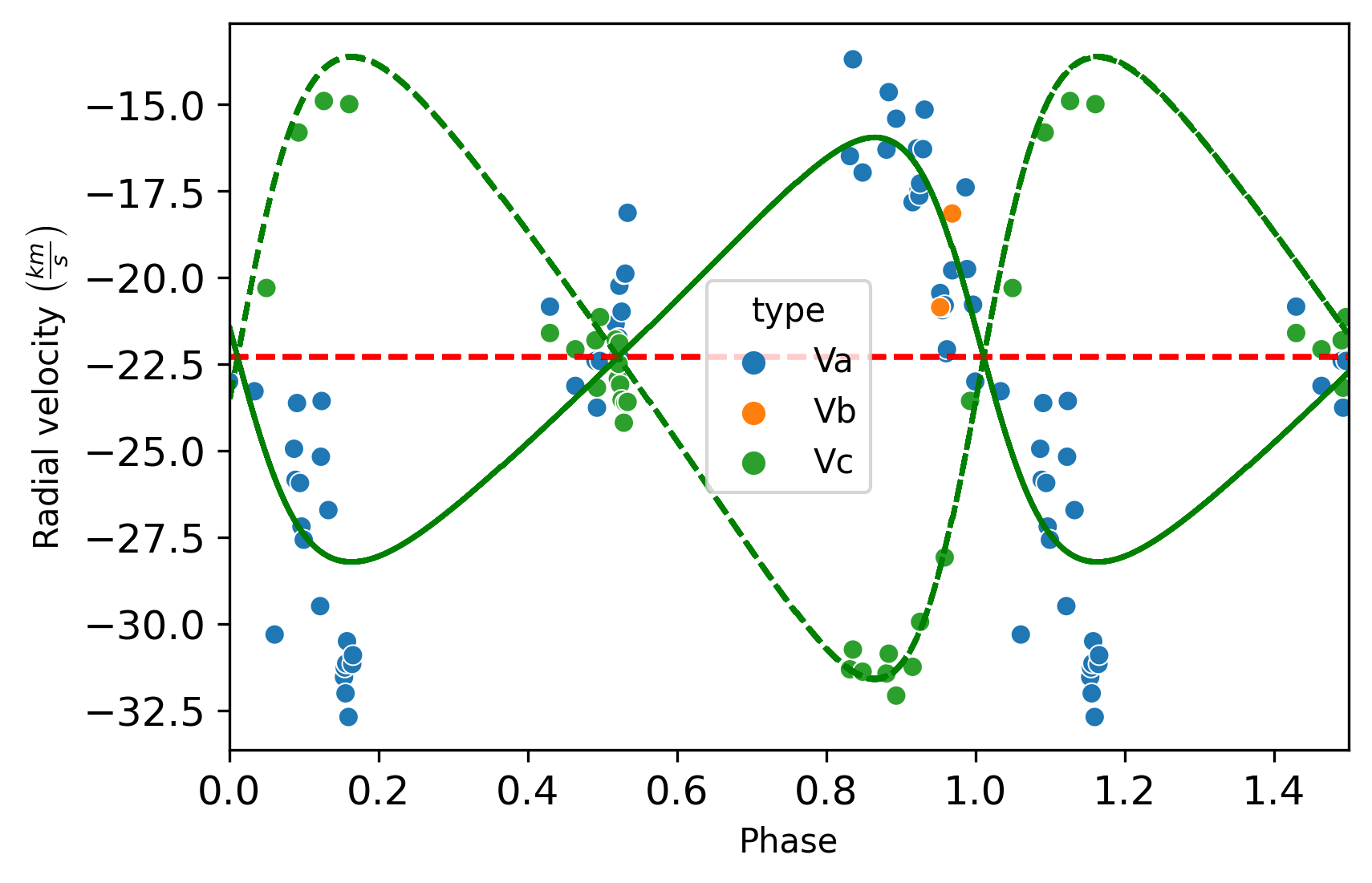}{0.45\textwidth}{}
    }  
\caption{Inner (left) and outer (right) RV curves of HIP111805. In the left panel the motion due to the presence of the external system has been subtracted, the solid and dashed lines for the primary and secondary components, respectively. The right panel shows the motion of the $CoM$ of the inner $BaBb$ pair (solid line) and A (dashed line). \label{fig:HIP111805rv}}
\end{figure}

The best 1000 inner (outer) orbits can be seen in the left (center and right) panel of Figure~\ref{fig:HIP111805as}, while the MAP's inner (outer) RV curve are shown in the left (right) panel of Figure~\ref{fig:HIP111805rv}, they both indicate reasonable overall fits.

\subsection{Analysis of benchmark results} \label{sec:anaben}

The best orbital elements obtained, along with their confidence intervals for our benchmark systems, can be seen in Table~\ref{tab:map}. Based on these parameters, the mutual inclination and the sum/individual masses were also obtained (see Table~\ref{tab:derived}), adopting the trigonometric parallax indicated in the second column of that table.

From the point of view of evaluating our methodology, it is interesting to compare in certain detail our results with those from the comprehensive recent studies of these three systems by Tokoving and collaborators, as follows:

\paragraph{LHS1070:} In general there is quite good agreement between our orbital elements and those presented in Table~2 of \citet{tokovinin2018dancing}. This can be appreciated graphically in Figures~\ref{fig:LHS1070inner} and~\ref{fig:LHS1070outer} where we indicate the corresponding purely parametric (i.e., non-Bayesian) best-fit values from Tokovinin, in comparison with our MAP value, obtained from the posterior distribution. For the inner system, our (quartile-based) uncertainty in the period is quite similar to the (1$\sigma$) value reported by Tokovinin, while the periods differ by less than 2$\sigma$. However, our semi-major axis seems better determined (by a factor of two) than Tokovinin's (measured in arcsec). This implies that, in principle, our mass sum should be better determined in our case. Indeed, our mass sum, according to Table~\ref{tab:derived} is  $0.15361^{+0.00017}_{-0.00034}M_\odot$, while Tokovinin reports a mass sum of $0.150 \pm 0.090M_\odot$ (note that he adopts the parallax by \citet{CostaMendezet2005} of $129.5 \pm 2.5$~mas, quite similar to the one adopted by us from eDR3, which has a formal error almost twenty times smaller, but the same value, namely $129.32 \pm 0.13$~mas, see Table~\ref{tab:derived}). It is difficult to perform a further analysis in this respect, since no uncertainties are provided for the individual masses by Tokovinin (see his Table~1, first row).  In any case, this confirms the very low-mass nature of the $AaAb$ pair (called the LEI 1 B and C components by discoverer \citet{Leinertetal2001}). Our orbit has a small, but significantly different from zero eccentricity (similar to Tokovinin), but we have adjusted for $\omega$ (see its PDF in  Figure~\ref{fig:LHS1070inner}), unlike in the case of Tokovinin, where he fixed it at $\omega = 202.53$. As can be seen from Table~\ref{tab:map}, it turns out that the difference between his adopted value and our fit is less than $2\sigma$. As for the outer system, the 1$\sigma$ uncertainty in the period from Tokovinin (reported at $77.6 \pm 2.1$~yr) is similar (or slightly larger) to our inter-quartile range, albeit our period is larger than Tokovinin's. This difference might be actually significant - and physically relevant (see below) - since, as we can see from Figure~\ref{fig:LHS1070inner}, the period seems to be actually quite well defined, and favouring a value above 81~yr. Indeed, as pointed out by Tokovinin, a dynamical analysis by \citet{kohler2012orbits} demonstrated that the system will be dynamically stable only when the period exceeds 80~yr, which seems indeed to be the case. Since we do not have RV for the system, we can not determine the $v_{CoM}$, but we can determine the wobble factor, which is similar to the one reported by Tokovinin (he gives $f=0.485 \pm 0.006$). A final point is that the period ratio derived from our MAP values is 4.74 (4.5 in Tokovinin), which, as pointed out by Tokovinin, is just above the lower stability limit of 4.7 predicted by \citet{MardAars2001} in the case of co-planar orbits (see the very small value of $\Phi$ in Table~\ref{tab:derived}).

\begin{deluxetable*}{ccccccccccc}
\tabletypesize{\scriptsize}
\tablecaption{Estimation of orbital elements for our benchmark systems (upper row) and quartiles (lower row). The upper two rows are for the compact (``inner'') system, the lower two rows are for for the external (``outer'') system. For each sub-system, the first row is the MAP value, while the second row shows the lower (bottom value) and upper (top value) quartiles.\label{tab:map}}
\tablewidth{\columnwidth}
\tablehead{
\colhead{Object} & \colhead{$T$} & \colhead{$P$} & \colhead{$e$} & \colhead{$a$} & \colhead{$\omega$} & \colhead{$\Omega$} & \colhead{$i$} & \colhead{$f$/$v_{CoM}$\tablenotemark{\tiny a}} & \colhead{$K_1$} & \colhead{$K_2$}\\
\colhead{WDS} & \colhead{(yr)} & \colhead{(yr)} &  & \colhead{($''$)} & \colhead{($^{\circ}$)} & \colhead{($^{\circ}$)} & \colhead{($^{\circ}$)} & \colhead{(kms$^{-1}$)} &  \colhead{(kms$^{-1}$)} & \colhead{(kms$^{-1}$)} 
}
\startdata
00247-2653 & 2005.30 & 17.271 & 0.0150 & 0.46268 & 179 & 14.684 & 62.984 & -0.5000 & - & - \\
inner ($B_a, B_b$) & $_{0.65}^{0.53}$ & $_{0.016}^{0.013}$ & $_{0.0017}^{0.0019}$ & $_{0.00032}^{0.00031}$ & $_{14}^{11}$ & $_{0.086}^{0.080}$ & $_{0.059}^{0.064}$ & $_{ 0.0000}^{0.0035}$  \\
00247-2653 & 2041.15 & 81.87 & 0.010 & 1.5532 & 163.3 & 14.70 & 61.86 & - & - & -\\
outer ($A, B$) & $_{1.40}^{0.36}$ & $_{2.24}^{0.36}$ & $_{ 0.000}^{0.025}$ & $_{0.0338}^{0.0045}$ & $_{1.9}^{1.3}$ & $_{0.15}^{0.51}$ & $_{0.61}^{0.18}$  \\
\\
20396+0458 & 1985.475 & 2.50881 & 0.5970 & 0.1199 & 104.6 & 153 & 14.9  & 0.446 & 3.45 & 6.79 \\
inner ($A_a, A_b$) & $_{0.020}^{0.017}$ & $_{0.00098}^{0.00140}$ & $_{0.0090}^{0.0105}$ & $_{0.0020}^{0.0018}$ & $_{9.9}^{14.8}$ & $_{14}^{10}$ & $_{7.6}^{4.5}$ & $_{0.012}^{0.011}$ & $_{0.12}^{0.14}$ & $_{0.15}^{0.36}$\\
20396+0458 & 2015.60 & 38.754 & 0.1083 & 0.8526 & 228.3 & 127.56 & 87.455 & -41.14 & 2.888 & 8.65 \\
outer ($A, B$) & $_{0.16}^{0.12}$ & $_{0.055}^{0.067}$ & $_{0.0037}^{0.0033}$ & $_{0.0036}^{ 0.0036}$ & $_{1.7}^{1.2}$ & $_{0.20}^{0.18}$ & $_{0.115}^{0.098}$ & $_{0.16}^{0.30}$ & $_{0.277}^{0.059}$ & $_{1.85}^{0.70}$\\
\\
22388+4419 & 1901.89 & 1.503 & 0.020 & 0.0408 & 54.80 & 157.6 & 88 & -0.329 & 12.99 & 20.7 \\
inner  ($B_a, B_b$) & $_{0.49}^{0.29}$ & $_{0.012}^{0.012}$ & $_{ 0.000}^{0.015}$ & $_{0.0050}^{0.0071}$ & $_{4.95}^{0.15}$ & $_{10.2}^{9.8}$ & $_{16}^{15}$ & $_{0.028}^{0.035}$ & $_{0.26}^{0.28}$ & $_{2.9}^{2.2}$  \\
22388+4419 & 1919.754 & 30.0905 & 0.3237 & 0.3338 & 83.81 & 154.15 & 88.14 & -22.30 & 6.12 & 8.97 \\
outer  ($A, B$) & $_{0.049}^{0.059}$ & $_{0.0126}^{0.0082}$ & $_{0.0056}^{0.0077}$ & $_{0.0012}^{0.0011}$ & $_{0.23}^{0.46}$ & $_{0.10}^{0.18}$ & $_{0.13}^{0.13}$ & $_{0.26}^{0.26}$ & $_{0.31}^{0.30}$ & $_{0.53}^{0.28}$\\
\enddata
\tablenotetext{\tiny a}{In this column column we have the wobble factor for the inner systems or the $v_{CoM}$ for the outer system} 
\end{deluxetable*}

\begin{deluxetable*}{ccccc}
\tabletypesize{\scriptsize}
\tablecaption{Adopted and derived quantities for our benchmark systems (see Appendix~\ref{sec:relevant_quant}).\label{tab:derived}}
\tablewidth{0pt}
\tablehead{
\colhead{Object} & \colhead{$\bar{\omega}$} & \colhead{$\Phi$} & \colhead{$M_A$} & \colhead{$M_B$} \\
\colhead{WDS} & \colhead{(mas)} & \colhead{($^{\circ}$)} & \colhead{($M_\sun$)} & \colhead{($M_\sun$)}
}
\startdata
00247-2653 & 129.32$\pm$0.13\tablenotemark{\tiny a} & 2.40$_{0.21}^{0.00}$  & 0.1221$_{0.0034}^{0.0048}$ & 0.15361$_{0.00034}^{0.00017}$\\
inner &  &   &  & $_{0.07834_{0.00010}^{0.00058}}^{0.07527_{0.00070}^{0.00013}}$\\
&&&&\\
20396+0458 & 59.80$\pm$3.42\tablenotemark{\tiny b} & 71.5$_{1.4}^{7.0}$  & 1.28$_{0.13}^{0.16}$ & 0.660$_{0.074}^{0.090}$\\
inner &  &  & $_{0.709_{0.078}^{0.093}}^{0.574_{0.070}^{0.089}}$ &  \\
&&&&\\
22388+4419 & 24.1$\pm$2.0\tablenotemark{\tiny c} & 1$_{00}^{17}$ & 0.78$_{0.78}^{0.54}$  & 2.15$_{0.67}^{1.49}$ \\
inner &  &  &   & $_{1.44_{0.49}^{1.10}}^{0.71_{0.19}^{0.34}}$\\
\enddata
\tablenotetext{a}{From Gaia eDR3, \citet{eDR3Gaiae2020}.}
\tablenotetext{b}{From the Hipparcos re-reduction on \citet{vanLeeuwen2007}.}
\tablenotetext{c}{From \citet{tokovinin2017}, derived from the orbital parallax of the AB system.}
\end{deluxetable*}

\paragraph{HIP101955:} Again, by comparing our Table~\ref{tab:derived} with Table~2 in \citet{tokovinin2017}, we see a good agreement in all orbital elements, including the wobble factor, and the RV amplitudes (we emphasize that, for our final calculations, we decoupled the estimation of $K_1$ and $K_2$ from the astrometric solution for the reasons explained before). We also reproduce the fact that the two orbits are highly inclined to each other (see the value of $\Phi$ in Table~\ref{tab:derived}), and that the inner orbit has a large eccentricity. The derived individual masses agree reasonably with those estimated by Tokovinin, and one interesting point is that his (purely photometric) dynamical parallaxes predict a total mass for the inner system of $1.36 M_{\odot}$ (see his Table~5), which is within less than $1\sigma$ of what we find from our dynamical solution, whereas the astrometric mass sum reported by Tokovinin is slightly larger, at $1.49~M_{\odot}$.\\

\paragraph{HIP111805:} This is a particularly challenging system for astrometry, given its nearly edge-on orientation. While it is true that in these systems we loose only one direction of motion (for example up/down for an orbit
with PA=90deg), meanwhile the amplitude in the perpendicular direction is
not diminished, the astrometric wobble in these systems is in general less noticeable  (see Figure~\ref{fig:HIP111805as}), which is crucial to be able to link the internal to the external system and, in particular, to constrain the mass ratio. Despite this, we reproduce the wobble factor, and mostly all other orbital elements as in  \citet{tokovinin2017}, except for the values of $\omega$ and $\Omega$ of the tighter system that, after a correction of 180 degrees\footnote{See Section 6, third paragraph, on \citet{tokovinin2017}, "...to get the orbit of B around A, change the outer elements $\omega_A$ and $\Omega_A$ by 180 degrees...", we show our results before applying this change.} coincide actually quite well (note that the tighter orbit is almost circular), considering our quartile range, and the quoted $1\sigma$ uncertainties by Tokovinin. Dynamical masses computed by Tokovinin imply masses of 0.85 ($A$ component), 1.03 ($Bb$ component), and $1.14~M_{\odot}$ ($Ba$ component)  (see his Table~5). The total mass of the tighter system is quite similar to our derived value of $2.15~M_{\odot}$ in Table~\ref{tab:derived}. The total (purely photometric) dynamical mass of the system is estimated by Tokovinin to be $3.02~M_{\odot}$ (he obtains a similar mass from his astrometric solution), again, close to our value (we adopted the same value of the parallax). Finally, we obtain a more co-planar orbit than that derived by Tokovinin (he quotes $\Phi = 2.5 \pm 1.5$~degrees), but with a larger uncertainty.\\

Our detailed comparison of these well-studied systems against previous works shows that our methodology is robust, works well and that furthermore, by providing PDFs for all fitted parameters, allows us to make judicious statements regarding the plausible value (or range of values) for orbital elements that have an impact on the interpretation, e.g., about the dynamical stability of a hierarchical system.


\section{Discussion and conclusions}
\label{sec:conclusions}

In this paper we have applied a Bayesian MCMC-based methodology to the problem of estimating the orbital elements in triple hierarchical stellar systems, that already have a measured parallax. Graphical models were employed for modeling the probabilistic relationship between parameters and observations in the astrometry-alone, radial-velocity-alone and the combined scenarios. Thus, we factorized the joint distribution in terms of independent blocks and then performed the estimation in a two-stage process, combining different sets of observations sequentially. 

Our proposed framework provides MAP estimates (that also minimizes the $\chi^2$ statistic) along with the full joint posterior distribution of the parameters, given the observations. This feature is perhaps one the greatest advantages of our proposed methodology, since it allows us to assess the uncertainties of all the parameters in a robust way. While we require prior knowledge about the system, non-informative priors\footnote{For example, a uniform prior.} could also be used to get good results.

Regarding the radial velocities-alone scenario, we introduce a mathematical formalism motivated by the works of \citet{wright2009efficient} and \citet{mendez2017orbits} in the context of exoplanets and visual-and-spectroscopic binaries, respectively, and adapted to the specific case of triple systems. It consists of a dimensionality reduction (that takes us from 15 to 10 parameters) using weighted-least squares, which allows us to sample from a subset of the parameter space, hence reducing the computational cost. This is applied for systems of the form $A_a, A_b$ - $B$ and $B_a, B_b$ - $A$.

On the other hand, the methodology is useful for outer long periods, because, even though we have just a few (or no) measurements from the distant body B (or A), the algorithm allows us to constrain the mass ratio of the outer system $q_{AB}$. 

This scheme is tested with real measurements of astrometry and RV, where we determined the inner and outer orbital elements. By utilizing both kinds of measurements, our results allow us to determine the mutual inclination of the orbits and the individual stellar masses, solutions that are consistent with former results reported for these systems. It would have been interesting to be able to apply our methodology to other, previously unstudied systems. Unfortunately, reliable data for these objects (required to detect the astrometric wobble) is only slowly becoming available, through the use of high-precision interferometric measurements, and we could not identify other systems to which we could apply our scheme at this point. It is expected that, in the near future, with the accumulation of more data, other systems could be subject to these studies. Being hierarchical, it usually happens that the external orbit has a long period, and therefore modern (high-precision) data covers only a tiny arc of the orbit. In other cases, the inner orbit is too tight to be resolved by Speckle observations, even on 4m or 8m facilities. A dedicated observational program, targeting the most promising cases, could be interesting to significantly increase the number of well-studied triple systems in hierarchical configuration. A good starting point to select these targets will be the MSC, indeed, a recent effort in that direction has been reported by \citet{Toko2021}), where inner and outer orbits for thirteen triple hierarchical systems are presented.


\subsection{Technical comments and outlook} \label{sec:techdiscussion}

Working with sample-based schemes in high dimensions is quite challenging, both in terms of convergence and fine-tuning of the algorithm hyperparameters. In this work, we decided to perform the estimation using a scheme that is hybrid between MCMC and the Gibbs sampler, exploring the state space using a random walk. The estimations were performed successfully in all scenarios; however, at this point, a few considerations must be made.

\paragraph{Convergence}

Our algorithm reaches the target distribution in the limit of long runs (see Section~\ref{sec:benchmarks}). In particular, in some cases, our algorithm takes many iterations to converge to stable values, specially when the data are of marginal quality, or if there is a sparse coverage of the phase space or orbital space. For that reason, other options should be evaluated like, e.g., simulated tempering or parallel tempering \citep{marinari1992simulated, earl2005parallel}. On the other hand, methods that employ the gradient of the prior and the likelihood could also be explored, such as MALA \citep{roberts1996exponential, robert2004monte}, Hamiltonian Monte Carlo \citep{neal2011mcmc}, proximal MCMC algorithms \citep{combettes2011proximal, parikh2014proximal}, or diffusion-based MCMC algorithms \citep{herbei2017applying}, but in all these schemes, care must be taken to avoid getting stuck in local (as opposed to global) optimal solutions. Finally, Bayesian methods different to MCMC could be also considered, such as the rejection-sampling method described by, e.g., \citet{PWet2017} and \citet{blunt2017orbits}.

\paragraph{Fine tuning of proposals and priors}

The hand-tuning of the priors and proposals is a demanding task, and it is important because those hyperparameters are directly related to the algorithm's rate of convergence.

Regarding the proposals, the variance (or range) for each one of the dimensions involved must be chosen. If it is too wide, a lot of particles where the target distribution is nearly zero could be chosen; if it is too small, most of the particles are accepted, so the chain moves slowly. Additionally, the problem could worsen because of possible correlations between parameters \citep{sharma2017markov}, which we are not considering at the moment. Despite some general guidelines regarding the variance of the random walk could be applied, some calculations are not easy for highly non-linear functions such as the ones present in this work. Hence, other alternatives, such as adaptive MCMC methods, could be used, which choose the new particle based on the earlier history of the chain \citep{haario2001adaptive, haario2004markov, haario2005componentwise}.

Concerning the priors, we found that lots of particles were rejected due to physical restrictions. This could be known beforehand by considering this information when adjusting the priors, and stop rejecting samples due to infeasible masses. Later on, this could be applied in unstable zones or another type of physical restrictions.

\paragraph{Physical constraints}

There are several restrictions related to the dynamics of triple hierarchical star systems that are included in the implementation of our algorithm. First of all, all of the parameters are bounded within a certain range, which could make the code to stick in the boundaries. Thus, the state space was adapted as a circular loop, keeping the sampling procedure as if there were no restrictions. 

An additional problem arises when dimensionality reductions are performed: The minimization of  $\chi^2$ statistic is conducted using a weighted least squares, so the particles are rejected after that process, which increases the computational cost and keeps the chain from converging. This issue could be fixed by solving a non-linear optimization problem with restrictions from the beginning.

Finally, the imputation framework also rejects particles that violate the physical constraints. We could be aware of that before sampling the posterior from the last step (the astrometric inner orbit), and thus prevent this situation earlier on the calculations.

\paragraph{Inconsistent data}

Finally, we have the issue of some inconsistencies between the astrometry observations and the RV. This could be solved by just not considering the ambiguous information or by adjusting the weights associated with those observations. In our case, we have shown how for HIP 101955 and HIP 111805, if we insist on self consistency, we are not able to obtain a good simultaneous fit to the astrometry and RV data, and we have to, instead, decouple the calculation of the RV amplitudes from the rest of the problem. This is somewhat equivalent to the case of double-line spectroscopic binaries with astrometric orbits, which sometime render meaningless orbital parallaxes if the data sets are not consistent between them.

\section{Acknowledgements}

RAM acknowledges support from ANID/FONDECYT Grant Nr. 1190038 and collaborator Dr. Andrei Tokovinin (CTIO) for his constant intellectual support, and for running and maintaining the superb HRCam instrument at the SOAR 4.1m telescope, JS \& MO acknowledge support from the Advanced Center for Electrical and Electronic Engineering, AC3E, Basal Project FB0008, ANID. MO acknowledges support from ANID/FONDECYT Grant Nr. 1210031.  We are indebted to an anonymous referee who provided numerous suggestions that have significantly improved the readability of the paper.

This research has made use of the Washington Double Star Catalog maintained at the U.S. Naval Observatory by Brian Mason and collaborators, the SIMBAD database, operated at CDS, Strasbourg, France, the 9th Catalogue of Spectroscopic Binary Orbits maintained by Dimitri Pourbaix in Belgium, and data from the European Space Agency (ESA) mission
{\it Gaia} (\url{https://www.cosmos.esa.int/gaia}), processed by the {\it Gaia}
Data Processing and Analysis Consortium (DPAC,
\url{https://www.cosmos.esa.int/web/gaia/dpac/consortium}). Funding for the DPAC
has been provided by national institutions, in particular the institutions
participating in the {\it Gaia} Multilateral Agreement. We are very grateful for the continuous support of the Chilean National Time Allocation Committee under programs CN2018A-1, CN2019A-2, CN2019B-13, and CN2020A-19.


\appendix 
\section{Triple Hierarchical Stellar Systems Model Equations}
\label{app:triple}

Hierarchical stellar systems are a particular case of the general $n$-body problem, because it can be separated into $(n-1)$ subgroups, where each hierarchy level can be treated as a binary system separately \citep{leonard2000multiple}. Thus, we can approximate triple hierarchical stellar systems  with two Keplerian orbits on top of each other; where one represents the motion of the wide system and other that of the inner/tighter system.

More precisely, those systems consist of an (inner) binary ($A_{a}$ and $A_{b}$, with its $CoM$ denoted by $A$) orbited by an external body $B$. It is also possible to have a star $A$ orbited by an (outer) binary ($B_{a}$ and $B_{b}$). It is worth mentioning that the dynamical interaction between the inner and outer systems constantly change both orbits, the inclination and eccentricity are free to evolve in time \citep{steves2010extra} and, under certain conditions, the argument of the pericenter of the orbit oscillates around a constant value, which leads to a periodic exchange between its eccentricity ($e$) and its inclination ($i$), known as Kozai-Lidov cycles \citep{naoz2016eccentric}. However, the timescale of that evolution is much longer than the time span of the observations, so the orbital elements can be considered constant \citep{tokovinin2017}, which is one of the basic assumptions adopted in this paper. 

\subsection{General Dynamics}
\label{app:general_dyn}

\paragraph{Inner system:}
First of all, the bodies $A_a$ and $A_b$ keep Newton motion laws:
\begin{eqnarray}
    \ddot{\Vec{r}}_{A_a} &=& \frac{Gm_{A_b}}{r_{A_{a}A_{b}}^2} \hat{r}_{A_{a}A_{b}} \label{eq1}\\
    \ddot{\Vec{r}}_{A_b} &=& \frac{Gm_{A_a}}{r_{A_{a}A_{b}}^2} \hat{r}_{A_{a}A_{b}} \label{eq2}   
\end{eqnarray}

Subtracting Equation~(\ref{eq1}) and (\ref{eq2}), we obtain that:
\begin{equation}
    \ddot{\Vec{r}}_{A_{a}A_{b}} = \ddot{\Vec{r}}_{A_b} - \ddot{\Vec{r}}_{A_a} = -\frac{G}{r_{A_{a}A_{b}}^2} (m_{A_a} + m_{A_b}) \hat{r}_{A_{a}A_{b}}.
    \label{eq3}
\end{equation}

This represents the movement of the secondary around the primary, which is what the astrometric measurement portrays.

\paragraph{Outer system:}
Repeating the procedure for the outer system, the movement of $B$ around (the $CoM$ of) $A$ (considered as a single object) is represented by: 
\begin{equation}
    \ddot{\Vec{r}}_{AB} = -\frac{G}{r_{AB}^2} (m_A + m_B) \hat{r}_{AB}.
    \label{eq5}
\end{equation}

However, as for the inner system, it is a matter of interest to obtain the movement of $B$ around the primary $A_a$, given that this is what what is usually measured in differential astrometry of visual binaries (see Figure~\ref{fig:tripleSystemVectors} for details). Therefore, considering Equation~(\ref{eq6}), the relationship between $A_a$ and the $CoM$ of $A$ Equation~(\ref{eq7}), and the position of the $CoM$ $A$ Equation~(\ref{eq8}), we can rewrite vector ${\Vec{r}}_{A_{a}A}$ as shown in Equation~(\ref{eq9}).

\begin{eqnarray}
    {\Vec{r}}_{A_{a}B} &=& {\Vec{r}}_{A_{a}A} + {\Vec{r}}_{AB}	\label{eq6} \\
    {\Vec{r}}_{A_{a}A} &=& {\Vec{r}}_{A} - {\Vec{r}}_{A_{a}} 		\label{eq7} \\
    {\Vec{r}}_{A} &=& \frac{m_{A_a} {\Vec{r}}_{A_{a}} + {m_{A_b} {\Vec{r}}_{A_{b}}}}{m_{A_a} + m_{A_b}}	\label{eq8} \\
    {\Vec{r}}_{A_{a}A} &=& \frac{m_{A_a} {\Vec{r}}_{A_{a}} + {m_{A_b} {\Vec{r}}_{A_{b}}}}{m_{A_a} + m_{A_b}} - {\Vec{r}}_{A_{a}} = ({\Vec{r}}_{A_{b}} - {\Vec{r}}_{A_{a}}) \frac{m_{A_b}}{m_{A_{a}} + m_{A_{b}}}	\label{eq9} 
\end{eqnarray}

If we define the mass ratio $q_{A_{a}A_{b}} = \frac{m_{A_b}}{m_{A_a}}$ and using that ${\Vec{r}}_{A_{a}A_{b}} = {\Vec{r}}_{A_b} - {\Vec{r}}_{A_a}$, Equation~(\ref{eq9}) can be rewritten as:
\begin{equation}
     {\Vec{r}}_{A_{a}A} =  {\Vec{r}}_{A_{a}A_{b}} \left( \frac{q_{A_{a}A_{b}}}{1 + q_{A_{a}A_{b}}} \right).
\end{equation}

Thereby, Equation~(\ref{eq6}) can be rewritten as:
\begin{equation}
    {\Vec{r}}_{A_{a}B} = {\Vec{r}}_{AB} + {\Vec{r}}_{A_{a}A_{b}} \left( \frac{q_{A_{a}A_{b}}}{1 + q_{A_{a}A_{b}}} \right).
    \label{eq11}
\end{equation}

Defining the wobble factor \citep{tokovinin2017} (also known as fractional mass \citep{heintz1978double}) as $f_{A_{a}A_{b}} = \frac{q_{A_{a}A_{b}}}{1 + q_{A_{a}A_{b}}}$, it is finally obtained that the movement of $B$ with respect to the primary $A_a$ is given by:
\begin{equation}
    {\Vec{r}}_{A_{a}B} = {\Vec{r}}_{AB} + f_{A_{a}A_{b}}\cdot {\Vec{r}}_{A_{a}A_{b}} .
    \label{eq12}
\end{equation}

It is important to note that Equation~(\ref{eq12}) clearly shows that if ${\Vec{r}}_{A_{a}A_{b}}$ and ${\Vec{r}}_{AB}$ satisfy Kepler's equations, the combined orbit ${\Vec{r}}_{A_{a}B}$ is not Keplerian and that, in particular, it is not a closed orbit. On the other hand, when the tight binary corresponds to $B$ and it is formed by $B_a, B_b$, it is easy to show that we obtain a negative wobble factor  \citep{lane2014orbits, tokovinin2017, tokovinin2018dancing}:
\begin{equation}
    {\Vec{r}}_{B_{a}A} = {\Vec{r}}_{BA} - f_{B_{a}B_{b}}\cdot {\Vec{r}}_{B_{a}B_{b}} 
    \label{eq13}
\end{equation}

\subsection{True Anomaly}
\label{app:trueanomaly}

Having obtained the orbital elements, the value of the relative position between stars can be computed at any given epoch  $\tau$. Let be $T$ the time of periastron passage, the epoch when the separation between primary and its companion reaches its minimum value. Then, Kepler's equation can be written as:
\begin{equation}
    M = \frac{2\pi(\tau - T)}{P} = E - e \sin(E),
\label{eq:keplerseq}    
\end{equation}

where the terms $M$ and $E$ are the mean anomaly and eccentric anomaly, respectively. As Equation~(\ref{eq:keplerseq}) has no analytic solution, it must be solved using numerical methods. Once $E$ is obtained, the true anomaly $\nu$ can be computed through:
\begin{equation} \label{eq:meantotrue}
    \tan \left( \frac{\nu}{2} \right) = \sqrt{\frac{1 + e}{ 1 - e}} \tan\left( \frac{E}{2}\right).
\end{equation}

The true anomaly corresponds to the angle between the main focus of the ellipse and the companion star, provided that the periastron is aligned with the X axis, and the primary star occupies the main focus of the ellipse. We note that the quadrant for $E$  and $\nu$  are the same, i.e., Equation~(\ref{eq:meantotrue}) allows to compute $\nu$ without any ambiguity.

\subsection{Cartesian Coordinates}
\label{app:as_eq}

\paragraph{Inner System}
Regarding the inner system, the movement of the secondary $A_b$ around the primary $A_a$ can be described in cartesian coordinates by:
\begin{equation}
\left[
\begin{matrix}
X_{A_{a}A_{b}}(\tau) \\
Y_{A_{a}A_{b}}(\tau)
\end{matrix}
\right]
=
\left[
\begin{matrix}
{r}_{A_{a}A_{b}}(\tau) \cdot \cos(\nu_{A_{a}A_{b}}(\tau)) \\
{r}_{A_{a}A_{b}}(\tau) \cdot \sin(\nu_{A_{a}A_{b}}(\tau))
\end{matrix}
\right].
\label{eq14}
\end{equation}

Then, to project the orbit in the plane of the sky, we make use of the Thiele-Innes constants $\{A_{A_{a}A_{b}}, B_{A_{a}A_{b}}, F_{A_{a}A_{b}}, G_{A_{a}A_{b}}\}$, which are a function of the orbital elements $\{  a_{A_{a}A_{b}}, \omega_{A_{a}A_{b}}, \Omega_{A_{a}A_{b}}, i_{A_{a}A_{b}} \}$:
\begin{equation}
\left[
\begin{matrix}
x_{A_{a}A_{b}}(\tau) \\
y_{A_{a}A_{b}}(\tau)
\end{matrix}
\right]
=
\left[
\begin{matrix}
A_{A_{a}A_{b}} X_{A_{a}A_{b}}(\tau)  + F_{A_{a}A_{b}} Y_{A_{a}A_{b}}(\tau)  \\
B_{A_{a}A_{b}} X_{A_{a}A_{b}}(\tau)  + G_{A_{a}A_{b}} Y_{A_{a}A_{b}}(\tau)
\end{matrix}
\right].
\label{eq15}
\end{equation}

\paragraph{Outer System}
On the other hand, concerning the outer system, the movement of the secondary $B$ around the \textit{fictional} primary $A$ can be described in Cartesian coordinates by:
\begin{equation}
\left[
\begin{matrix}
X_{AB}(\tau) \\
Y_{AB}(\tau)
\end{matrix}
\right]
=
\left[
\begin{matrix}
{r}_{AB}(\tau) \cdot \cos(\nu_{AB}(\tau)) \\
{r}_{AB}(\tau) \cdot \sin(\nu_{AB}(\tau))
\end{matrix}
\right].
\label{eq16}
\end{equation}

Then, to project the orbit in the plane of the sky, we make use of the Thiele-Innes constants $\{A_{AB}, B_{AB}, F_{AB}, G_{AB}\}$, which are a function of the orbital elements $\{ a_{AB}, \omega_{AB}, \Omega_{AB}, i_{AB} \}$:
\begin{equation}
\left[
\begin{matrix}
x_{AB}(\tau) \\
y_{AB}(\tau)
\end{matrix}
\right]
=
\left[
\begin{matrix}
A_{AB} X_{AB}(\tau) + F_{AB} Y_{AB}(\tau)  \\
B_{AB} X_{AB}(\tau) + G_{AB} Y_{AB}(\tau)
\end{matrix}
\right].
\label{eq17}
\end{equation}

Finally, we rewrite Equation~(\ref{eq12}) in Cartesian coordinates:
\begin{eqnarray}
{\Vec{r}}_{A_{a}B} &=& {\Vec{r}}_{AB} + f_{A_{a}A_{b}}\cdot {\Vec{r}}_{A_{a}A_{b}} \\
\label{eq18}
\left[
\begin{matrix}
X_{A_{a}B}(\tau) \\
Y_{A_{a}B}(\tau)
\end{matrix}
\right]
&=&
\left[
\begin{matrix}
X_{AB}(\tau)\\
Y_{AB}(\tau)
\end{matrix}
\right]
+ f_{A_{a}A_{b}} \cdot 
\left[
\begin{matrix}
X_{A_{a}A_{b}}(\tau)  \\
Y_{A_{a}A_{b}}(\tau)
\end{matrix}
\right].
\nonumber
\end{eqnarray}

As the projection in the plane of the sky can be represented as consecutive rotation matrices, it is a linear transformation. So, the projection of a weighted sum is the weighted sum of the projections. Then, the Cartesian coordinates of the position of $B$ with respect to $A_{a}$ can be written as:
\begin{equation}
\left[
\begin{matrix}
x_{A_{a}B}(\tau) \\
y_{A_{a}B}(\tau)
\end{matrix}
\right]
=
\left[
\begin{matrix}
x_{AB}(\tau) \\
y_{AB}(\tau)
\end{matrix}
\right]
+ f_{A_{a}A_{b}} \cdot 
\left[
\begin{matrix}
x_{A_{a}A_{b}}(\tau)  \\
y_{A_{a}A_{b}}(\tau)
\end{matrix}
\right].
\label{eq19}
\end{equation}

The procedure is analogous when the binary corresponds to $B$, and it is formed by $B_a, B_b$:
\begin{equation}
\left[
\begin{matrix}
x_{B_{a}A}(\tau) \\
y_{B_{a}A}(\tau)
\end{matrix}
\right]
=
\left[
\begin{matrix}
x_{AB}(\tau) \\
y_{AB}(\tau)
\end{matrix}
\right]
- f_{B_{a}B_{b}} \cdot 
\left[
\begin{matrix}
x_{B_{a}B_{b}}(\tau)  \\
y_{B_{a}B_{b}}(\tau)
\end{matrix}
\right].
\label{eq20}
\end{equation}

\subsection{Radial velocity} \label{app:rv_eq}

Conversely, it can be noted that the velocity of $A_b$ with respect to the $CoM$ $A$ can be written as:
\begin{equation}
    {u}_{A,A_{b}}(\tau) = {u}_{A_{b}}(\tau) - {u}_{A}(\tau).
\end{equation}

Considering that $A$ and $B$ are moving in an elliptic orbit around the $CoM$ of the system, then the the velocity of $A$ can be described as:
\begin{eqnarray}
    {u}_{A}(\tau) &=& {v}_{CoM} + K_3 \left( e_{AB}\cos(\omega_{AB}) + \cos(\omega_{AB} + \nu_{AB}(\tau) \right) \label{eq:triple31}\\
    {u}_{B}(\tau) &=& {v}_{CoM} - K_4 \left( e_{AB}\cos(\omega_{AB}) + \cos(\omega_{AB} + \nu_{AB}(\tau) \right)
    \label{eq:triple32}
\end{eqnarray}

If we introduce the mass ratio of the outer system $q_{AB} = \frac{m_B}{m_A}$, we can rewrite Equations \ref{eq:triple31} and \ref{eq:triple32} as:
\begin{eqnarray}
    {u}_{A}(\tau) &=& {v}_{CoM} + q_{AB}\cdot K_4 \left( e_{AB}\cos(\omega_{AB}) + \cos(\omega_{AB} + \nu_{AB}(\tau) \right)\label{eq:triple61}\\
    {u}_{B}(\tau) &=& {v}_{CoM} -   K_4 \left( e_{AB}\cos(\omega_{AB}) + \cos(\omega_{AB} + \nu_{AB}(\tau) \right) \label{eq:triple62}
\end{eqnarray}

Besides, $A_a$ and $A_b$ are also moving in an elliptic orbit around $A$, and therefore:
\begin{eqnarray}
    {u}_{A_{a}}(\tau) &=& {u}_{A} + K_1 \left( e_{A_{a}A_{b}} \cos(\omega_{A_{a}A_{b}}) + \cos(\omega_{A_{a}A_{b}} + \nu_{A_{a}A_{b}}(\tau) \right) \label{eq:triple41}\\
    {u}_{A_{b}}(\tau) &=& {u}_{A} - K_2 \left( e_{A_{a}A_{b}} \cos(\omega_{A_{a}A_{b}}) + \cos(\omega_{A_{a}A_{b}} + \nu_{A_{a}A_{b}}(\tau) \right) \label{eq:triple42}
\end{eqnarray}

If we now introduce the mass ratio $q_{A_{a}A_{b}}$, we can rewrite Equations \ref{eq:triple41} and \ref{eq:triple42} as:
\begin{eqnarray}
    {u}_{A_{a}}(\tau) &=& {u}_{A} + K_1 \left( e_{A_{a}A_{b}} \cos(\omega_{A_{a}A_{b}}) + \cos(\omega_{A_{a}A_{b}} + \nu_{A_{a}A_{b}}(\tau) \right) \label{eq:triple43}\\
    {u}_{A_{b}}(\tau) &=& {u}_{A} - \frac{K_1}{q_{A_{a}A_{b}}} \left( e_{A_{a}A_{b}} \cos(\omega_{A_{a}A_{b}}) + \cos(\omega_{A_{a}A_{b}} + \nu_{A_{a}A_{b}}(\tau) \right) \label{eq:triple44}
\end{eqnarray}

We choose to express the above Equations as a function of $K_1$ and $K_4$, because it is more likely to obtain measurements from the inner primary $A_a$ than from the inner secondary $A_b$. In addition, the object $A$ is fictional, so only $B$'s RV measurements could be available.

\subsection{Other relevant quantities}
\label{sec:relevant_quant}

There are some derived quantities that are relevant to compute after the orbital elements have been estimated, mainly, the individual stellar masses and the mutual inclination of the system. In the following subsections we derive explicit expressions for them, in terms of the orbital elements.

\subsubsection{Stellar masses}

To obtain the stellar masses for each one of the three bodies involved, we make use of relationships for binary systems, given the hierarchical approximation. For the inner system, given that
\begin{eqnarray}
    a_{A_{a}A_{b}} &=& a_{A_a} + a_{A_b}, \label{eq:another1}\\
    q_{A_{a}A_{b}} &=& \frac{a_{A_a}}{a_{A_b}} = \frac{m_{A_b}}{m_{A_a}}, \text{  and} \label{eq:another2}\\
    m_{A_a} + m_{A_b} &=& \frac{1}{\bar{\omega}^3} \cdot \frac{a_{A_{a}A_{b}}^3}{P_{A_{a}A_{b}}^2}, \label{eq:another3}
\end{eqnarray}

then, the individual masses correspond to:
\begin{eqnarray}
    m_{A_a} &=& \frac{a_{A_{a}A_{b}}^3}{\bar{\omega}^3 \cdot P_{A_{a}A_{b}}^2} \cdot \frac{1}{(1+q_{A_{a}A_{b}})}\\
    m_{A_b} &=& \frac{a_{A_{a}A_{b}}^3}{\bar{\omega}^3 \cdot P_{A_{a}A_{b}}^2} \cdot \frac{q_{A_{a}A_{b}}}{(1+q_{A_{a}A_{b}})}
\end{eqnarray}

The procedure is analogous for the outer system.

\subsubsection{Mutual inclination}

Given the orbital elements $i_{A_{a}A_{b}}$, $i_{AB}$, $\Omega_{A_{a}A_{b}}$ and $\Omega_{AB}$, the mutual inclination $\Phi$ can be obtained as \citep{muterspaugh2010phases, lane2014orbits}:
\begin{equation}
    \cos(\Phi) = \cos(i_{A_{a}A_{b}}) \cdot \cos(i_{AB}) + \sin{i_{A_{a}A_{b}}} \cdot \sin{i_{AB}} \cdot \cos(\Omega_{AB} - \Omega_{A_{a}A_{b}}).
\end{equation}

\section{Dimensionality Reduction}
\label{rvdimred}

\subsection{Preliminaries}

Here, we propose a dimensionality reduction in the \textit{radial-velocity-alone} scenario to reduce the state space from 20 to 15 dimensions. Even though it is just $25 \%$ reduction, in sample-based schemes like MCMC any decrease in the computational cost is appreciated. 

The reduction consists on the separation of the parameter vector into two lower dimension vectors: one containing non-linear components ($\theta_{NL}$) and the other components (that are linearly dependent with respect to $\theta_{NL}$) can be obtained through a weighted least-squares procedure ($\theta_L$). The method is inspired by \citep{wright2009efficient} where they reformulate the RV equations in such a way that they get a linear relationship with some parameters, allowing for an analytic calculation of weighted least-square solutions.  

Therefore, the search of the state space is focused on 
\begin{equation*}
    \theta_{NL} = [ T_{A_{a}A_{b}}, P_{A_{a}A_{b}}, e_{A_{a}A_{b}}, q_{A_{a}A_{b}}, i_{A_{a}A_{b}}, T_{AB}, P_{AB}, e_{AB}, q_{AB},  i_{AB}  ],
\end{equation*}
and then the vector of parameters $ \theta_L = [ a_{A_{a}A_{b}}, \omega_{A_{a}A_{b}}, a_{AB}, \omega_{AB}, v_{CoM} ]$ is obtained.

\begin{figure}[htp]
\centering
        \resizebox{0.9\textwidth}{!}{
                \begin{tikzpicture}
        \tikzstyle{main}=[circle, minimum size = 10mm, thick, draw =black!80, node distance = 16mm]
        \tikzstyle{connect}=[-latex, thick]
        \tikzstyle{box}=[rectangle, draw=black!80, minimum size = 10mm]
        
          \node[box] (z) [] { $T_{A_{a}A_{b}}, P_{A_{a}A_{b}}, e_{A_{a}A_{b}}, q_{A_{a}A_{b}},  i_{A_{a}A_{b}}, T_{AB}, P_{AB}, e_{AB}, q_{AB}, i_{AB}$};        
          \node[box] (z1) [right = of z] { $\{{z_3}(\tau_k)_{k=1}^{N_3}, {z_4}(\tau_k)_{k=1}^{N_4},  {z_5}(\tau_k)_{k=1}^{N_5}\}$}; 
          \node[box] (cp) [below = of z] {Weighted Least Squares};
		\node[box] (p1) [below = of cp] {$a_{A_{a}A_{b}}, \omega_{A_{a}A_{b}}, a_{AB}, \omega_{AB}, v_{CoM}$};          
		
          \path (z) edge [connect] (cp);
          \path (z1) edge [connect] (cp);
          \path (cp) edge [connect] (p1);          		         
                  
        \end{tikzpicture}
        }
    \caption{Dimensionality reduction in RV alone scenario.}
    \label{fig:dimensionalityreduction}
\end{figure}
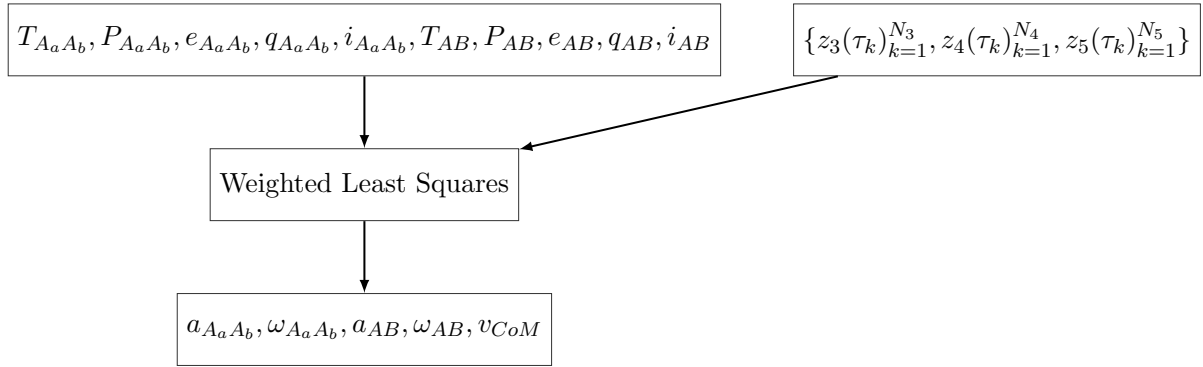

\subsection{Method} 

Following the procedure described in Appendix \ref{app:triple}, the RV equations can be formulated as \ref{eq:triple611}. This modeling does not consider linear trends $d(t-t_0)$, which are used to account for unmodeled noise sources and to notice the presence of massive objects in wide orbits around the star \citep{wright2009efficient, retired2010stars}.
\begin{eqnarray}
    {u}_{A}(\tau) &=& {v}_{CoM} + q_{AB}\cdot K_4 \left( e_{AB}\cos(\omega_{AB}) + \cos(\omega_{AB} + \nu_{AB}(\tau) \right)\label{eq:triple611}\\
    {u}_{B}(\tau) &=& {v}_{CoM} -   K_4 \left( e_{AB}\cos(\omega_{AB}) + \cos(\omega_{AB} + \nu_{AB}(\tau) \right)\\
    {u}_{A_{a}}(\tau) &=& {u}_{A} + K_1 \left( e_{A_{a}A_{b}} \cos(\omega_{A_{a}A_{b}}) + \cos(\omega_{A_{a}A_{b}} + \nu_{A_{a}A_{b}}(\tau) \right)\\
    {u}_{A_b}(\tau) &=& {u}_{A} - \frac{K_1}{q_{A_{a}A_{b}}} \left( e_{A_{a}A_{b}} \cos(\omega_{A_{a}A_{b}}) + \cos(\omega_{A_{a}A_{b}} + \nu_{A_{a}A_{b}}(\tau) \right) 
\end{eqnarray}

However, we must remember that the amplitudes $K_1$ and $K_4$ can be calculated as a function of the orbital elements $P_{A_{a}A_{b}}$, $e_{A_{a}A_{b}}$, $a_{A_{a}A_{b}}$, $i_{A_{a}A_{b}}$, $P_{AB}$, $e_{AB}$, $a_{AB}$ and $i_{AB}$ as follows:
\begin{eqnarray}
    K_1 &=& \frac{2\pi \sin(i_{A_{a}A_{b}})}{P_{A_{a}A_{b}} \sqrt{(1 - e_{A_{a}A_{b}}^2)}} \cdot \frac{a_{A_{a}A_{b}}''}{\bar{\omega}}\cdot\frac{q_{A_{a}A_{b}}}{1 + q_{A_{a}A_{b}}} \cdot \lambda\\
    K_4 &=& \frac{2\pi \sin(i_{AB})}{P_{AB} \sqrt{(1 - e_{AB}^2)}} \cdot \frac{a_{AB}''}{\bar{\omega}}\cdot\frac{1}{1 + q_{AB}} \cdot \lambda,
\end{eqnarray}

where $\lambda$ is a constant designed to convert from $\frac{arcsec}{yr}$ to $\frac{km}{s}$. Then, if we consider the variables $K_1'$, $K_4'$, $m_1$ and $m_2$ from Equations~(\ref{eq:K1p}) to~(\ref{eq:m2}) , the RV equations can be reformulated as shown in Equations~(\ref{eq:triple612}) to~(\ref{eq:triple615}).
\begin{eqnarray}
    K_1' &=& \frac{\sin(i_{A_{a}A_{b}})}{P_{A_{a}A_{b}} \sqrt{(1 - e_{A_{a}A_{b}}^2)}} \cdot\frac{q_{A_{a}A_{b}}}{1 + q_{A_{a}A_{b}}} \label{eq:K1p}\\
    K_4' &=& \frac{\sin(i_{AB})}{P_{AB} \sqrt{(1 - e_{AB}^2)}} \cdot\frac{1}{1 + q_{AB}} \label{eq:K2p} \\
    m_1 &=& 2\pi \cdot \frac{a_{A_{a}A_{b}}''}{\bar{\omega}} \cdot \lambda \label{eq:m1} \\
    m_2 &=& 2\pi \cdot \frac{a_{AB}''}{\bar{\omega}} \cdot \lambda \label{eq:m2}
\end{eqnarray}

\begin{eqnarray}
    {u}_{A}(\tau) &=& {v}_{CoM} + q_{AB}\cdot K_4'\cdot m_2 \cdot \left( e_{AB}\cos(\omega_{AB}) + \cos(\omega_{AB} + \nu_{AB}(\tau) \right)\label{eq:triple612}\\
    {u}_{B}(\tau) &=& {v}_{CoM} -   K_4'\cdot m_2 \cdot \left( e_{AB}\cos(\omega_{AB}) + \cos(\omega_{AB} + \nu_{AB}(\tau) \right) \label{eq:triple613}\\
    {u}_{A_{a}}(\tau) &=& {u}_{A} + K_1'\cdot m_1 \cdot \left( e_{A_{a}A_{b}} \cos(\omega_{A_{a}A_{b}}) + \cos(\omega_{A_{a}A_{b}} + \nu_{A_{a}A_{b}}(\tau) \right) \label{eq:triple614}\\
    {u}_{A_{b}}(\tau) &=& {u}_{A} - \frac{K_1'}{q_{A_{a}A_{b}}} \cdot m_1 \cdot \left( e_{A_{a}A_{b}} \cos(\omega_{A_{a}A_{b}}) + \cos(\omega_{A_{a}A_{b}} + \nu_{A_{a}A_{b}}(\tau) \right) \label{eq:triple615}
\end{eqnarray}

Finally, considering the following auxiliary variables $\alpha_1$, $\beta_1$, $\alpha_2$ and $\beta_2$:
\begin{eqnarray}
    \alpha_1 &=& m_1 \cdot \cos{(\omega_{A_{a}A_{b}}),}\\
    \beta_1 &=& -m_1 \cdot \sin{(\omega_{A_{a}A_{b}})},\\
    \alpha_2 &=& m_2 \cdot \cos{(\omega_{AB})}, and\\
    \beta_2 &=& -m_2 \cdot \sin{(\omega_{AB})},
\end{eqnarray}

we end up with:
\begin{eqnarray}
    \Vec{u}_{A}(\tau) &=& \Vec{v}_{CoM} + q_{AB}\cdot K_4'\cdot \alpha_2 \cdot \left( e_{AB} + \cos(\nu_{AB}) \right) + q_{AB}\cdot K_4'\cdot \alpha_2 \cdot \sin(\nu_{AB}(\tau)) \\
    \Vec{u}_{B}(\tau) &=& \Vec{v}_{CoM} -   K_4'\cdot \alpha_2 \cdot \left( e_{AB} + \cos(\nu_{AB})\right) - K_4'\cdot \beta_2 \cdot \cos(\nu_{AB}(\tau))\\
    \Vec{u}_{A_{a}}(\tau) &=& \Vec{u}_{A} + K_1'\cdot \alpha_1 \cdot \left( e_{A_{a}A_{b}} + \cos(\nu_{A_{a}A_{b}}) \right) + K_1'\cdot \beta_1 \cdot \sin(\nu_{A_{a}A_{b}}(\tau))\\
    \Vec{u}_{A_{b}}(\tau) &=& \Vec{u}_{A} - \frac{K_1'}{q_{A_{a}A_{b}}} \cdot \alpha_1 \cdot \left( e_{A_{a}A_{b}} + \cos(\nu_{A_{a}A_{b}}) \right) - \frac{K_1'}{q_{A_{a}A_{b}}} \cdot \beta_1 \cdot \sin(\nu_{A_{a}A_{b}}(\tau))  
\end{eqnarray}

The RV equations depicted above can be represented in a matrix form if we define the parameter's vector $\Vec{\theta} = [ \alpha_1, \beta_1, \alpha_2,  \beta_2, v_{CoM} ]^T$:
\begin{equation}
    u_{A_a}(\tau) = \Vec{\theta}^T \cdot \begin{bmatrix} 
    K_1' \cdot (\cos{(\nu_{A_{a}A_{b}}(\tau)) + e_{A_{a}A_{b}})} \\
    K_1' \cdot \sin{(\nu_{A_{a}A_{b}}(\tau))} \\
    q_{AB} \cdot K_4' \cdot (\cos{(\nu_{AB}(\tau))} + e_{AB}) \\
    q_{AB} \cdot K_4' \cdot \sin{(\nu_{AB}(\tau))} \\
    1  \end{bmatrix} = \Vec{\theta}^T \cdot F_{A_a}(\tau)
\end{equation}

\begin{equation}
    u_{A_b}(\tau) = \Vec{\theta}^T \cdot \begin{bmatrix} 
    -\frac{K_1'}{q_{A_{a}A_{b}}} \cdot (\cos{(\nu_{A_{a}A_{b}}(\tau)) + e_{A_{a}A_{b}})} \\
    -\frac{K_1'}{q_{A_{a}A_{b}}} \cdot \sin{(\nu_{A_{a}A_{b}}(\tau))} \\
    q_{AB} \cdot K_4' \cdot (\cos{(\nu_{AB}(\tau))} + e_{AB}) \\
    q_{AB} \cdot K_4' \cdot \sin{(\nu_{AB}(\tau))} \\
    1  \end{bmatrix} = \Vec{\theta}^T \cdot F_{A_a}(\tau)
\end{equation}

\begin{equation}
    u_B(\tau) = \Vec{\theta}^T \cdot \begin{bmatrix} 
    0 \\ 
    0 \\ 
    -K_4' \cdot (\cos{(\nu_{AB}(\tau))} + e_{AB}) \\
    -K_4' \cdot \sin{(\nu_{AB}(\tau))} \\
    1  \end{bmatrix} = \Vec{\theta}^T \cdot F_B(\tau)
\end{equation}

Then, considering a matrix $F$ with the matrices $F_{A_a}, F_{A_b}$ and $F_B$ for all the epochs of measurement for the three bodies involved $
F = [F_{A_a}(\tau_{r_{a}(0)}) \dotso F_{A_a}(\tau_{r_{a}(N_{a})}) | F_{A_b}(\tau_{r_{b}(0)}) \dotso F_{A_b}(\tau_{r_{b}(N_{b})}) | F_B(\tau_{r_{B}(0)}) \dotso F_B(\tau_{r_{B}(N_{B})})]$ and a vector with the modeled values for all the epochs of measurement for the three bodies involved $\Vec{u} = [u_{A_a}(\tau_{r_{a}(0)}) \dotso u_{A_a}(\tau_{r_{a}(N_{a})}), u_{A_b}(\tau_{r_{b}(0)}) \dotso u_{A_b}(\tau_{r_{b}(N_{b})}), u_B(\tau_{r_{B}(0)}) \dotso u_B(\tau_{r_{B}(N_{B})})]$, it can be written in the following compact form: $\Vec{u} = \Vec{\theta}^T \cdot F$.

Afterwards, the vector of parameters $\Vec{\theta}$ can be estimated from the data directly using least-squares and the figure of merit $\chi^2$ (Equation~(\ref{eq:chiRV})). If we define the matrix  $W$ as the diagonal with the weights associated to each observation $W_{kl} = \frac{\delta_{kl}}{\sigma_k^2}$, and $\Vec{v}$ a vector with all the observations from the three bodies concatenated, then the weighted least-squares is
\begin{equation}
    \chi^2 = \sum_{k=1}^{N_{a}} \frac{(v_k - u_{A_a}(\tau_{r_a(k)}))^2}{\sigma_{A_{a}}(\tau_{r_a(k)})^2} + \sum_{k=1}^{N_{b}} \frac{(v_k - u_{A_b}(\tau_{r_b(k)})^2}{\sigma_{A_{b}}(\tau_{r_b(k)})^2} + \sum_{k=1}^{N_{B}} \frac{(v_k - u_{B}(\tau_{r_B(k)})^2}{\sigma_{B}(\tau_{r_{B}(k)})^2} \label{eq:chiRV}\\
\end{equation}

and we find $\Vec{\theta}$ is the solution for
\begin{equation}
    \frac{\partial \chi^2}{\partial \Vec{\theta}} = -2(\Vec{v} - \Vec{\theta}^TF)WF^T = \Vec{0}.
\end{equation}

Consequently, the parameters' vector is obtained through $\Vec{\theta} = \Vec{v} W F^T (FWF^T)^{-1}$.

Finally, the original parameters can be recovered as follows:
\begin{eqnarray}
m_1 &=& \sqrt{\alpha_1^2 + \beta_1^2}, \\
m_2 &=& \sqrt{\alpha_2^2 + \beta_2^2}, \\
\omega_{A_aA_b} &=& \arctan{ \left( \frac{-\beta_1}{\alpha_1} \right)}, \\
\omega_{AB} &=& \arctan{ \left( \frac{-\beta_2}{\alpha_2} \right)}, \\
a_{A_aA_b}'' &=&  \frac{m_1 \cdot \bar{\omega}}{2\pi \lambda}, and \\
a_{AB}'' &=&  \frac{m_2 \cdot \bar{\omega}}{2\pi \lambda}.
\end{eqnarray}


\section{Computation of predictive models}
\label{app:mcmc}

All of the predictive distributions were numerically approximated using MCMC algorithms. Those sampling-schemes construct a Markov chain with $\Theta$ as state space and $\pi(\theta) = p(\theta | z)$ as the stationary distribution. They generate a sequence of parameter values $\theta_1, \theta_2 \cdots \theta_n$, by sampling from a proposal distribution and then accepting/rejecting the sample according to a criteria that depends on prior information and the likelihood. The resulting empirical distribution approaches the target distribution in the limit of long runs. 

Following the Bayes rule, $\pi$ can be written as:
\begin{equation}
\pi(\theta) = p(\theta | z) = \frac{p(z | \theta)p(\theta)}{\int p(z | \theta)p(\theta) d\theta}
\end{equation}

However, it is not necessary to compute the denominator, as MCMC algorithms base the acceptance/rejection of a sample $\theta'$ based on the posterior ratio:
\begin{equation}
\frac{\pi(\theta')}{\pi(\theta')} = \frac{p(z | \theta')p(\theta')}{p(z | \theta)p(\theta)}
\end{equation}

Due to the highly dimensional state space, we are working with an hybrid Gibbs sampler MCMC variant, which allows us to draw iteratively samples from the conditional posterior distribution for each variable given the remaining ones using one iteration. The state space is explored using a random walk with Gaussian proposal distributions. Besides, due to physical restrictions on the parameters we have a constrained state space, thus it is assumed to be circular for each dimension, so we avoid wasting several iterations on the boundaries. 

On the other hand, given that the observation model includes Gaussian additive noise (see Equation~(\ref{eq:obsmodel})), all likelihoods are proportional to $\exp(-\frac{1}{2}\chi^2)$, i.e,
\begin{equation}
	\mathcal{L}(\theta) = p(z | \theta)  \propto \exp{\left( -\frac{1}{2} \sum_{k=1}^{N} \frac{{\left\lVert z_k - f(\theta, \tau_k)  \right\rVert}^2}{\sigma_k^2}  \right)}  
\end{equation}

Finally, it is worth mentioning that as $T$ can be replaced by $T \pm nP$ ($n$ could be any integer), we choose $T \in (0, P)$. Then, we define the variable $T' = \frac{T}{P} \in (0,1)$ and we work with it along the estimations. After finishing, we return to the old variable $T$.

Here below, we will explain the details of each one of the scenarios presented in Section \ref{sec:orbitscalc}. 

\subsection{Astrometry alone}

As seen in Figure~\ref{fig:inference_process_as}, two processes are run consequently. First, we use the astrometric observations from the inner system $\{\vec{z_1}\}_{k=1}^{N_1}$ to estimate the parameters set $\Theta_1 = \{T_{A_{a}A_{b}}, P_{A_{a}A_{b}}, e_{A_{a}A_{b}}, a_{A_{a}A_{b}}, \omega_{A_{a}A_{b}}, \Omega_{A_{a}A_{b}}, i_{A_{a}A_{b}}\}$; however, a dimensional reduction is performed. For each iteration of the algorithm, the parameters $\theta_{NL} = \{T_{A_{a}A_{b}}, P_{A_{a}A_{b}}, e_{A_{a}A_{b}}\}$ are left free and the parameters $\theta_L = \{a_{A_{a}A_{b}}, \omega_{A_{a}A_{b}}, \Omega_{A_{a}A_{b}}, i_{A_{a}A_{b}}\}$ are obtained using a weighted least-squares procedure. The process is explained in detail in \citet{mendez2017orbits}.

Then, we use the astrometric observations from the outer system to estimate the parameters set $\Theta_2~=~\{q_{A_{a}A_{b}}, T_{AB}, P_{AB}, e_{AB}, a_{AB}, \omega_{AB}, \Omega_{AB}, i_{AB}\}$ employing an imputations framework within MCMC (see \citet{claveria2019visual} for more details). Here, for each iteration of the algorithm:
\begin{enumerate}
\item We sample $\Theta_2$ using the proposal distribution, and obtain $\{\vec{y_2} = f_1(T_{AB}, P_{AB}, e_{AB}, a_{AB}, \omega_{AB}, \Omega_{AB}, i_{AB}, \tau_k)  \}_{k=1}^{N_2}$.
\item We sample from the distribution $P_{\Theta_1 | \vec{Z_1}}(\cdot | \vec{z_1})$, obtained in the last step, and generate $\{\vec{y_1}~=~\frac{q_{A_{a}A_{b}}}{1 + q_{A_{a}A_{b}}}~\cdot~f_1(T_{A_{a}A_{b}}, P_{A_{a}A_{b}}, e_{A_{a}A_{b}}, a_{A_{a}A_{b}}, \omega_{A_{a}A_{b}}, \Omega_{A_{a}A_{b}}, i_{A_{a}A_{b}}, \tau_k)  \}_{k=1}^{N_2}$.
\item We compute $\vec{z} = \vec{y_2} + \vec{y_1}$ and use it as observation to continue with the algorithm.
\end{enumerate} 
Later, we add a physical restrictions step. Besides checking the support for each parameter, it is necessary to check:
\begin{itemize}
\item The hierarchical approximations in periods ($P_{A_{a}A_{b}} < P_{AB}$) and semi-major axes ($a_{A_{a}A_{b}} <  a_{AB}$) are valid.
\item The sum of masses has sense: $m_{A_a} + m_{A_b} < m_A + m_B \Leftrightarrow \frac{a_{A_{a}A_{b}}^3}{P_{A_{a}A_{b}}^2} < \frac{a_{AB}^3}{P_{AB}^2}  $.
\end{itemize}
This process is shown in detail in Algorithm~\ref{alg:as1}.  

\subsection{Radial velocities alone}

Unlike the last scenario, all observations are used at once, achieving the estimation of the parameter set $\Theta_{RV} = \{T_{A_{a}A_{b}}, P_{A_{a}A_{b}}, e_{A_{a}A_{b}}, a_{A_{a}A_{b}}, \omega_{A_{a}A_{b}},  i_{A_{a}A_{b}}, q_{A_{a}A_{b}}, T_{AB}, P_{AB}, e_{AB}, a_{AB}, \omega_{AB}, i_{AB}, q_{AB}, v_{CoM}\}$ in just one process, as seen in Figure~\ref{fig:inference_process_rv}. Nonetheless, since the dimension of the parameter space rises to fifteen, a dimensionality reduction is proposed. 

For each iteration of the algorithm, the parameters $\theta_{NL} = \{T_{A_{a}A_{b}}, P_{A_{a}A_{b}}, q_{A_{a}A_{b}},  i_{A_{a}A_{b}}, T_{AB}, P_{AB}, e_{AB}, q_{AB}, i_{AB}\}$ are left free and the parameters $\theta_L = \{ a_{A_{a}A_{b}}, \omega_{A_{a}A_{b}}, a_{AB}, \omega_{AB}, v_{CoM} \}$ are obtained using a weighted least-squares procedure, similar to the processes described in \citep{wright2009efficient, mendez2017orbits}, and explained in detail in Appendix \ref{rvdimred}. This allows us to sample from a reduced parameter space $\theta_{NL}$, while linearly deriving the rest of the parameters $\theta_L$. 

Later, we add a physical restrictions step. Besides checking the support for each parameter, it is necessary to check:
\begin{itemize}
\item The hierarchical approximation in periods $P_{A_{a}A_{b}} < P_{AB}$ and semi-major axes $a_{A_{a}A_{b}} < a_{AB}$ is valid.
\item The sum of masses has sense: $m_A = m_{A_a} + m_{A_b} \Leftrightarrow |\frac{a_{AB}^3}{P_{AB}^2} \cdot \frac{1}{(1+q_{AB})} - \frac{a_{A_{a}A_{b}}^3}{P_{A_{a}A_{b}}^2}| < \epsilon$.
\end{itemize}

The algorithm can be seen in detail in Algorithm~\ref{alg:rv}.

\section{Algorithms for parameter estimation}

At last, here we show the pseudocode of the MCMC algorithms mentioned in Section~\ref{sec:orbitscalc}:
\begin{itemize}
\item Algorithm~(\ref{alg:as1}) shows the MCMC-based method to perform the estimation of parameters $\{T_{A_{a}A_{b}}, P_{A_{a}A_{b}}, e_{A_{a}A_{b}}, a_{A_{a}A_{b}}, \omega_{A_{a}A_{b}}, \Omega_{A_{a}A_{b}}, i_{A_{a}A_{b}} \}$.
\item Algorithm~(\ref{alg:as2}) shows the MCMC and imputations-based framework to perform the estimation of parameters $\{q_{A_{a}A_{b}}, T_{AB}, P_{AB}, e_{AB}, a_{AB}, \omega_{AB}, \Omega_{AB}, i_{AB}\}$.
\item Algorithm~(\ref{alg:rv}) shows the MCMC-based method to perform the parameter estimation of orbital elements $\{T_{A_{a}A_{b}}, P_{A_{a}A_{b}}, e_{A_{a}A_{b}}, a_{A_{a}A_{b}}, \omega_{A_{a}A_{b}}, i_{A_{a}A_{b}}, q_{A_{a}A_{b}}, T_{AB}, P_{AB}, e_{AB}, a_{AB}, \omega_{AB}, i_{AB}\, q_{AB}, v_{CoM}\}$.
\end{itemize}  

\begin{algorithm}[H]
\caption{MCMC - Parameter estimation for inner parameters (astrometric alone scenario).}
\label{alg:as1}
\begin{algorithmic}[1]
\STATE $\theta_{NL} = \{T_{A_{a}A_{b}}, P_{A_{a}A_{b}}, e_{A_{a}A_{b}} \}$
\STATE $\theta_L = \{ a_{A_{a}A_{b}}, \omega_{A_{a}A_{b}}, \Omega_{A_{a}A_{b}}, i_{A_{a}A_{b}} \}$
\STATE Initialize $\theta^{(0)}$ sampling from priors
\FOR{$k=1 \dotso N_{steps}$}
    \STATE $\theta' = \theta^{(k)}$
    \FOR{$j=1 \dotso 3$ (All the non linear parameters)}  
        \STATE Sample $\theta_{j}' \sim \mathcal{N}(\theta_{j}', \sigma_j^2)$ (Apply additive gaussian perturbation on component $j$)
        \STATE Compute $\theta_L$
        \STATE Compute $\mathcal{L}(\theta')$
        \STATE Sample $u' \sim \mathcal{U}(0,1)$
        \STATE Compute acceptance ratio $\frac{\mathcal{L}(\theta')}{\mathcal{L}(\theta^{(k)})} $
            \IF{$u <$ ratio} 
                \STATE $\theta_j^{(k+1)} = \theta_{j}'$ 
            \ELSE
                \STATE $\theta_j^{(k+1)} =\theta_j^{(k)}$
            \ENDIF
    \ENDFOR
\ENDFOR
\end{algorithmic}
\end{algorithm}

\begin{algorithm}[H]
\caption{MCMC and Imputations - Parameter estimation for outer  orbital elements (astrometric alone scenario).}
\label{alg:as2}
\begin{algorithmic}[1]
\STATE $\theta_{1} = \{T_{A_{a}A_{b}}, P_{A_{a}A_{b}}, e_{A_{a}A_{b}}, a_{A_{a}A_{b}}, \omega_{A_{a}A_{b}}, \Omega_{A_{a}A_{b}}, i_{A_{a}A_{b}}\}$
\STATE $\theta_{2} = \{q_{A_{a}A_{b}}, T_{AB}, P_{AB}, e_{AB}, a_{AB}, \omega_{AB}, \Omega_{AB}, i_{AB}\}$
\STATE Initialize $\theta_{2}^{(0)}$ sampling from priors
\FOR{$k=1 \dotso N_{steps}$}
    \STATE {Generate imputations $\{\vec{y_1}\}_{k=1}^{N_2}$, using $P_{\Theta_1 | \vec{Z_1}}(\cdot | \vec{z_1})$.}
    \STATE $\theta_{2}' = \theta_{o}^{(k)}$
    \FOR{$j=1 \dotso 8$}
        \STATE Sample $\theta_{2_j}' \sim \mathcal{N}(\theta_{2_j}', \sigma_j^2)$ (Apply additive gaussian perturbation on component $j$)
        \STATE Compute physical restrictions         
        \STATE Compute $\Vec{y}_{2}(\theta_{2}', \tau)$ and $\mathcal{L}(\theta_2')$
        \STATE Sample $u' \sim \mathcal{U}(0,1)$
        \STATE Compute acceptance ratio $\frac{\mathcal{L}(\theta')}{\mathcal{L}(\theta^{(k)})} $
            \IF{$u <$ ratio} 
                \STATE $\theta_{2_j}^{(k+1)} = \theta_{2_j}'$
            \ELSE
                \STATE $\theta_{2_j}^{(k+1)} =\theta_{2_j}^{(k)}$
            \ENDIF
    \ENDFOR
\ENDFOR
\end{algorithmic}
\end{algorithm}

\begin{algorithm}[H]
\caption{MCMC - Parameter estimation in the RV alone scenario.}
\label{alg:rv}
\begin{algorithmic}[1]
\STATE $\theta_{NL} = \{T_{A_{a}A_{b}}, P_{A_{a}A_{b}}, e_{A_{a}A_{b}}, q_{A_{a}A_{b}}, i_{A_{a}A_{b}}, T_{AB}, P_{AB}, e_{AB}, q_{AB}, i_{AB}\}$
\STATE $\theta_L = \{ a_{A_{a}A_{b}}, \omega_{A_{a}A_{b}}, a_{AB}, \omega_{AB}, v_{CoM} \}$
\STATE Initialize $\theta^{(0)}$ sampling from priors
\FOR{$k=1 \dotso N_{steps}$}
    \STATE $\theta' = \theta^{(k)}$
    \FOR{$j=1 \dotso 10$ (All the non linear parameters)}
        \STATE Sample $\theta_{NL_j}' \sim \mathcal{N}(\theta_{NL_j}', \sigma_j^2)$ (Apply additive gaussian perturbation on component $j$)
        \STATE Compute $\theta_L$
        \STATE Compute physical restrictions 
        \STATE Compute $\mathcal{L}(\theta')$
        \STATE Sample $u' \sim \mathcal{U}(0,1)$
        \STATE Compute acceptance ratio $\frac{\mathcal{L}(\theta')}{\mathcal{L}(\theta^{(k)})} $:
            \IF{$u <$ ratio} 
                \STATE $\theta^{(k+1)} = \theta_{j}'$ and save $\theta'$
            \ELSE
                \STATE $\theta^{(k+1)} =\theta^{(k)}$
            \ENDIF
    \ENDFOR
\ENDFOR
\end{algorithmic}
\end{algorithm}

\bibliography{main} 
\bibliographystyle{aasjournal}

\facilities{CTIO: SOAR 4.0m}

\end{document}